%% file: ms.tex
\documentclass[longbibliography,twocolumn,preprintnumbers,amsmath,amssymb,rmp,floatfix]{revtex4-1}

\usepackage{natbib}

\usepackage{graphicx}
\usepackage{dcolumn}
\usepackage{bm}
\usepackage{hyperref}
\hypersetup{backref=true, pagebackref=true, hyperindex=true, breaklinks=true,colorlinks=true,urlcolor=blue, linkcolor=black,  citecolor=blue,pagecolor=red, bookmarks=true, bookmarksopen=true}

\newcommand{\myeq}{\overset{\mathrm{D_s=D_l}}{=\joinrel=}}

\newcommand{\dmu}{\ {\rm cm^{-3} \ pc}}

\newcommand{\aj}{Astron. J.}
\newcommand{\apjl}{Astrophys. J.}
\newcommand{\apjs}{Astrophys. J.}
\newcommand{\aap}{Astron. Astrophys.}
\newcommand{\mnras}{Mon. Not. R. Astron. Soc.}
\newcommand{\araa}{Ann. Rev. Astron. Astrophys.}

\newcommand{\pasj}{Pubs. Astron. Soc. Japan}

\newcommand{\sovast}{Soviet Astronomy}
\newcommand{\aapr}{Astronomy and Astrophysics Reviews}

\newcommand{\physrep}{Phys. Rep.}
\newcommand{\solphys}{Sol. Phys.}
\newcommand{\jcap}{J. Cos. AstroPart. Phys.}
\newcommand{\aplett}{Astro. Phys. Lett.}

\include{jdefs}

\begin{document}

\title{THE PHYSICS OF FAST RADIO BURSTS}

\author{Bing Zhang}
\affiliation{Nevada Center for Astrophysics and Department of Physics and Astronomy, University of Nevada Las Vegas, Nevada 89154 USA}

\date{published 25 September 2023}

\begin{abstract}
Fast radio bursts (FRBs),  millisecond-duration bursts prevailing in the radio sky, are the latest big puzzle in the universe and have been a subject of intense observational and theoretical investigations in recent years. The rapid accumulation of the observational data has painted the following sketch about the physical origin of FRBs: They predominantly originate from cosmological distances so that their sources produce the most extreme coherent radio emission in the universe; at least some, probably most, FRBs are repeating sources that do not invoke cataclysmic events; and at least some FRBs are produced by magnetars, neutron stars with the strongest magnetic fields in the universe. Many open questions regarding the physical origin(s) and mechanism(s) of FRBs remain. This article reviews the phenomenology and possible underlying physics of FRBs. Topics include: a summary of the observational data, basic plasma physics, general constraints on FRB models from the data, radiation mechanisms, source and environment models, propagation effects, as well as FRBs as cosmological probes. Current pressing problems and future prospects are also discussed.
\end{abstract}

\maketitle

\tableofcontents


\section{Introduction}

Fast radio bursts (FRBs), milliseconds-duration radio bursts predominantly originating from cosmological distances, are one of the few remaining unsolved puzzles in contemporary astrophysics. The study of these mysterious events has a relatively short history. The first reported FRB was detected on July 24th, 2001 (now called FRB 20010724, for FRB naming conventions, see \S\ref{sec:naming}), by the Parkes 64-m telescope in Australia. It was not discovered until later by Duncan Lorimer and collaborators during an archival search for burst-like events. The burst was located 3$^{\rm o}$ from the Small Magellanic Cloud (SMC), had a peak flux density $S_\nu \gtrsim 30$ Jy at $\sim$1.4 GHz, a duration (also called ``width'', see \S\ref{sec:temporal}) $W \sim 5$ ms, and a dispersion measure (DM, see definition is \S\ref{sec:DM}, which is a proxy of distance from the source to Earth) $\sim 375 \dmu$, which is  in great excess of the value expected from Milky Way or SMC, suggesting that it likely originated from a cosmological distance. The discovery was published in 2007 in the journal Science \cite{lorimer07}, so  2007 was widely regarded as the birth time of the FRB research field. Note that there was an unconfirmed report about some repeating bursts from the nearby galaxy M87 back in 1980 \citep{linscott80} that were regarded by most as radio frequency interferences (RFIs). However, the inferred energy ($\sim 10^{40} \ {\rm erg}$) and luminosity ($\sim 10^{37} \ {\rm erg \ s^{-1}}$) of those bursts fall into the range of typical known FRBs. If confirmed, those bursts could be the earliest detected repeating FRB bursts.

Like the studies of other cosmological puzzles (the closest analogy being gamma-ray bursts, GRBs), the study of FRBs went through several phases from not being sure about whether they are even genuinely astronomical to getting to the bottom of the emitting source(s) and physical mechanisms. While it took half a century (from 1967 to 2017) to solve the full puzzle of GRBs, the pace of studying FRBs is much faster. In a merely fifteen-year period, observations have led to the answers or partial-answers to the following four questions: 1. Are they astronomical? 2. Are there multiple types? 3. Where are they? 4. What make(s) them?

Fully addressing the first question took 5-8 years. After the detection of the ``Lorimer burst'', no similar events were detected until several years later. On the other hand, there were many somewhat similar events that were detected by the Parkes telescope that appeared artificial. These so-called ``perytons''  \cite{burke-spolaor11} differed from genuine FRBs by being detected by all 13 beams of the Parkes telescope and clustering in time. Their existence cast doubt on the astronomical origin of the ``Lorimer burst'' itself. In 2012, \citet{keane12} reported another highly dispersed burst-like event (later termed as FRB 20010621A) with $S_\nu \sim 400$ mJy at $\sim$1.4 GHz, $W \sim 7.8$ ms, and DM $\sim 746 \dmu$. Since the burst was close to the Galactic plane, the excess DM is not significant. The possibility that the burst was a giant pulse of an underlying pulsar or from a Rotating RAdio Transient (RRAT) -- a type of part-time pulsars \cite{mclaughlin06} -- was not ruled out. 
A strong support to the existence of extragalactic/cosmological FRBs was established the next year, when \citet{thornton13} reported four more FRBs discovered by the Parkes telescope. It was shown that all the events were detected in one or a few beams of the telescope, different from the perytons. They were from high Galactic latitudes, had large DM values in great excess of the MW values in those directions, similar to the Lorimer burst. \citet{thornton13} also estimated that the event rate of FRBs is very high, about $10^4$ per day all sky above $\sim 3 \ {\rm Jy \ ms}$ fluence density threshold at 1.4 GHz. 
Finally, the ``perytons'' were eventually identified as artificial signals caused during the magnetron shut-down phase of a microwave oven when a person impatiently opens the oven before heating is over \cite{petroff15c}. Since those seemingly genuine bursts all happened not during the dining time when perytons were generated, this development finally separated perytons from true FRBs and suggested that FRBs are indeed of an astronomical origin.

After the initial detection of the Lorimer burst, the source direction was intensively monitored for 90 additional hours but no detection of repeated bursts was made \citep{lorimer07}. Later detected FRBs were all one-off events until 2016 when \citet{spitler16} first reported that one FRB source, named FRB 20121102A (also called FRB 121102, ``R1'', or ``Spitler burst''), emitted repeated bursts with a similar DM as detected by the Arecibo 305-m radio telescope. This source remained the sole detected repeater for not long before the Canadian Hydrogen Intensity Mapping Experiment (CHIME) discovered a few more repeating sources \cite{chime-2ndrepeater,chime-repeaters}. More repeaters were discovered through deep monitoring with the the Australian Square Kilometre Array Pathfinder (ASKAP) \cite{kumar19} and the Five-hundred-meter Aperture Spherical radio Telescope (FAST) in China \cite{luo20b,niu22}. On the other hand, most detected FRBs are still one-off. So, at least observationally one can say that there are two apparent types: repeaters and non-repeaters, but it is unclear whether all non-repeaters will eventually repeat. 

The repeating nature of FRB 20121102A allowed targeted observations using the Karl G. Jansky Very Large Array (VLA) and the Arecibo telescope to detect additional bursts and  eventually localize the source using the interferometric technique \cite{chatterjee17}. This enabled the detection of a compact persistent radio source in association with the burst source \cite{chatterjee17}. Further very-long-baseline  radio interferometric observations using the European VLBI Network and the Arecibo telescope refined the persistent radio source to milliarcsecond scale, which corresponds to $\leq 70$ pc at the source \cite{marcote17}. It also led to direct identification of the source host galaxy in the optical band, which is a dwarf star-forming galaxy at redshift $z=0.19$ \cite{tendulkar17}. This finally answered the ``where'' question and established the cosmological origin of FRBs. Localizations of FRBs, both repeaters and non-repeaters, have been later made via interferometry by the ASKAP collaboration, Deep Synotic Array (DSA) collaboration, and several other groups, which revealed a gallery of host galaxy types and positions of the FRBs within the hosts \citep{bannister19,ravi19,prochaska19,marcote20,macquart20,bhandari22,xuh22} and the confirmation of the theoretically expected ${\rm DM_{IGM}}-z$ correlation \cite{macquart20}. 

The question ``What make(s) them?'' is the most difficult to answer. Shortly after the reports of the discovery of the first FRBs, especially the four more FRBs reported by \citet{thornton13},  dozens of theoretical models were proposed \citep[e.g.][for a summary]{platts19}. The bright persistent radio source \cite{chatterjee17}, the actively star forming host galaxy \cite{tendulkar17}, as well as an extremely large Faraday rotation measure (RM, which is a proxy of the strength of magnetic field and density near the FRB source, see \S\ref{sec:RM} for definition) of FRB 20121102A \cite{michilli18} suggested that young magnetars might be sources of active repeaters. Even though a twin-source, FRB 20190520B, was later detected by FAST \cite{niu22}, most other sources, including both repeating and non-repeating FRBs, display diverse emission and host galaxy properties that are inconsistent with such a simple picture.  

A definite clue on the magnetar origin of at least some FRBs came from the detection of the Galactic FRB 20200428. 
The identification of cosmological origin of FRBs suggests that if an FRB would occur in the Milky Way Galaxy, it should be extremely bright. This expectation was realized on April 28th, 2020, when an extremely high fluence, FRB-like event with two pulses was detected by CHIME \cite{CHIME-SGR} and the Survey for Transient Astronomical Radio Emission 2 (STARE2) \cite{STARE2-SGR}, which only detected one of the two pulses. The radio burst was associated with a hard X-ray burst (XRB) from a Galactic magnetar named Soft Gamma-ray Repeater (SGR) J1935+2154 during one of its active phases \cite{HXMT-SGR,Konus-SGR,Integral-SGR,AGILE-SGR}. This established a long-speculated connection between FRBs and magnetars. Deep monitoring of the magnetar by FAST, on the other hand, suggests that the majority of X-ray bursts emitted by the magnetar are actually {\em not} associated with FRBs \cite{lin20}, suggesting the rarity of the magnetar FRB-XRB associations. Deeper monitoring by FAST and European radio telescopes discovered fainter radio pulses from this source \cite{zhangcf20,kirsten21}.

Despite this breakthrough discovery, the mystery of cosmological FRBs still remains. Some recent discoveries pose more clues and in the meantime more confusion to the big picture. 
An apparent $\sim 16$ day periodicity of a repeating source, FRB 20180916B (also called FRB 180916.J0158+65), was reported from the CHIME observations \cite{chime-periodic}. Follow-up observations suggest that the active window is ``chromatic'', with bursts detected in higher frequencies appearing at somewhat earlier phases than those detected in lower frequencies \cite{paster-marazuela21,pleunis21}. A tentative $\sim 157$-day period was also suggested for FRB 20121102A \cite{rajwade20}. Bursting activities during the active windows are actually very sporadic. For FRB 20121102A, $>1600$ bursts were detected by FAST in a total of 59.5 observing hours spanning 47 days during one active window \cite{lid21}, but there were no active bursts detected during some projected active windows later. 


A repeating source FRB 20200120E discovered by the CHIME FRB collaboration was found to be associated with a nearby spiral galaxy M81 at a distance of 3.6 Mpc \citep{bhardwaj21}. Follow-up observations surprisingly localized the source to a globular cluster in the host galaxy \citep{kirsten22}. The bursts from the source have lower luminosities than typical cosmological FRBs. Some bursts have rapid temporal structures as short as 60 nanoseconds \citep{nimmo21}. 

The FAST-detected repeating FRB source, FRB 20190520B \citep{niu22}, besides showing similar properties as FRB 20121102A, also showed some unique properties. For example, its very large RM showed an extreme sign change in a month timescale \citep{anna-thomas22,dais22}. Being located at $z=0.241\pm 0.001$, its estimated host-contribution of DM exceeds $\sim 1000 \ {\rm pc \ cm^{-2}}$, which is the largest among known FRBs \cite{niu22}. 

The polarization properties of FRBs have been studied closely over the years, which bring clues in understanding FRB sources, environments, and radiation mechanisms. Evidence of a large rotation measure (RM $ \simeq186 \ {\rm rad \ m^{-2}}$) in excess of the Galactic value was first reported for FRB 20110523A, which suggested a dense magnetized  plasma associated with the FRB \cite{masui15}. 
More extreme values (of the order $10^5 {\rm rad \ m^{-2}}$) were detected from FRB 20121102A \cite{michilli18} and  FRB 20190520B \citep{anna-thomas22,dais22}. FRB 20121102A showed an essentially non-varying polarization angle across each burst during individual bursts \cite{michilli18}. An opposite case was observed in another active repeating source FRB 20180301A, which showed diverse polarization angle swings among different bursts \cite{luo20b}. 
Intense follow-up observations of the CHIME-discovered repeating source FRB 20201124A by the FAST telescope \citep{xuh22} revealed peculiar short-term polarization property variations, including un-predictable RM evolution and non-evolution and oscillations of circular and linear polarization degrees and linear polarization angles as a function of wavelength in a small fraction of bursts. Significant circular polarization was discovered from the source \citep{kumar22,xuh22}. Extreme RM variations, including a reversal of RM \citep{anna-thomas22,dais22}, was observed from FRB 20190529B. All these suggest a dynamically evolving magnetized environment around repeating FRB sources. Frequency-dependent polarization degree was noticed in a sample of repeating FRBs, which may be interpreted as a scatter of RM due to the multi-path propagation effect of radio emission \citep{feng22}.

One special source detected by CHIME, FRB 20191221A, was identified to show a 216.8(1) ms periodicity with a significance of $6.5\sigma$ \cite{chime-period}. It has a roughly 3 s long duration, making it an outlier in the FRB population. However, this periodicity offers a strong support to a magnetar (or pulsar) origin of this special event.

With the rapid accumulation of observational data, the physical understanding of FRBs also enjoyed a steady advancement in recent years, from knowing essentially nothing to painting a rough sketch of the FRB production mechanism. Similar to the field of gamma-ray bursts \cite{nemiroff94}, the early years of the FRB study also witnessed a large number of theoretical papers dedicated to guessing the origin of FRBs based on very limited observational data \cite{platts19}. Not surprisingly, most of these ideas are quickly disfavored or completely rejected as data are accumulated. Rather than surveying all the proposed models (such a task has been carried out, see \citet{platts19} and an online FRB theory Wiki page\footnote{ https://frbtheorycat.org.}), this article focuses on a critical assessment of the leading ideas of interpreting FRBs that currently under active investigations. 

In the following, I will discuss the topics related to the physical nature of FRBs. I will first concisely summarize observational facts in \S\ref{sec:data} for the preparation of later discussion and refer the readers to more comprehensive observational reviews  \cite{petroff19,cordes19,petroff22,bailes22} and references therein\footnote{On the other hand, this review includes the most updated observational progress not included in the previous reviews.}. After reviewing the basic plasma physics relevant to the FRB mechanisms (\S\ref{sec:basic}), I will discuss some generic theoretical arguments that pose constraints on any FRB models (\S\ref{sec:general}). The next section (\S\ref{sec:radiation}) discusses possible mechanisms for generating the extremely coherent radiation of FRBs, with two general types of models (magnetospheric vs. relativistic shock models) discussed and compared in detail. This is followed by a survey of the source models (\S\ref{sec:source}) for repeating FRBs and some attractive ideas of generating genuinely non-repeating FRBs. The environmental models of FRBs are discussed un \S\ref{sec:environment} and the propagation effects of FRBs are discussed in \S\ref{sec:propagation}. FRBs as various cosmological probes are summarized in \S\ref{sec:probes}. The review ends with a discussion of the problems and prospects in the field in \S\ref{sec:prospects}. Early theory reviews on the surveys of many theoretical models can be found in \citet{katz18,popov18,platts19}. Concise theory reviews on the physical mechanisms of FRBs can be found in \citet{zhang20b}, \citet{lyubarsky21} and \citet{xiao21}. 

It is worth noting that the FRB field is a rapidly evolving field. For the topics discussed in this reivew, I have tried to separate the parts that involve robust physics (\S\S\ref{sec:basic}, \ref{sec:general} and \ref{sec:propagation}) from those that are undergoing intense investigations (\S\S\ref{sec:data}, \ref{sec:radiation}, \ref{sec:source}, and \ref{sec:environment}). For the latter part, I attempt to describe both sides of the debate for controversial topics and critically comment on the pros and cons of various models.  
It is my hope that at least the former part and most of the latter part will have a long shelf life.

\section{FRB phenomenology}\label{sec:data}

\subsection{Arrival times, coordinates, and naming convention}\label{sec:naming}

A detected FRB is characterized by the time when it is detected on Earth (corrected to the barycentric time) and the spatial coordinate of the source. There have been different conventions to name FRBs. Since they are bursting events in nature, a widely adopted scheme is to name them based on the time when the burst was detected similar to gamma-ray bursts (GRBs), i.e. FRB YYMMDD. However, since some (probably most) FRB sources emit repeated bursts, one has to adopt the time when the first burst was detected to name the source. For example, the first repeater is widely named as FRB 121102 or now officially FRB 20121102A. When CHIME came online, many detected FRBs flooded in. Since multiple FRBs could be detected on the same day and some of them could be the repeating ones, the CHIME/FRB collaboration adopted a more informative/complicated name by combining time information and spatial information (right ascension [R.A.] and declination [dec]) of the source. For example, the second repeater detected by CHIME was named as FRB 180814.J0422+73 (now officially FRB 20180814A). There was also a suggestion to call repeaters as `R\#', where `\#' is an assigned number based on the sequence of their discoveries. For example, FRB 121102 and FRB 180814.J0422+73 are also called `R1' and `R2', respectively. The 16-d periodic repeater FRB 20180916B is `R3'. Another possibility was that one can add a prefix `r' before the FRB name if a source is discovered to repeat. For example, `R1', `R2', and `R3' may be also called `rFRB 20121102A', `rFRB 20180814A' and `rFRB 20180916B', respectively. There was an unofficial voting for the preferred naming convention among the attendees of the 2019-February FRB Workshop at Amsterdam, the Netherlands, but no consensus was reached. The commonly adopted naming convention in the literature now follows the Transient Name Server (TNS) convention `FRB YYYYMMDDabc'. Personally, I think the information whether the source is a repeater is important. Throughout the review, I will follow the official TNS convention, but will add the `r' prefix for repeating sources in the rest of the review. Other nicknames are also used occasionally. Note that the prefix `r' is not a universally accepted convention but rather my personal preference.

\begin{figure*}
\includegraphics[width=6in]{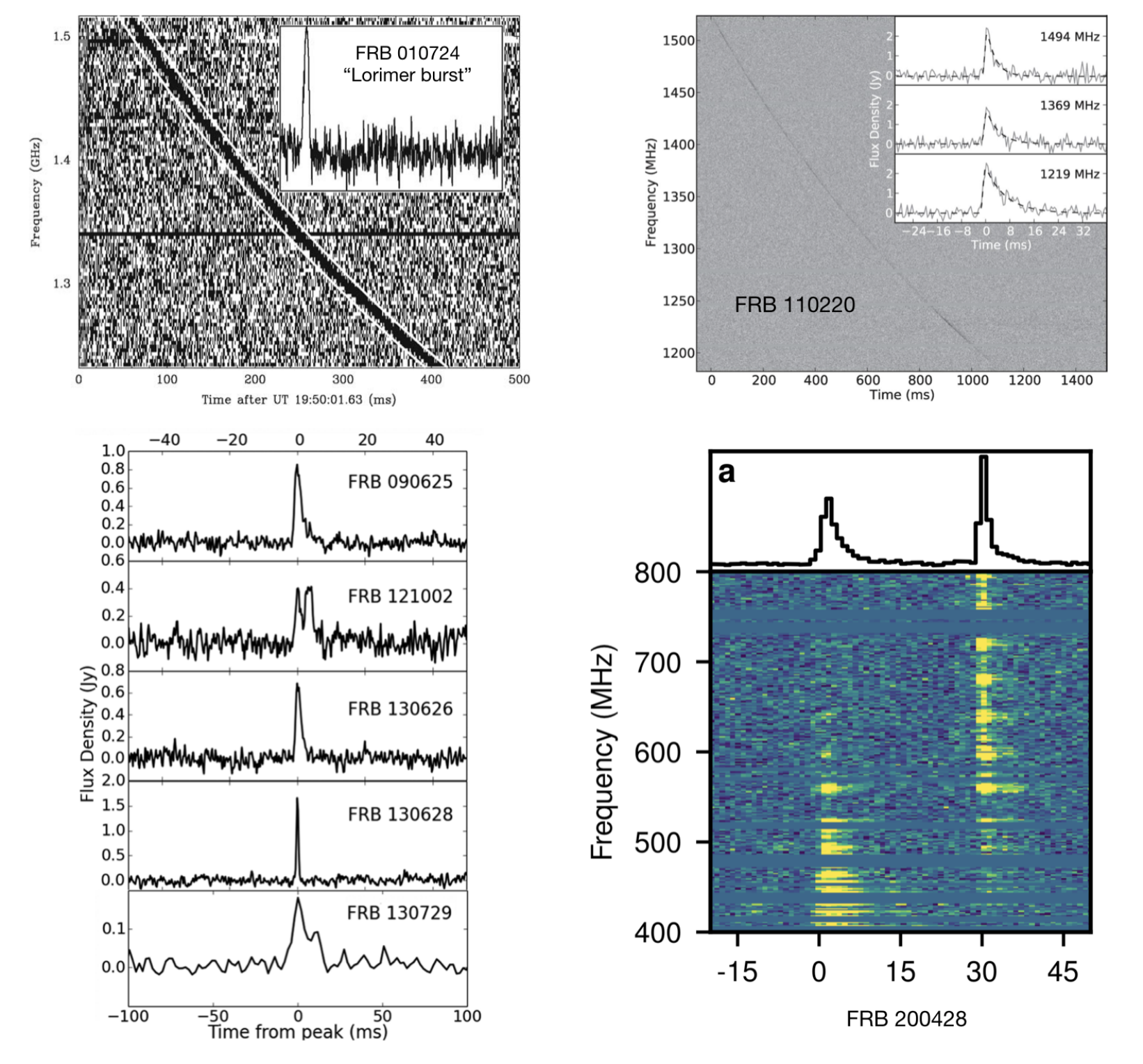}
\caption{Diverse lightcurves of FRBs. {\em Upper left:} The first reported FRB: FRB 20010724 or the ``Lorimer burst''. The main panel shows the frequency vs. arrival time of the radio burst as a result of dispersion. The inset panel shows the lightcurve of the burst after correcting for the effect of dispersion. From \citet{lorimer07}. Reprinted with permission from AAAS; {\em Upper right:} FRB 20110220 that shows frequency-dependent widths. The convention is the same as the upper left panel, but the lightcurves are constructed for three central frequencies. The decaying tail is wider in lower frequencies as a result of plasma scattering along the line of sight. From \citet{thornton13}. Reprinted with permission from AAAS; {\em Lower left:} Five more FRBs detected with the Parkes 64-m telescope. From \citet{champion16}; {\em Lower right:} The Galactic FRB 20200428 from SGR J1935+2154 as detected by the CHIME telescope. The upper panel is the lightcurve and the lower panel shows the two-dimensional frequency-time distribution (also called the ``dynamic spectrum'') of the emission after correcting for dispersion.  From \cite{CHIME-SGR}. Reproduced with permission from Springer Nature.}
\label{fig:FRBs}
\end{figure*}

\begin{figure*}
\includegraphics
[width=5in]{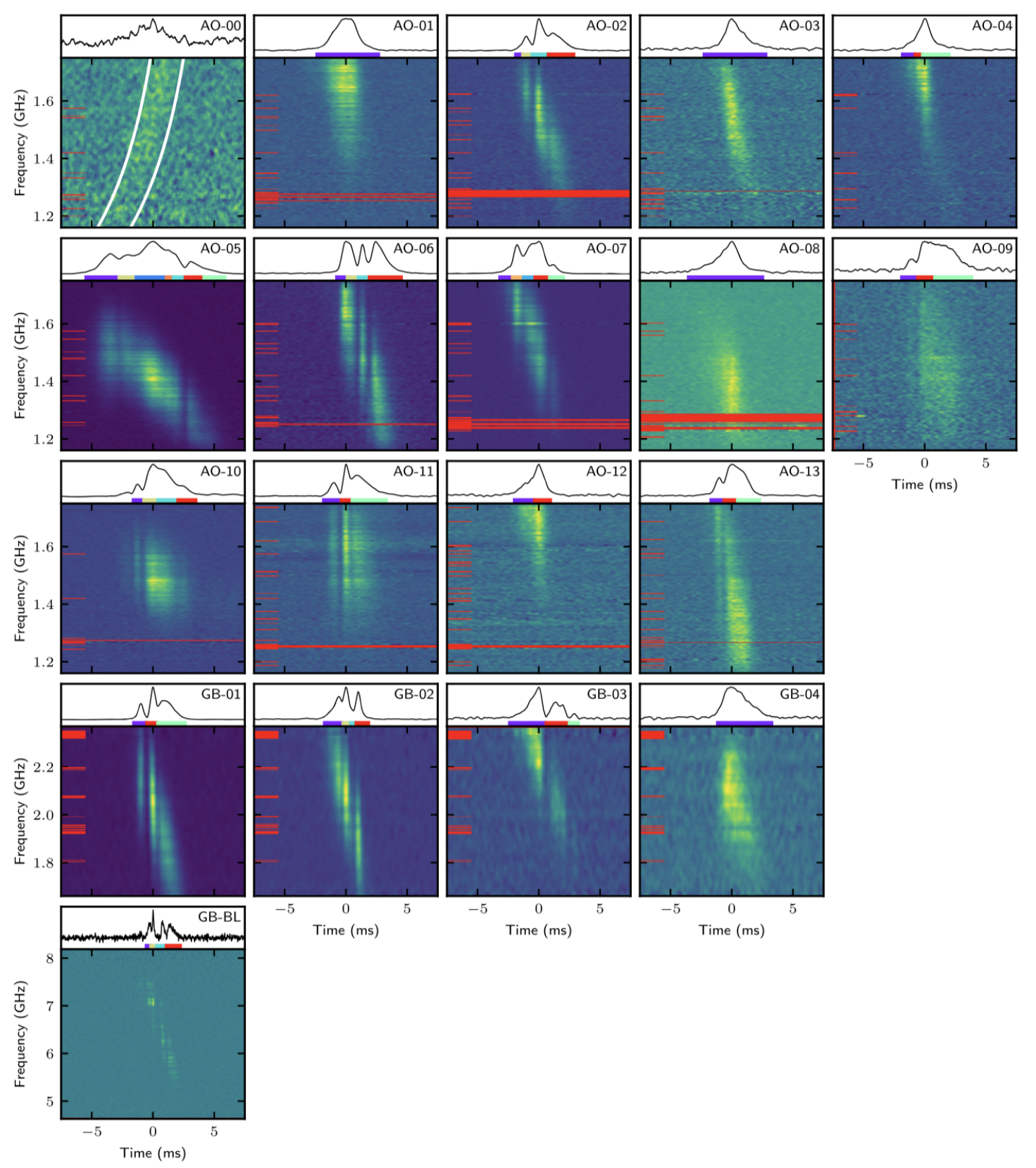}
\caption{An example of the dynamic spectra of individual bursts from rFRB 20121102A (R1) that show down-drifting of pulses with frequency, also called the ``sad trombone'' effect. Horizontal solid (red) bars denote RFI excision. From \citet{hessels19}. @AAS. Reproduced with permission.}
\label{fig:down-drifting}
\end{figure*}

\subsection{Temporal properties}\label{sec:temporal}

The typical observed duration (also known as width $W$) of an FRB is milliseconds. This duration is believed to be the convolution of the intrinsic pulse duration at the source ($W_i$), plasma scattering broadening ($\tau_{\rm sc}$) during the propagation of the pulse, as well as instrumental broadening by the radio telescope ($t_{\rm tel}$). Assuming uncorrelated Gaussian profiles of these components, one may write the observed width as \citep[e.g.][]{lorimer12,cordes03}
\begin{equation}
    W = \left[W_i^2 (1+z)^2 + \tau_{\rm sc}^2 + t_{\rm ins}^2 \right]^{1/2},
\end{equation}
where $W_i$ is the intrinsic duration of the FRB pulse in the source frame (the observed duration is longer by a factor of $(1+z)$ due to the cosmological time-dilation effect);
\begin{equation}
    \tau_{\rm sc} = \left[ \tau_{\rm MW}^2 + \tau_{\rm IGM}^2 + \tau_{\rm HG}^2 (1+z)^2 \right]^{1/2}
\end{equation}
is the scattering time, which includes the contributions from the Milky Way, intergalactic medium (IGM), and the  FRB host galaxy (see \S\ref{sec:scattering} for a detailed discussion on scattering); and 
\begin{equation}
    t_{\rm ins} = \left(t_{\rm samp}^2 + \Delta t_{\rm DM}^2 + \Delta t_{\rm \delta DM}^2 + \Delta t_{\rm \delta \nu}^2 \right)^{1/2}
\end{equation}
is the instrumental broadening \citep[e.g.][]{cordes03,petroff19}, which includes the data sampling interval, $t_{\rm samp}$, the frequency-dependent smearing due to dispersion measure (DM),
\begin{equation}
    \Delta t_{\rm DM} = (8.3 \ {\rm \mu s}) \ {\rm DM}  \Delta\nu_{\rm MHz}  \nu_{\rm GHz}^{-3},
\end{equation}
the smearing due to the error of DM, $\Delta t_{\rm \delta DM}$, and the smearing due to the bandwidth, $\Delta t_{\delta \nu} \sim (\Delta\nu)^{-1} = 1 \ {\rm \mu s} \ (\Delta\nu_{\rm MHz})^{-1}$.

As shown in Figure \ref{fig:FRBs}, the lightcurves of FRBs show diverse behaviors. Many FRBs have one single pulse (or indistinguishable multiple pulses). However, some FRBs (e.g. FRB 20121002) show an apparent temporal structure \cite{champion16}. The Galactic FRB 20200428 had two pulses separated by roughly 30 ms, which may be also regarded as a repeating source that emitted two bursts. Some bursts clearly show an asymmetric pulse profile, with a longer decaying wing than the rising phase. This decaying wing is frequency dependent, with a longer tail at a lower frequency (e.g. FRB 20111220, upper right panel, \cite{thornton13}). The frequency-dependent scattering tail of these FRBs is consistent with $\tau_{\rm sc} \propto \nu^{-4}$ or $\tau_{\rm sc} \propto \nu^{-4.4}$ as predicted by the plasma scattering effect \cite{luan14,cordes16b,xu16}.

One interesting temporal feature of some FRBs is down-drifting of pulses with frequency  \cite{hessels19,chime-2ndrepeater,chime-repeaters}, also called the ``sad trombone'' effect (Fig. \ref{fig:down-drifting}). This is after correcting the standard dispersive delay due to propagation and is likely related to the intrinsic radiation physics of FRBs. Such a behavior is often seen in repeating FRB bursts. The down-drifting is predominating. The opposite trend (up-drifting) is much rarer \cite{chime-catalog,zhou22}, even though the two apparently separated pulses in FRB 20200428 indeed showed a higher peak frequency in the second pulse \cite{CHIME-SGR}. 

The morphology of FRBs, especially for repeaters, has been studied extensively. \citet{pleunis21b} studied 536 bursts from 492 sources from the CHIME first catalog and identified four observed archetypes of burst morphology, namely ``simple broadband,'' ``simple narrowband,''``temporally complex,'' and ``downward drifting''. \citet{zhou22} studied more than 700 bursts from one repeating source rFRB 20201124A detected by FAST, and identified five morphological types based on the drifting patterns: downward drifting, upward drifting (a small fraction), complex, no drifting, and no evidence for drifting. Subtypes are introduced as needed based on the emission frequency range in the band (low, middle, high, and wide), and also the number of sub-pulses in the burst (1, 2, or multiple). Altogether, 18 morphological sub-types are identified. The longest burst includes 11 pulses lasting 124 ms. There are no apparent correlations among duration, bandwidth, central frequency and flux. 

\subsection{Spectral properties}

FRBs have been detected from 110 MHz \cite{pleunis21} to at least 8 GHz \cite{gajjar18}. Non-detection at higher frequencies could be due to limited sensitivity \cite{law17} or the difficulty to achieve strong coherence. The lack of dispersion at high frequencies makes it difficult to differentiate RFIs from true signals, which might also contribute to the deficit. The non-detection at lower frequencies, especially with LOFAR at 145 MHz, may suggest an intrinsic hardening of spectrum at low frequencies probably due to a certain absorption process \cite{karastergiou15}. 
    
The spectral shape of some early FRBs was not well measured. If one approximates the spectral shape as a power law function $F_\nu \propto \nu^{-\alpha}$, the power law index $\alpha$ was observed to vary significantly from case to case. For example, the Lorimer burst had $\alpha=4 \pm 1$  \cite{lorimer07}, while FRB 20110523A had $\alpha = 7.8 \pm 0.4$ \cite{masui15}. Even for different bursts from the same repeating source, $\alpha$ can be very different. For example, the $\alpha$ values of rFRB 20121102A bursts ranged from $-10.4$ to $+13.6$ \cite{spitler16}. Such a large variation may be the indication that the intrinsic spectrum of FRBs is narrow. Multi-telescope studies of some repeater bursts often show that the bursts detected in one band are not detected in another, e.g. for rFRB 20121102A \cite{law17} and rFRB 20180916B \cite{paster-marazuela21}. This suggests that the spectra of these bursts are not simple power laws. Indeed, the dynamical spectra of FRBs (Figs.\ref{fig:FRBs} and \ref{fig:down-drifting}) often show that the bursts are bright only in part of the whole observing bandpass. The Galactic magnetar burst FRB 20200428 had two pulses as detected by CHIME \cite{CHIME-SGR}, but only the second pulse that had a higher peak frequency was detected by STARE2 \cite{STARE2-SGR}, which has a higher bandpass than CHIME. This again suggests that the FRB spectra could be quite narrow. A systematic study of the spectral properties of more than 700 bursts from rFRB 20201124A detected by FAST \citep{zhou22} suggests that the majority of repeating FRBs have narrow spectra, with the typical spectral band width of $\sim 275$ MHz in the FAST band.

\subsection{Repetition \& Periodicity}\label{sec:repetition}

More than 20 FRBs have been reported to repeat \cite{spitler16,chime-2ndrepeater,chime-repeaters,kumar19,luo20b,niu22}. Since a repeating FRB is identified whenever one more burst is detected from the same source, it is essentially impossible to claim that an FRB source is NOT a repeater. In fact, it is quite possible that all FRB sources repeat but with a wide range of repetition rate. Since the observed FRB rate density exceeds the rate density of supernovae, the most common catastrophic events, it is immediately inferred that the majority of the FRBs have to be from repeating sources \cite{ravi19b,luo20}. The remaining question is whether all FRB sources repeat and whether there exists a minority population of FRBs that do originate from catastrophic events \cite{palaniswamy18,caleb19a}. 

Some differences in the observational properties between repeaters and apparent one-off FRBs have been noticed, but no conclusive results have been drawn. 
\begin{itemize}
    \item The CHIME/FRB Collaboration \cite{chime-repeaters,chime-catalog,pleunis21b} reported that repeaters tend to have wider widths than one-off FRBs. They also tend to have narrower spectra than one-off bursts.  
    However, the two populations have overlapping parameter spaces, so that it is difficult to definitely tell whether an apparent one-off burst actually belongs to the repeater population.  
    \item The frequency down-drifting feature has been observed in several repeating sources \citep{hessels19,chime-2ndrepeater}. However, not all bursts from these sources and not all repeating sources show such a behavior. On the other hand, some apparently one-off FRBs show such a behavior, which may be regarded as candidate repeating FRBs. 
    \item Both supervised \cite{luojw23} and unsupervised \cite{zhuge23} machine learning algorithms applied on the first CHIME FRB catalog reached the consensus that repeaters and most non-repeaters seem to belong to different categories. Including both observed and derived parameters, both algorithms recognize brightness temperature and rest-frame spectral width as the two dominant traits to differentiate between the two categories. Some common candidate repeaters can be identified from these two independent categories of machine learning methods \cite{luojw23,zhuge23}.  However, the accuracy of the predicted repeaters is not high in comparison with the latest repeater catalog reported by the CHIME/FRB Collaboration \cite{chime-new-repeaters} as more high-luminosity FRBs turn into repeaters. 
\end{itemize}
It is worth noting that some polarization properties, e.g. varying polarization angle (PA) \cite{cho20} or circular polarization  \cite{dais21}, had once been proposed to be the unique properties of non-repeaters. However, later observations showed that some repeaters also possess these properties \cite{luo20b,xuh22}. It is now clear that polarization properties cannot be used to differentiate between the two categories.

If all FRBs are repeaters, then at least some apparent one-off FRBs must have a very low repetition rate. \citet{palaniswamy18} and \citet{caleb19a} suggested that most FRBs cannot have a similar repetition rate as rFRB 20121102A. Otherwise, many of them should have been observed to repeat. Indeed, extensive follow-up observations of some bright FRBs such as the ``Lorimer burst'' have so far failed to detect any repeated bursts \cite{lorimer07,petroff15b}, suggesting that they might have a different origin. \citet{katz19} pointed out that the duty factor defined as $D \equiv \left<S\right>^2/ \left< S^2 \right>$ ($S$ is flux density) may be used to differentiate repeaters from non-repeaters, with active repeaters such as rFRB 20121102A having $D \sim 10^{-5}$ while non repeaters having $D \sim (10^{-8} - 10^{-10})$. 

With detailed simulations, \citet{ai21} suggested that tracking the evolution of observed repeater fraction $F_{\rm r,obs}$ may shed light into the existence of genuinely non-repeating FRBs. This is because if genuinely non-repeating FRBs indeed exist, their numbers will linearly increase as a function of time. The number of repeaters, on the other hand, may approaching a limit with time. As a result, $F_{\rm r,obs}$ is expected to reach a peak and then decline. Therefore, detecting such a peak would strongly suggests the existence of genuinely non-repeating FRBs. In reality, however, depending on parameters and possible evolution of source populations, the time to reach the peak could be long and the the duration at the peak could be also long. Long term monitoring of the sky using CHIME-like wide-field survey telescopes will hold the key to place constraints on the existence of genuinely non-repeating FRBs. It is interesting to note that the recent CHIME observations suggested that $F_{\rm r,obs}$ stays constant for a few years already, which is consistent with the hypothesis that genuinely non-repeating FRBs do exist (Z. D. Pleunis, 2022, talk at the Cornell FRB workshop).

Searches for periodicity of repeating FRB sources have been carried out extensively. The early targeted periods in the searches were in the milliseconds to seconds range, similar to the periods of known pulsars and magnetars. Deep searches of periodicity in this period range for rFRB 20121102A (\citet{lid21,zhangy18,hewitt22}, see also an independent search by \citet{katz22b}) and rFRB 20201124A \citep{xuh22,niujr22} using thousands of bursts all led to null results, suggesting that FRB bursts are likely not giant pulses of rotating neutron stars. On the other hand, unexpected, very long periods (or active cycles) were found in some repeating sources. The most robust case is the CHIME-discovered rFRB 20180916B, which shows a $\sim 16$-d period with a $\sim 5$-d active window \cite{chime-periodic}. The duration and phase of the active window seems to be frequency-dependent, with the windows in higher frequencies appearing earlier in phase and being narrower than the windows in lower frequencies \cite{pleunis21,paster-marazuela21}. Long-term monitoring of rFRB 20121102A also revealed a possible long-term $\sim 160$-d periodicity \cite{rajwade20,cruces21}. Long-term monitoring of rFRB 20121102A with FAST suggests that bursts are often missing during the predicted active window and the duty cycle of the periodicity becomes greater than 50\% (P. Wang et al. 2023, in preparation). This casts a shadow to the claimed periodicity. Finally, the ``oddball'' source FRB 20191221A was detected to have a 0.2168-s period with a significance of $6.5\sigma$ \cite{chime-period}. Since the total duration ($\sim 3$ s) is much longer than other FRBs, this event likely has a different origin from the bulk of the FRB population. On the other hand, a deep periodicity search of rFRB 20201124A bursts \cite{niujr22} suggested that even though no global periodicity was found, fake local periodicity in adjacent burst clusters can be found with a significance up to $3.9\sigma$. This cautions against claiming any periodicity from clustered bursts with significance $\lesssim 4\sigma$. 

\begin{figure*}
\includegraphics
[width=\columnwidth]{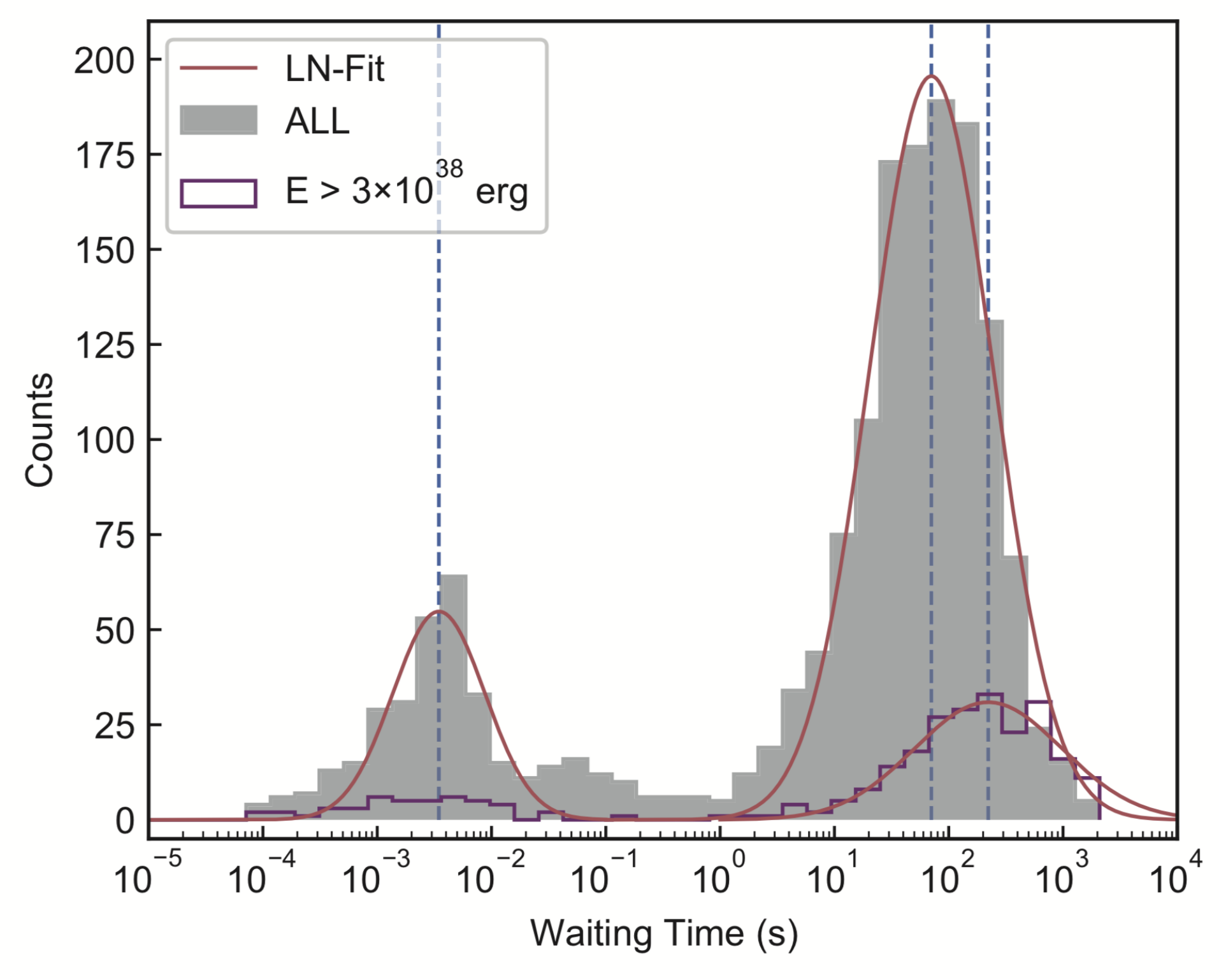}
\includegraphics
[width=\columnwidth]{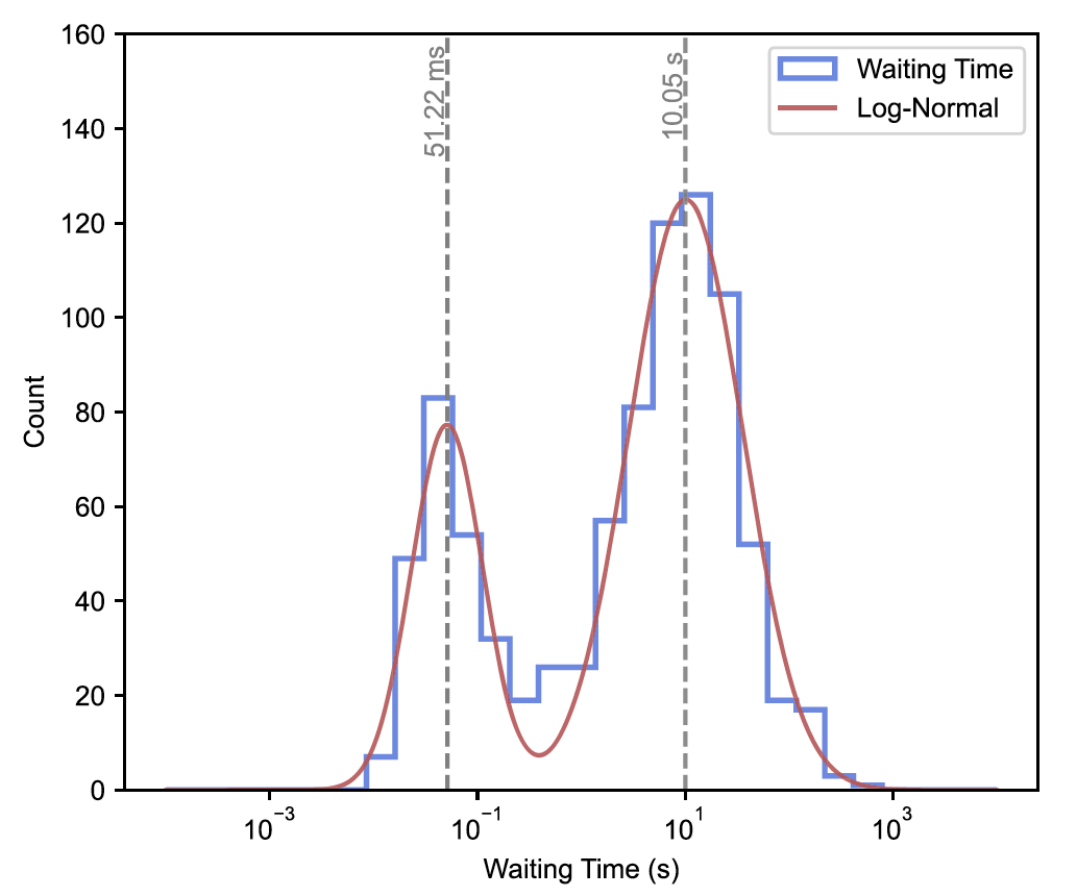}
\caption{The waiting time distributions of repeating FRBs. {\em Left:} The case of rFRB 20121102A during the 2019 active episode. From \citet{lid21}. The two peaks are at a few milliseconds and $\sim 100$ s, respectively. {\em Right:} The case of rFRB 20201124A during a four-day very active episode in September 2021. The second peak is at $\sim 10$ s, suggesting a very active episode. From \citet{zhangyk22}. In an earlier episode of the same source in April 2021, the second peak is $\sim 100$ s, suggesting that the same source can have very different activity levels, and hence, different waiting time distributions. }
\label{fig:waitingtime}
\end{figure*}

One interesting common feature of active repeaters is that the waiting time distributions of their bursts show two distinct peaks \citep{lid21,xuh22,zhangyk22,niujr22,zhou22}. As shown in Figure \ref{fig:waitingtime}, the first peak is around milliseconds and the exact value depends on how distinct bursts are defined. The second peak actually depends on the activity level of the source, ranging from 10s of seconds to 100s of seconds, even for the same source at different epochs. The bridge between the two peaks lie around 10s of milliseconds. Since some FRB bursts show multiple peaks, the short separations of bursts in the first component of waiting time distribution can be regarded as due to the similar origin as multi-peaks, which may be related to the continuous activity of the FRB source from one emission episode. \citet{zhou22} defined ``burst clusters'' that include all the bursts whose relative waiting times fall onto this first waiting time peak. The second peak apparently scales with the global activity level of the source. More observations are needed to see whether the dip between the two components may carry information about the periodicity of the underlying engine.

\subsection{Dispersion measure and distance}\label{sec:DM}

Radio waves in a plasma are dispersed, with waves with lower frequencies delayed with respect to waves with higher frequencies. The {\em dispersion measure} (DM) (see \S\ref{sec:DM2} for details) describes the degree of such delay. The best-fit DM is obtained for each FRB when it is discovered\footnote{The FRB search algorithm scans through a range of DM values to correct for such a delay. The DM of FRB is assigned either for the highest signal-to-noise ratio (S/N) or the finest burst temporal structure \cite{hessels19}.}, and it carries the physical meaning of the column density of free electrons along the line of sight from the source to the observer (with the units of $\rm pc \ cm^{-3}$). Since FRBs are from cosmological distances, the DM can be most generally written as 
\begin{equation}
    {\rm DM} = \int_0^{D_z} \frac{n_e(l)}{1+z(l)} \ d l,
\label{eq:DM}
\end{equation}
where $n_e$ (a function of location denoted by $l$) is the local electron number density, $z$ is the redshift at that location, $l$ is the comoving distance from the observer to a location along the path of propagation, and 
\begin{equation}
     D_z = \frac{c}{H_0} \int_0^{z} \frac{d z'}{E(z')},
\label{eq:Dz}
\end{equation}
is the comoving distance from the observer to the source, where 
\begin{equation}
     E(z)=    \sqrt{\Omega_m(1+z)^3+\Omega_k(1+z)^2+\Omega_{\rm DE} f(z)},
     \label{eq:E(z)}
\end{equation}
\begin{equation}
    f(z) = \exp \left[3 \int_0^z \frac{(1+w(z')) dz'}{1+z'} \right],
    \label{eq:f(z)}
\end{equation}
$H_0$ is Hubble constant, $\Omega_m$, $\Omega_k$ and $\Omega_\Lambda$ are the energy density fraction of matter, curvature and dark energy, respectively, and $w(z) \equiv p(z)/\rho(z)$ is the dark energy equation of state parameter. For the concordance $\Lambda$CDM cosmological model, one has $\Omega_k=0$, $\Omega_{\rm DE} = \Omega_\Lambda$, $w=-1$, and $f(z) = 1$.

The observed DM is usually split into multiple terms (e.g. \citet{thornton13}, \citet{deng14}, \citet{prochaska19b})
\begin{equation}
    {\rm DM = DM_{\rm MW} + DM_{halo} + DM_{IGM}} + \frac{\rm DM_{host} + DM_{src}}{1+ z},
\label{eq:DMterms}
\end{equation}
where $\rm DM_{MW}$, $\rm DM_{halo}$, $\rm DM_{IGM}$, $\rm DM_{host}$, and $\rm DM_{src}$ are the contributions from the Milky Way, its halo, the inter-galactic medium (IGM), the host galaxy, and the immediate environment of the source, respectively. Notice that the observed contributions from the last two components are smaller by a factor of $(1+z)$, where $z$ is the source redshift. The Milky Way term $\rm DM_{MW}$ can be obtained using the MW electron density models derived from the radio pulsar data \cite{cordes02,yao17} (with a $>50\%$ uncertainty). The extended Milky Way halo contributes to an additional $\rm DM_{halo} \sim (30-80) \ {\rm pc \ cm^{-3}}$ beyond $\rm DM_{MW}$ \citep[e.g.][]{dolag15,prochaska19b}.  

\begin{figure*}
\includegraphics[width=5in]
{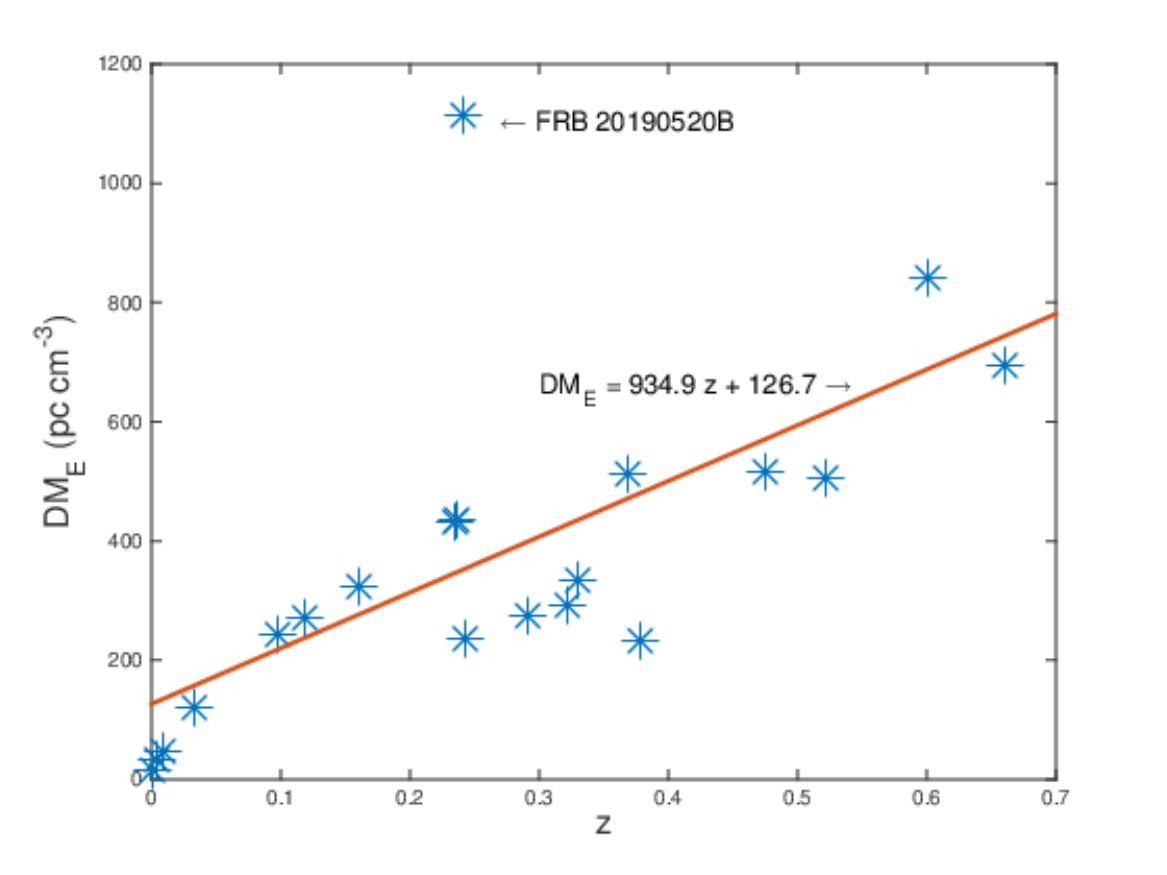}
\caption{The DM-$z$ relation of 21 FRBs with known redshifts, which is an updated version of the results of \citet{macquart20}. The NE2001 electron density model and $\rm DM_{halo} = 30 pc \ cm^{-3}$ are adopted. The best-fit linear regression line is plotted. rFRB 20190520B has an abnormally large $\rm DM_{host}$ \cite{niu22}, which is marked separately.}
\label{fig:DM-z}
\end{figure*}

The IGM component of DM is a function of redshift \cite{ioka03,inoue04}.
The full expression reads \cite{deng14,gao14,zhou14,macquart20}:
\begin{equation}
    \left< {\rm DM_{IGM}}(z) \right> = \frac{3 c H_0 \Omega_b f_{\rm IGM}}{8\pi G m_p} \int_0^z \frac{\chi(z')(1+z') dz'}{E(z')},
\label{eq:DM-z}
\end{equation}
where
\begin{equation}
    \chi(z) \simeq \frac{3}{4} \chi_{\rm e,H} (z) + \frac{1}{8} \chi_{\rm e,He} (z)
    \label{eq:chiz}
\end{equation}
noticing that the cosmological mass fractions of H and He are $\sim 3/4$ and $\sim 1/4$, respectively, $\Omega_b$ is the energy density fraction of baryons, $f_{\rm IGM}$ is the fraction of baryons in the IGM, and $\chi_{\rm e,H}(z)$ and $\chi_{\rm e,He}(z)$ are the fractions of ionized electrons in hydrogen (H) and helium (He), respectively, as a function of redshift. The DM-$z$ relation is roughly linear at low redshifts \cite{ioka03,inoue04}. With the standard cosmological parameters as measured by the Planck mission \cite{planck}, one can derive a rough linear relation at $z<3$ (\citet{zhang18a}, see also \citet{pol19,cordes21})
\begin{eqnarray}
    \left< {\rm DM_{IGM}} \right> & \simeq & (855 \ {\rm pc \ cm^{-3}}) \ z \ \left(\frac{H_0}{67.74 \ {\rm km \ s^{-1} \ kpc^{-1}}}\right) \nonumber \\ &\times & \left(\frac{\Omega_b}{0.0486}\right) \left(\frac{f_{\rm IGM}}{0.83}\right) \left(\frac{\chi}{7/8}\right), 
\label{eq:DM-z2}
\end{eqnarray}
where $f_{\rm IGM}$ is normalized to $\sim 0.83$ \cite{fukugita98,lizx20}. In the literature, the DM-$z$ relation is also called the ``Macquart-relation'' to honor J-P Macquart's leadership in the ASKAP collaboration to precisely localize a sample of FRBs and measure their redshifts to prove the theoretically motivated relation (\ref{eq:DM-z}). Notice that Equations (\ref{eq:DM-z}) and (\ref{eq:DM-z2}) apply to average values. For individual FRBs, the measured DM can be either greater or smaller than the theoretical value due to the inhomogeneity of the IGM caused by large scale structures \cite{ioka03,mcquinn14}. 

The redshifts of the localized FRBs (Table I) 
indeed follow the theoretical expectations \cite{tendulkar17,bannister17,ravi19,marcote20,prochaska19,macquart20}. After deducting the Milky Way contribution, the external component of DM indeed shows a rough linear relation with $z$, with the best-fit line consistent with the prediction of the $\Lambda$CDM model \cite{macquart20}. Using the \citet{macquart20} sample and systematically deducting an average $\rm DM_{host}$ value, the DM-$z$ relation could give a constraint on $f_{\rm IGM} \sim 0.85$ \cite{lizx20}, which is consistent with previous results \cite{fukugita98}.

\begin{table*}\label{tab:FRB-z}
\caption{Published FRBs with measured redshifts, their observed DM values and the MW contributions.}
\begin{ruledtabular}
\begin{tabular}{lccccc}
FRB & $z$ & $\rm DM$\footnote{All DMs have the units of $\rm pc \ cm^{-3}$. } & $\rm DM_{MW}(NE2001)$\footnote{Calculated from the NE2001 model \cite{cordes02}. Data provided by Ye Li who ran the script provided from https://pypi.org/project/pyne2001/.} & $\rm DM_{MW}(YMW16)$\footnote{Calculated from the YMW17 model \cite{yao17} using the website interface https://www.atnf.csiro.au/research/pulsar/ymw16/.} & References \\
\hline
rFRB 20121102A & 0.19273 & $\sim557$ & $\sim 188$ & $\sim 287$ & \citet{tendulkar17} \\
~FRB 20171020A & 0.0087 & $\sim 114$ & $\sim 37$ & $\sim 25$ & \citet{mahony18} \\
rFRB 20180301A & 0.3304 & $\sim 517$ & $\sim 152$ & $\sim 254$ & \citet{luo20b} \\
rFRB 20180916B    & 0.0337 & $\sim 349$ & $\sim 199$ & $\sim 325$ & \citet{marcote20} \\
rFRB 20180924C   & 0.3214 & $\sim 362$ & $\sim 41$ & $\sim 28$ & \citet{bannister19} \\
~FRB 20181030A   & 0.0039 & $\sim 104$ & $\sim 41$ & $\sim 33$ & \citet{bhandari22} \\
~FRB 20181112A & 0.4755 & $\sim 589$ & $\sim 42$ & $\sim 29$ &  \citet{prochaska19} \\
~FRB 20190102C  & 0.2913 & $\sim 363$ & $\sim 57$ & $\sim 43$ & \citet{macquart20} \\
rFRB 20190520B & 0.241 & $\sim 1205$ & $\sim 60$ & $\sim 50$ & \citet{niu22} \\
~FRB 20190523A   & 0.6600 & $\sim 761$ & $\sim 37$ & $\sim 30$ & \citet{ravi19} \\
~FRB 20190608B  & 0.1178 & $\sim 339$ & $\sim 37$ & $\sim 27$ & \citet{macquart20} \\
~FRB 20190611B & 0.3778 & $\sim 321$ & $\sim 58$ & $\sim 44$ & \citet{macquart20} \\
~FRB 20190614D & 0.60 & $\sim 959$ & $\sim 88$ & $\sim 109$ & http://frbhosts.org/ \\
rFRB 20190711A & 0.5220 & $\sim 593$ & $\sim 56$ & $\sim 43$ & \citet{macquart20} \\
~FRB 20190714A & 0.2365 & $\sim 504$ & $\sim 39$ & $\sim 31$ & \citet{bhandari22}\\
~FRB 20191001A & 0.2340 & $\sim 508$ & $\sim 44$ & $\sim 31$ & \citet{bhandari22}\\
~FRB 20191228A & 0.2432 & $\sim 298$ & $\sim 32$ & $\sim 20$ & \citet{bhandari22}\\
rFRB 20200120E & 0.0008 & $\sim 88$ & $\sim 41$ & $\sim 33$ & \citet{kirsten22}\\
~FRB 20200430A & 0.1608 & $\sim 380$  & $\sim 27$ & $\sim 26$ & \citet{bhandari22}\\
~FRB 20200906A & 0.3688 & $\sim 578$ & $\sim 36$ & $\sim 38$ & \citet{bhandari22}\\
rFRB 20201124A & 0.0979 & $\sim 414$ & $\sim 140$ & $\sim 197$ & \citet{ravi22}\\
\end{tabular}
\end{ruledtabular}
\end{table*}

Figure \ref{fig:DM-z} gives the updated ${\rm DM_E}-z$ relation with the 21 redshift-known FRBs listed in Table I. 
The vertical axis is $\rm DM_E = DM - DM_{MW} - DM_{halo}$, where NE2001 model \cite{cordes02} and $\rm DM_{halo} = 30 \ {\rm pc \ cm^{-3}}$ have been adopted. A simple linear regression best fit is presented. Using the YMW17 \cite{yao17} model or the average NE2001/YMW17 model lead to similar results, with slightly different regression results:
\begin{eqnarray}
    {\rm DM_E} = 934.9 z + 126.7, & ~~~{\rm NE2001}, \\
    {\rm DM_E} = 979.7 z + 103.1, & ~~~{\rm YMW17}, \\
    {\rm DM_E} = 957.3 z + 114.9, & ~~~{\rm average}.
\end{eqnarray}
Here the slope can be compared with the prediction in Eq.(\ref{eq:DM-z2}) and the $y$-intersection may be regarded as the average $({\rm DM_{host}+DM_{src}})/(1+z)$. Comparing the fitting results to Eq.(\ref{eq:DM-z2}), one may tentatively draw the conclusion that $f_{\rm IGM} > 0.9$, which is greater than the estimate in the past \citep{fukugita98}. Considering the outlier rFRB 20190520B with huge $\rm DM_{host}+DM_{src}$ \citep{niu22} might have leveraged the $y$-intersection, an average value of $\rm DM_{host}+DM_{src} \sim 100 \ pc \ cm^{-3}$ would be reasonable. A systematically lower $\rm DM_E$ than the linear fit is noticeable at low redshifts, but this may be a result of large scale density fluctuations. More data are needed to judge whether there is a systematic deficit of $\rm DM_E$ at low redshifts. 

\subsection{Luminosity, energy and brightness temperature}\label{sec:lum}

With measured redshifts, the isotropic-equivalent energy and peak luminosity of FRBs can be measured precisely. Because the ${\rm DM}-z$ relation has been confirmed from the data, for most FRBs without redshift measurements, the measured DM values can be used to estimate the redshift, and hence, the energetics of the FRBs. Lacking the geometric beaming information of FRBs, one can only estimate the isotropic-equivalent values of the peak luminosity and energy. The best estimates depend on the spectral shape of the FRB. If the FRB spectra are narrow-band with emission contained within the telescope observing band (which is the case for most bursts from repeaters, e.g. \citet{zhou22}), it is more appropriate to multiply the bandwidth $\Delta\nu$ by the specific flux to obtain luminosity. On the other hand, if the FRB spectra are broad-band (which is relevant to some non-repeating FRBs, e.g. the Lorimer burst, \citet{lorimer07}) with emission extending beyond the telescope observing band, it would be more appropriate to multiply the band central frequrncy $\nu_c$ by the specific flux to obtain luminosity \citep{zhang18a}. So, in general, one may write
    \begin{eqnarray}
     L_{\rm p,iso} & \simeq & 4\pi D_{\rm L}^2 {\cal S}_{\nu,p} \cdot \left\{
     \begin{array}{cc}
         \Delta\nu, & {\rm narrow~spectrum,}  \\
         \nu_c,  & {\rm broad~spectrum,}
     \end{array} \right. \nonumber \\
     & = & (4\pi \cdot 10^{42} \ {\rm erg \ s^{-1}}) \left( \frac{D_{\rm L}}{10^{28} \ {\rm cm}} \right)^2 \frac{{\cal S}_{\nu,p}} {\rm Jy} \frac{(\Delta\nu~ {\rm or}~\nu_c)} {\rm GHz}, \nonumber \\
     \label{eq:Liso}\\
     E_{\rm iso} & \simeq & \frac{4\pi D_{\rm L}^2}{1+z} {\cal F_\nu}  \cdot \left\{
     \begin{array}{cc}
         \Delta\nu, & {\rm narrow~spectrum,}  \\
         \nu_c,  & {\rm broad~spectrum,}
     \end{array} \right. \nonumber \\
     & = & \frac{4\pi \cdot 10^{39} \ {\rm erg} }{1+z} \left( \frac{D_{\rm L}}{10^{28} \ {\rm cm}} \right)^2 \frac{{\cal F}_{\nu}} {\rm Jy \cdot ms} \frac{(\Delta\nu~ {\rm or}~\nu_c)} {\rm GHz}, \nonumber \\
     \label{eq:Eiso}
    \end{eqnarray}
    where ${\cal S}_{\nu,p}$ is the specific peak flux density, ${\cal F}_\nu$ is the specific fluence, and $D_{\rm L}
    = (1+z) D_z$ 
    is the luminosity distance.
    The isotropic peak luminosities of known FRBs vary from \cite{STARE2-SGR,ravi19} $\sim 10^{38} {\rm erg \ s^{-1}}$ to a few $10^{46} \ {\rm erg \ s^{-1}}$. The corresponding isotropic energies vary from a few $10^{35}$ erg to a few $10^{43}$ erg. The luminosity is extremely high by the radio pulsar standard, but is minuscule by the GRB standard. The true energetics of FRBs should be reduced by a beaming factor $f_b = {\rm max} (\Delta\Omega/4\pi, 1/4\gamma^2) \leq 1$, where $\Delta\Omega$ is the solid angle of the geometric beam, and $\gamma$ is the Lorentz factor of the FRB emitter ($1/\gamma$ is the half kinetic beaming angle for an FRB emitter traveling close to speed of light). For an one-off FRB, a successful FRB engine should at least generate a luminosity and an energy of the order of $f_b L_p$ and $f_b E$, respectively. Observationally, the majority of hard X-ray bursts from SGR J1935+2154 were not associated with FRBs \cite{lin20}. One possibility is that FRB emitters (at least those produced by magnetars) are narrowly beamed. If so, one would also expect to detect less-luminous but longer-duration radio bursts (``slow radio bursts'') with line of sight outside the emission beam \citep{zhang21}.

The combination of high luminosity and short variability timescale of an FRB defines an extremely high brightness temperature $T_b$. One may derive this by noticing that the observed specific intensity $I_\nu = S_\nu/\Delta \Omega$, where $S_\nu$ is the observed specific flux, $\Delta\Omega=\pi (c \Delta t_0)^2/D_{\rm A}^2$ is the solid angle of the source viewed at the observer location ($\Delta t_0 = \Delta t/(1+z)$ is the rest-frame duration of the burst, $c \Delta t_0$ is adopted as the transverse scale, which is true for a non-relativistic, spherical, transparent emitter), and $D_{\rm A} =D_z/(1+z) = D_{\rm L}/(1+z)^2$ is the angular diameter distance of the source. Considering an imaginary blackbody emitter with temperature $T_b(\nu_0)$ at the rest frame frequency $\nu_0=(1+z)\nu$, and noticing $I_\nu (\nu_0) \simeq 2 k_B T_b(\nu_0) (\nu_0^2/c^2)$ in the Rayleigh-Jeans regime ($k_B$ is the Boltzmann constant) and $I_{\nu}(\nu_0)=I_\nu(\nu)(1+z)^3$ (i.e. $I_\nu/\nu^3$ is constant), one finally obtains the brightness temperature at the source frequency $\nu_0$ \cite{luojw23}\footnote{If the emitter is moving relativistically towards earth with a Lorentz factor $\Gamma$, the transverse size in the comoving frame would be $\Gamma c \Delta t$, so that $T'_b$ is smaller by a factor of $\Gamma^2$ with respect to Eq.(\ref{eq:Tb}). The observer-frame $T_b$ is boosted up by a factor of $\sim \Gamma$, so the overall $T_b$ is smaller by a factor of $\Gamma$ than Eq.(\ref{eq:Tb}) (see also \citet{lyubarsky21}). Here we define $T_b$ solely based on observables without assuming whether the source has relativistic motion.}
    \begin{eqnarray}
        T_b(\nu_0)& = &\frac{{\cal S}_{\nu} D_{\rm A}^2(1+z)^3}{2 \pi k_B (\nu \Delta t)^2} = \frac{{\cal S}_{\nu} D_{\rm L}^2}{2 \pi k_B (\nu \Delta t)^2(1+z)} \nonumber \\
        &\simeq& (1.2\times 10^{36}  \ {\rm K} ) \frac{{\cal S}_{\nu}}{\rm Jy} \left(\frac{ \nu}{\rm GHz}\right)^{-2} \left(\frac{\Delta t}{\rm ms}\right)^{-2} \nonumber \\
        &\times& \left\{
         \begin{array}{cc}
            (1+z)^3 \left( \frac{D_{\rm A}}{10^{28} \ {\rm cm}} \right)^2,& \\
            \frac{1}{1+z} \left( \frac{D_{\rm L}}{10^{28}\ {\rm cm}} \right)^2.& ~
         \end{array}
        \right.
        \label{eq:Tb}
    \end{eqnarray}
The physical meaning of $T_b$ is the imaginary temperature of the emitter if the photons and the electrons that emit the photons were in thermal equilibrium. This is apparently not the case for FRBs. The gigantic $T_b$ ($\sim 10^{36}$ K for nominal FRB parameters) is much greater than any temperature allowed for incoherent radiation (see \S\ref{sec:Tb} for details).
This demands that the radiation mechanism for FRB emission must be ``coherent'', i.e. the radiation by relativistic electrons must not only not be absorbed but also greatly enhanced with respect to the total expected emission if electrons radiate independently (or incoherently). Before the discovery of FRBs, radio pulsars have been the only known sources of producing extremely high $T_b$'s (typically $\sim (10^{25}-10^{30})$ K). FRBs further push the limit of the degree of coherent radiation in the universe.

\begin{figure*}
\includegraphics
[width=5in]{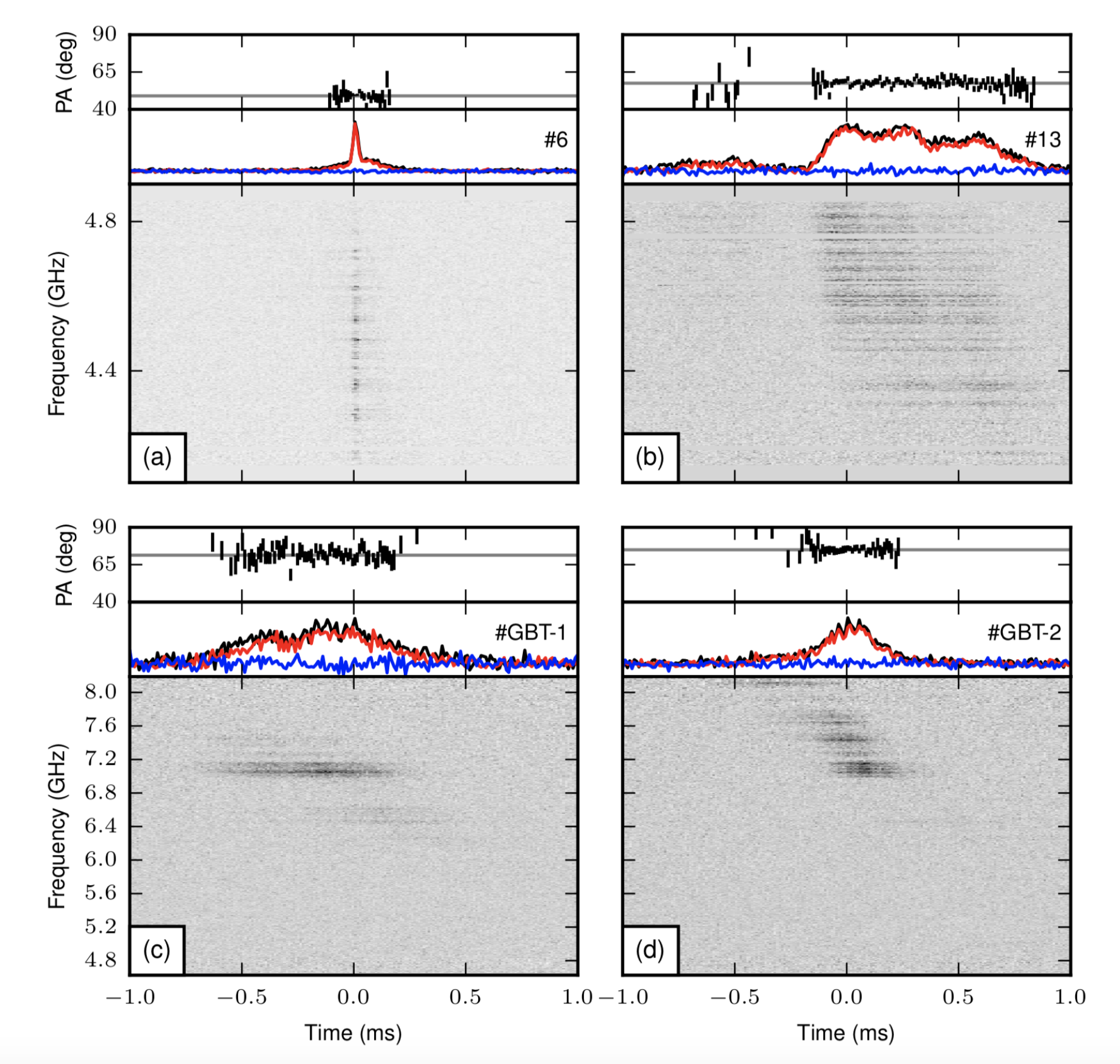}
\includegraphics
[width=6in]{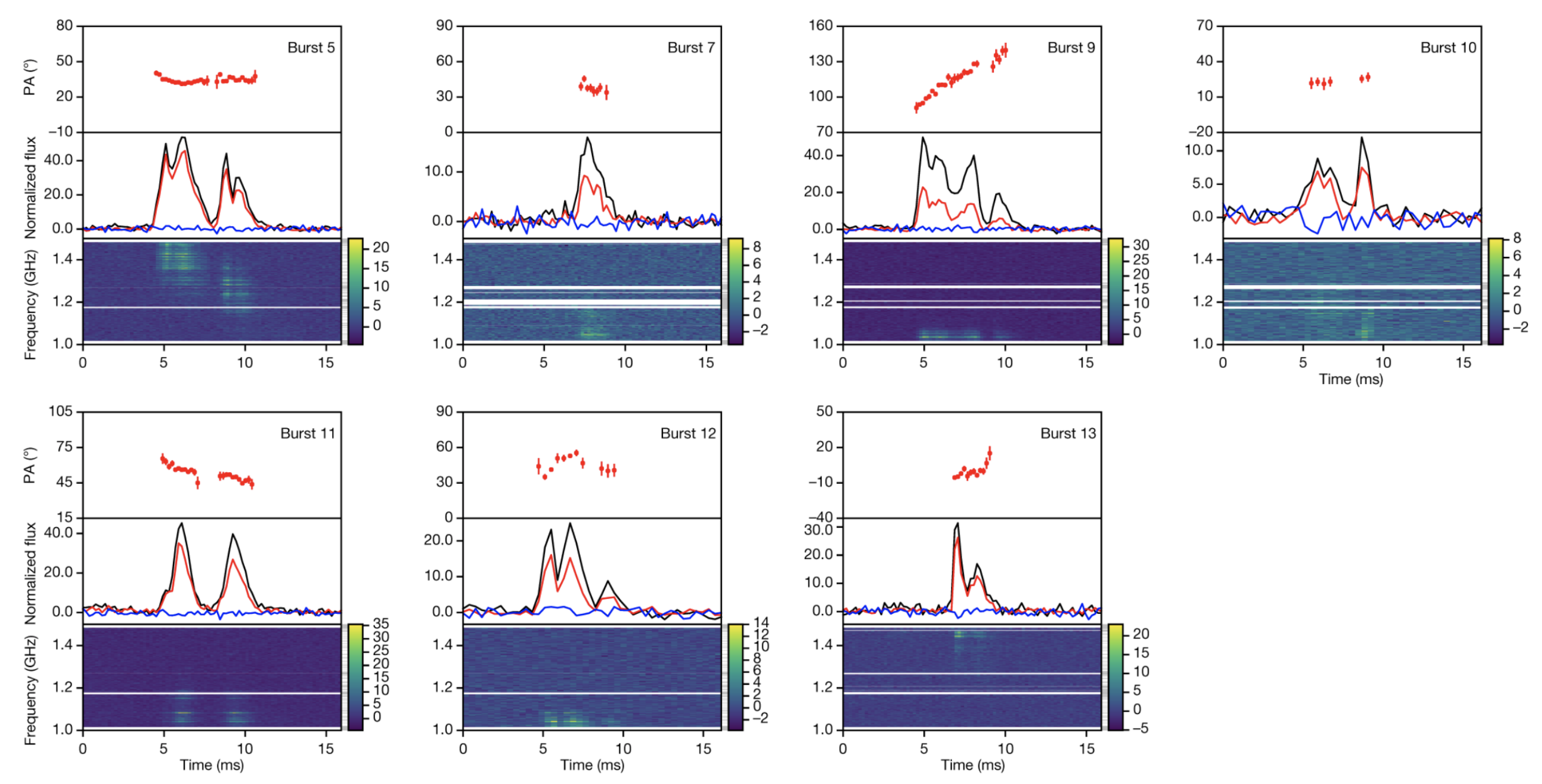}
\caption{Examples of polarization angle (PA) variations across individual bursts from FRBs. For each sub-figure, the upper panel is the polarization angle, the middle panel is the lightcurve, and the lower panel is the dynamic spectrum. {\em Upper figure:} constant PA in rFRB 20121102A bursts. From \citet{michilli18}. Reproduced with permission from Springer Nature; {\em Lower figure:} diverse PA swing patterns in rFRB 20180301A. From \citet{luo20b}.  }
\label{fig:poln}
\end{figure*}

\begin{figure*}
\includegraphics
[width=7in]{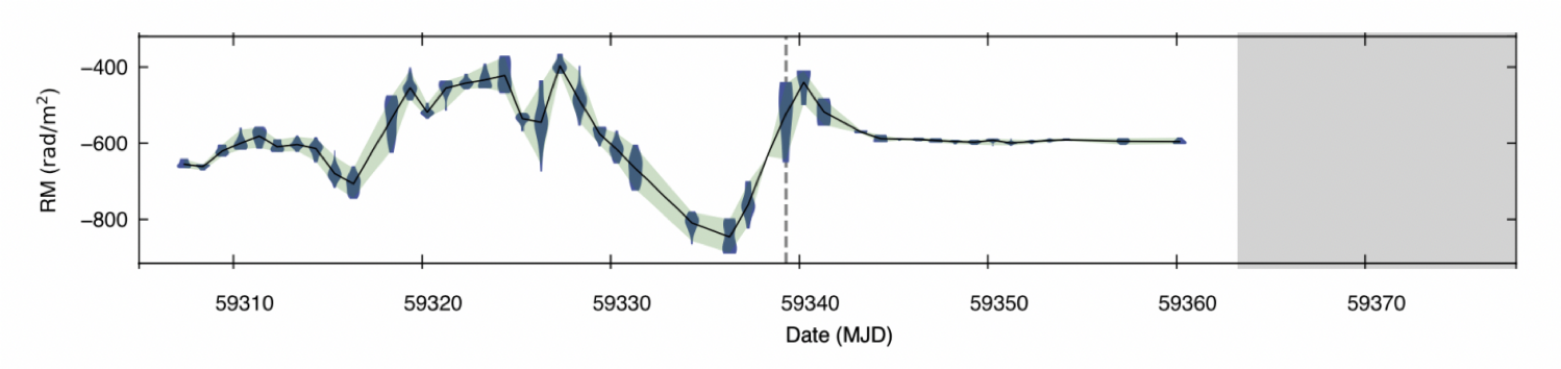}
\includegraphics
[width=5in]{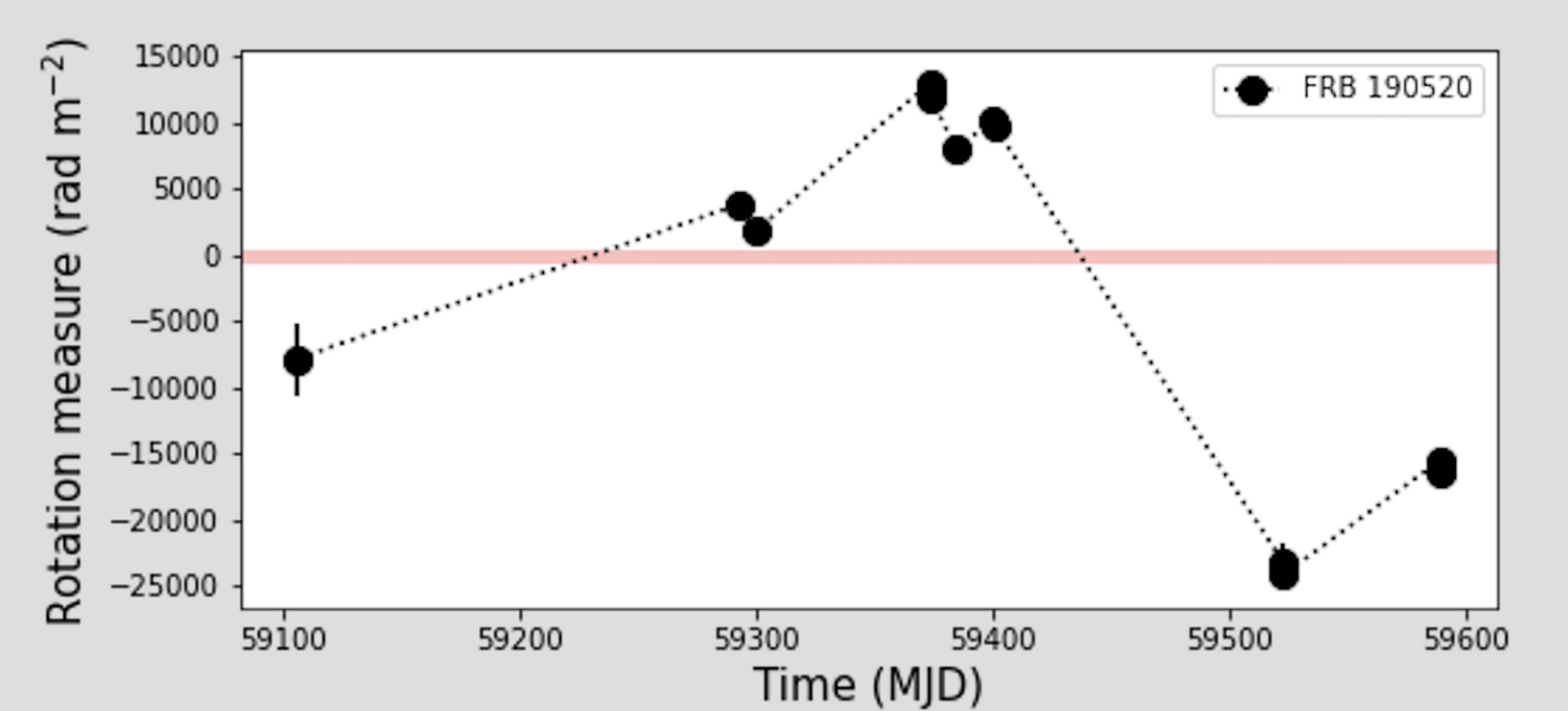}
\caption{Examples of short-term rotation measure (RM) variations in active repeating FRBs. {\em Upper:} irregular RM variations observed during a 54-day compaign with FAST. From \citet{xuh22}; {\em Lower:} surprising RM reversal from rFRB 201890520B. From \citet{anna-thomas22}.  }
\label{fig:RM}
\end{figure*}

 \subsection{Polarization properties and rotation measure}\label{sec:RM}

According to \citet{petroff19}, early polarization measurements indicated a puzzling, heterogeneous picture: the polarization properties can vary significantly among bursts. The high-quality polarization data accumulated later suggested a more consistent picture: it seems that most FRBs have strongly polarized emission. The linear polarization degree is typically $\Pi_{\rm L} > 30\%$, sometimes nearly 100\% \cite{michilli18,luo20b,cho20,day20}. The apparent low polarization observed in some FRBs might be intrinsic, but could be also due to the large {\em Faraday rotation measure} (RM, see Eq.(\ref{eq:RM}) below) in these sources, as is the case of rFRB 20121102A \cite{michilli18}. A frequency-dependent linear polarization degree has been observed in some FRBs, but it could be understood within a picture that the multi-path propagation effect introduces a scatter of RM so that the intrinsically strong polarization is smeared at low frequencies \citep{feng22}. Strong circular polarization has been observed in both apparently non-repeating FRBs \cite{petroff15,masui15,caleb18} and repeating FRBs \cite{kumar22,xuh22}. For linear polarization, the polarization angle (PA) remains constant across each burst for some FRBs (e.g. rFRB 20121102A, \citet{michilli18}, see Fig.\ref{fig:poln} upper panel). However, in some other FRBs, both apparent one-off ones \cite{cho20} and repeating ones \cite{luo20}, swings of PA across each burst are clearly observed, and the swing patterns are quite diverse among bursts (Fig.\ref{fig:poln} lower panel). For the most detailedly studied repeater rFRB 20201124A, even though most of bursts are consistent with non-varying PAs, significant PA variations above $5\sigma$ are observed in $\sim 33\%$ of bursts \citep{jiangjc22}.

Linearly polarized radio waves propagating in a magnetized medium would have the polarization angle undergoing a frequency-dependent variation known as ``Faraday rotation''. The degree of rotation is measured by the rotation measure defined by
    \begin{equation}
        {\rm RM} = (-0.81 \ {\rm rad \ m^{-2}}) \ \int_0^{D_z}  \frac{ [B_\parallel(l) / {\rm \mu G}] n_e(l)}{[1+z(l)]^2} dl,
        \label{eq:RM}
    \end{equation}
where $B_\parallel(l)$ is the $l$-dependent magnetic field strength along the line of sight (in units of micro-Gauss), $n_e$ is the number density of the medium along the line of sight in units of $\rm cm^{-3}$, and $l$ is in units of pc. FRBs have a wide range of measured RM absolute values: whereas some of them have sizeable RMs ranging from a few hundreds to $\sim 10^5 \ {\rm rad \ m^{-2}}$ in the case of FRB 20121102A \cite{michilli18}, some others have RMs consistent with being close to zero and could be used to place a  constraint on the magnetic field strength in the intergalactic medium (IGM) \cite{ravi16}. The distribution of $\rm RM/DM$ of FRBs, which gives a rough estimate of $|B_\parallel|$, is slightly larger but not inconsistent with the distribution of Galactic pulsars \cite{wangwy20}. 

The observed RM values of active repeaters show interesting variations. The first repeater rFRB 20121102A \citet{michilli18}) showed a secular decaying trend in RM. Short-term RM variation was observed in rFRB 20180301A \citet{luo20} and more clearly in rFRB 20201124A \cite{xuh22}. As shown in Figure \ref{fig:RM} upper panel, during an active episode of rFRB 20201124A, the RM of the source showed irregular variations during the first 36 days and turned to essentially invariant for another 18 days before the source quenched \cite{xuh22}. Another active repeater, rFRB 20190520B \cite{niu22}, showed an even weirder behavior. Its very large RM value of the order of $10^4 \ {\rm rad \ m^{-2}}$ underwent an unexpected reversal within 6 months (\citet{anna-thomas22,dais22}, see Fig.\ref{fig:RM} lower panel).

\subsection{Global properties} 

\begin{figure}
\includegraphics
[width=\columnwidth]{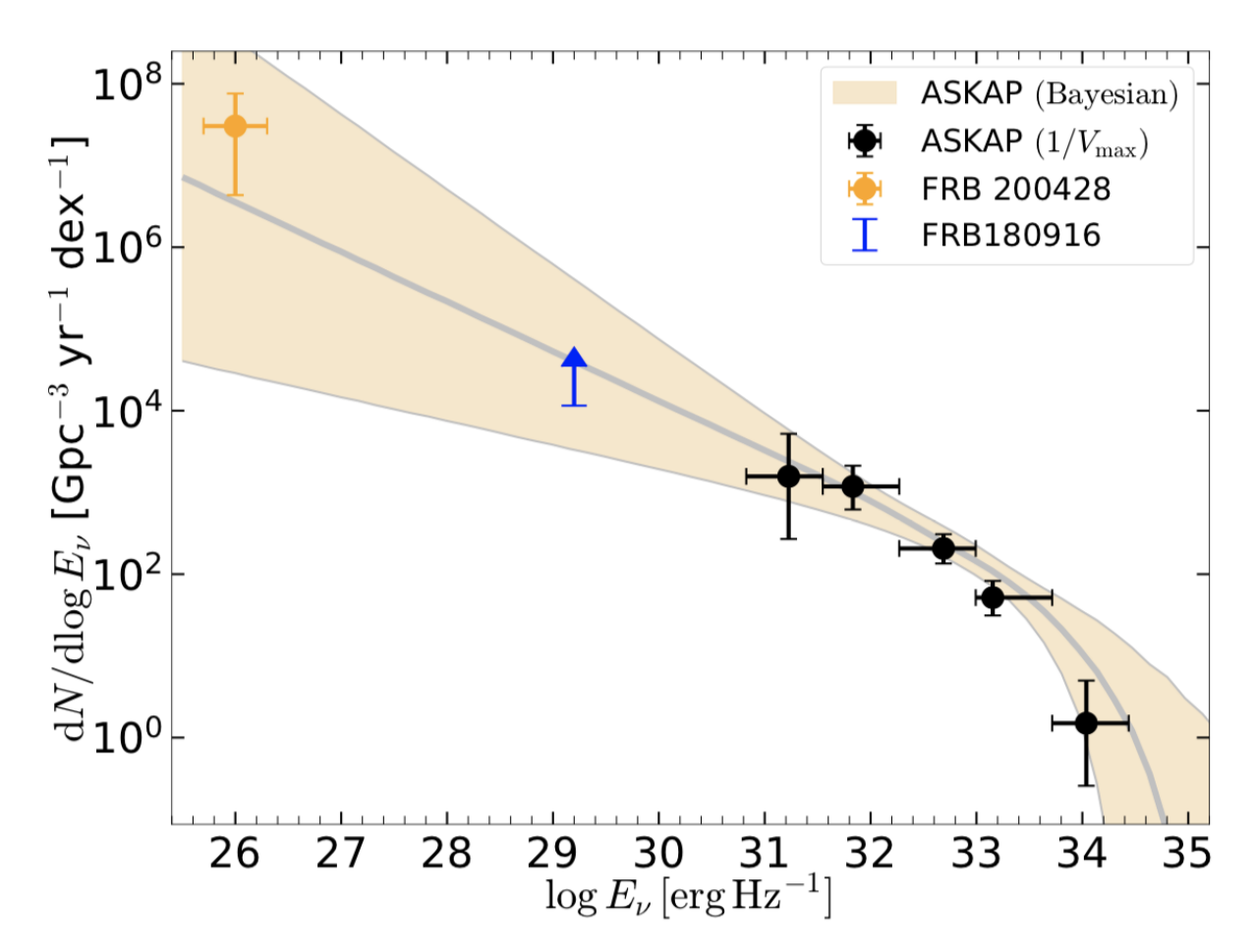}
\includegraphics
[width=\columnwidth]{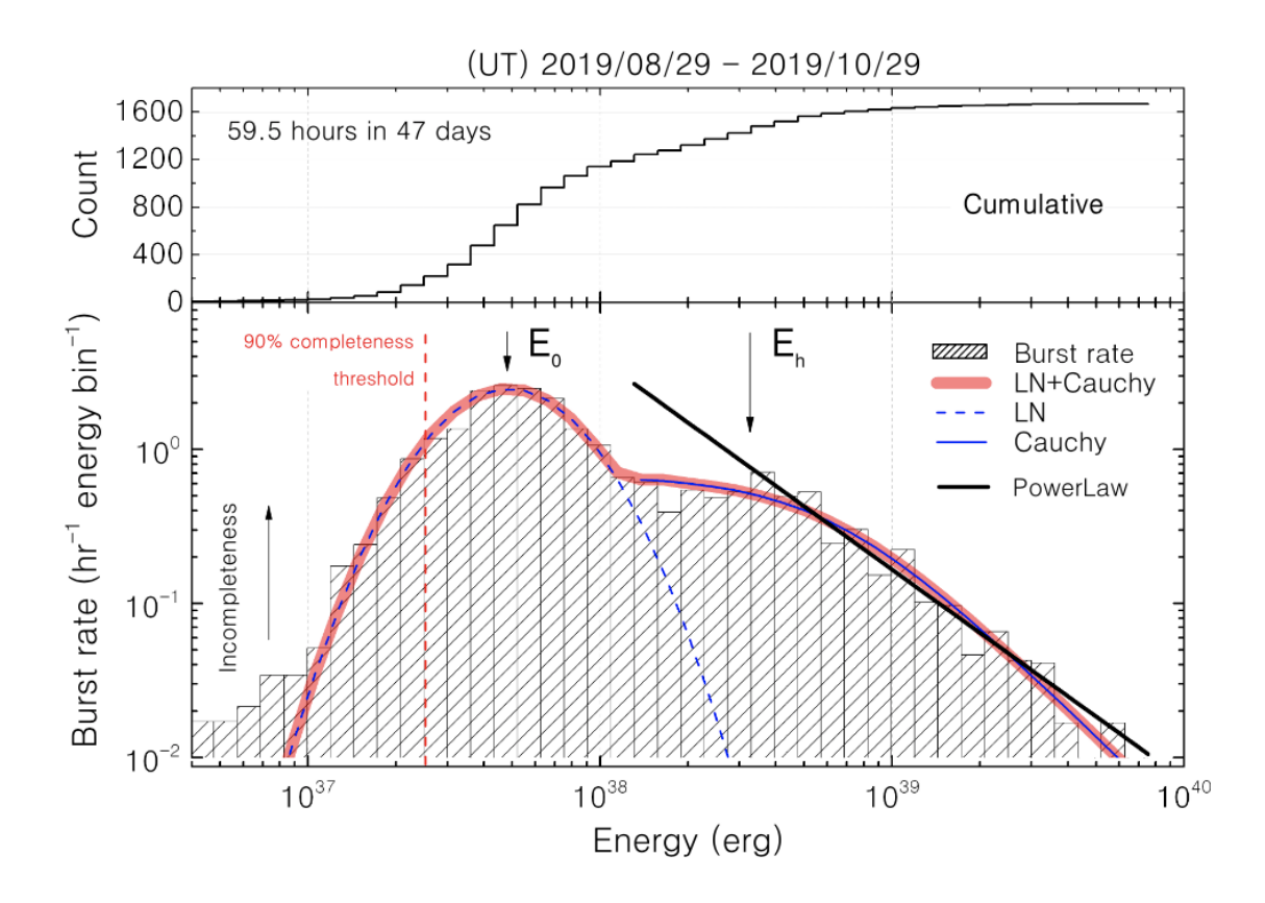}
\caption{Energy distribution of FRBs. {\em Upper:} The FRB isotropic energy distribution among different sources that shows a rough $-1.8$ power law distribution covering at least 7 orders of magnitude. From \citet{lu20}. {\em Lower:} The energy distribution of 1652 bursts detected from rFRB 20121102A, which shows a bimodal distribution. From \citet{lid21}.}
\label{fig:FRB-LF}
\end{figure}

The ${\rm DM}-z$ relation allows one to estimate the isotropic peak luminosity and energy of FRBs. For individual sources, the estimated luminosity/energy can have a large error because of the uncertainty of the correlation. When a large sample of FRBs is considered, the uncertainties can be averaged out, so that the luminosity/energy function of FRBs can be reasonably studied. Independent groups \cite{luo18,luo20,lu19b,lu20,zhangrc21,zhangzhang22,hashimoto20,hashimoto22} reached the consistent conclusion that the bulk of the energy/luminosity function can be fit with a power law distribution:
\begin{equation}
    N(E) dE \propto E^{-\gamma_{\rm E}} dE, ~~ N(L) dL  \propto L^{-\gamma_{\rm L}} dL.
\end{equation}
The index $\gamma_{\rm E} \sim \gamma_{\rm L}$ is not well constrained, e.g. 1.3-1.9 \citep{lu19b} or 1.5-2.1 \citep{luo20}, but a central value 1.8 seems to be able to accommodate FRBs in at least 7 orders of magnitude, extending from $\sim 10^{26} \ {\rm erg \ Hz^{-1}}$ for the Galactic FRB 20200428 to $\sim 10^{33} \ {\rm erg \ Hz^{-1}}$, above which a possible exponential cutoff may exist \cite{luo20,lu20}, see Fig.\ref{fig:FRB-LF} upper panel. 

Besides global energy/luminosity distributions among FRB sources, for active repeaters one can derive detailed energy/luminosity distributions for individual sources. The most comprehensive analysis has been done for a few active repeaters using FAST data. \citet{lid21} reported the detection of more than 1600 bursts detected from rFRB 20121102A in 47 days and found that there exist two components in the energy distribution. Whereas the high-energy part is consistent with a power law distribution, a distinct log-normal distribution component peaking at $E_0 \sim 4.8 \times 10^{37}$ erg at 1.25 GHz is observed (Fig.\ref{fig:FRB-LF} lower panel). The energy distributions of rFRB 20201124A \citep{xuh22,zhangyk22} and rFRB 20190520B \citep{niu22} show somewhat different shapes, but all require more complicated functions than the simple power law function. 

With the observed DM distribution, one can in principle constrain the redshift distribution of FRBs. The observed DM distribution is the convolution of the intrinsic redshift distribution, FRB energy/luminosity function, and the instrumental fluence/flux sensitivity threshold, so inferring it is not straightforward. One needs to apply a uniform sample (e.g. FRBs detected with the same telescope) to place the constraints. With the pre-CHIME data,  \citet{zhangrc21} tested several astrophysically-motivated redshift distribution models, from a model assuming FRBs tracking star-forming history to a model assuming FRBs tracking compact star merger events, which have a significant delay with respect to star formation. They found that the available Parkes or ASKAP FRBs are not inconsistent with either model. \citet{james22} showed that the simple non-evolution model is inconsistent with the data and found that the star formation model is consistent with the ASKAP data. However, they did not test models invoking delays with respect to star formation. \citet{hashimoto20} suggested that the limited data are consistent with no evolution with redshift.

With the first CHIME-catalog, the FRB redshift distribution can be further constrained. \citet{zhangzhang22} pointed out that the DM distribution peaks at a value lower than predicted by the star formation history model and suggested that the CHIME FRB data are consistent with a redshift model with a significant delay with respect to star formation. The conclusion was confirmed by \citet{hashimoto22} and \citet{qiang22}, with the former group also claiming that the data are consistent with FRBs tracking the stellar mass rather than star formation rate. Using a reduced sample from the CHIME catalog, \citet{shin22} found that the CHIME bursts are still consistent with following the star formation history. However, this might be because \citet{shin22} have adopted  criteria to remove low DM and low S/N bursts, which have removed a significant number of nearby low-luminosity FRBs. However, those removed FRBs are the dominant population that demands a delayed distribution from star formation. The existence of rFRB 20200120E in a globular cluster in M81 \cite{kirsten22} suggests that such burst sources should be in abundance, which require significant delay from star formation. 

\subsection{Host galaxies}
The first identified FRB host galaxy, that of rFRB 20121102A, is a low-metallicity, dwarf star-forming galaxy, which is quite analogous to those of long-duration gamma-ray bursts (LGRBs) and superluminous supernovae (SLSNe) \cite{tendulkar17,nicholl17}. On the other hand, the later identified host galaxies, mostly for apparently non-repeating FRB sources, are typically Milky-Way-like massive spiral galaxies \cite{bannister19,ravi19,marcote20,bhandari20,heintz20}. The positions of FRBs within the host galaxies also carry clues for the origin of FRB sources. Even though rFRB 20121102A is located in an active star formation region of the host galaxy \cite{tendulkar17,nicholl17}, most other FRBs, especially apparently non-repeating ones, are not. Instead, many of them lie in the outskirt of the host galaxies with not particularly high star formation rate \cite{bhandari20,heintz20}. The active repeater rFRB 20201124A has a Milky Way-like massive host galaxy with high star formation rate \cite{fong21,ravi22,piro21}. Detailed observations with the Keck telescopes suggested that the host galaxy is a metal-rich, barred spiral galaxy, with the FRB source residing in a low stellar density, interarm region at an intermediate galactocentric distance \citep{xuh22}. This is inconsistent with the environment expected for long GRBs and superluminous supernovae. 
Cross comparing the host galaxy and FRB position properties with other astronomical transients, \citet{lizhang20} showed that the global properties of FRBs are inconsistent with those of LGRBs and SLSNe, but are more consistent with Type II SNe and even compact object mergers. Overall, FRBs are not inconsistent with being all produced by magnetar engines, even though multiple formation channels are also possible. \citet{bochenek21} compared the host properties of FRBs and core-collapse supernovae and reached the conclusion that the FRB environments are consistent with core collapse supernovae making magnetars. 

\subsection{Counterparts}

Most FRBs do not have counterparts detected in other bands or other messenger channels (e.g. gravitational waves and neutrinos). Searches have been conducted, and some putative counterparts were reported but not confirmed \citep[e.g.][]{keane16,delaunay16,williams16,sakamoto21}. So far, only two confirmed multi-wavelength counterparts have been observed for a few sources:

First, both rFRB 20121102A \citep{chatterjee17,marcote17} and rFRB 20190520B \citep{niu22} are found to be associated with a point-like persistent radio source (PRS). Incidently, these two sources are also active repeaters with relatively large RMs. It is suspected that all repeaters may have an associated synchrotron-emitting PRS (either a supernova remnant, a magnetar wind nebula, or a mini-AGN) but only the ones with a dense and highly magnetized environment (so a large RM) could be detectable \citep{yang20b,yang22}. 

Second, the Galactic FRB 20200428 \citep{CHIME-SGR,STARE2-SGR} detected from the magnetar SGR J1935+2154 was associated with a contemporary hard X-ray burst \citep{HXMT-SGR,Integral-SGR,Konus-SGR,AGILE-SGR}. Searches for  X-ray/$\gamma$-ray emission in association with cosmological FRBs have been carried out for multiple sources with null results \citep[e.g.][]{zhangzhang17,yangyh19,cunningham19,guidorzi20,piro21,laha22a,laha22b,xuh22}. The non-detection is expected since the predicted X-ray flux is below the sensitivity threshold of the detectors for cosmological FRBs even if the X-ray-to-radio luminosity ratio is the same as FRB 20200428. It is worth noting that there was a stringent optical upper limit (Z-equivalent 17.9 mag in a 60-s exposure) during the prompt epoch of FRB 20200428 \citep{lin20}. Since the prompt optical flux is very low even for the Galactic FRB, the chance of detecting a prompt optical counterpart for cosmological FRBs is slim. Recent searches have set up more upper limits in the optical band for some nearby FRBs before, during and after the bursts \citep{niino22,hiramatsu22}. The non-detection is consistent with the expectation that the optical counterparts of FRBs are faint \citep[e.g.][]{yang19}.

Searches for FRBs following some GRBs or superluminous supernovae in the timescale of years have been carried out but with null results \citep{men19,law19}. Searches for progenitor explosions years prior to some FRBs have been also carried out, with some candidates reported \citep{wangxg20,lil22}.

Searches for gravitational waves (GWs) temporarily coincident with CHIME FRBs have been carried out, which led to tight upper limits on the GW fluxes \cite{LIGO-FRB22,wangnitz22}. The null results imply at most ${\cal O} (0.01)\% - {\cal O} (1)\%$ of FRBs are associated with compact binary coalescences (CBCs), which is consistent with the much higher rate density of FRBs than CBCs. Allowing a time difference between FRBs and GW events, a potential association pair between the NS-NS merger event GW190425 and a bright CHIME burst FRB 20190425A, with the FRB delayed by 2.5 hours with respect to the GW event, has been suggested \cite{moroianu22}. Its candidate host galaxy and the FRB environment are consistent with those expected for an NS-NS merger \cite{panther23}. 

\section{Basic plasma physics}\label{sec:basic}

A plasma is a gas that contains a significant fraction of charged particles, usually with charge balance between negatively charged species (free electrons) and positively charged species (positive ions or positrons). An FRB is likely produced in a plasma and radio waves need to propagate through  plasmas before reaching Earth. The discussion of the physics of FRBs inevitably involves plasma physics, which we briefly review in this section. 

\subsection{Plasma physics in the FRB context}

The most important property of a plasma is the double reaction between particles and electromagnetic (EM) fields. While the EM fields would control the motion of the plasma, the motion of the plasma would generate currents and alter EM fields. The description of the physical behavior of a plasma is therefore complicated \citep[e.g.][]{kulsrud05}. In general, one needs to solve the evolution of the particle component (i.e. each species of the plasma) in six-dimensional $(\vec r, \vec v)$ phase space in the form of the Fokker-Planck equation, and to solve the EM field component in three dimensions in the form of Maxwell equations. For each particle and field component, one also needs to consider the physics in three scales: the large scale of smooth particle distribution and EM fields, the small scale of particle distribution and EM field variations due to particle collisions, and the intermediate scale variation of particle distribution and EM fields dictated by various plasma waves. 

For the FRB problem, the most relevant scale is the intermediate one related to plasma waves. In many FRB radiation models, the observed FRB emission is related to certain types of plasma waves in the emission region to begin with. The microscopic particle collisional/collisionless interaction processes are usually not important in interpreting FRB observations and we will not discuss them in the rest of the review. The largest macroscopic scale, on the other hand, could be important. This is particularly true if the emission region is from the magnetosphere of a rotating object (e.g. a magnetar), in which case the global magnetic field configuration and plasma density distribution play an important role in defining FRB emission properties. For models invoking relativistic shocks, the globally ordered magnetic fields also play an important role in reproducing some properties (e.g. high brightness temperature, high linear polarization degree) of FRB observations. More generally, radio waves associated with FRBs need to go through the plasmas between the source and the observer, undergoing dispersion, absorption, scattering, scintillation, and Faraday rotation and conversion for polarized emission. In the rest of the section, we discuss the basics of dispersion and Faraday rotation and conversion, and leave more complicated multi-path effects (e.g. scattering, scintillation and plasma lensing effects) to Section \ref{sec:environment}. 

\subsection{Radio wave propagation in a non-magnetized plasma}\label{sec:DM2}

Electromagnetic waves are oscillations of electromagnetic fields in both space and time in the form of $\exp i(\vec k \cdot \vec r - \omega t)$. When waves with a particular frequency go through a stationary plasma, even though their oscillations in time (represented by angular frequency $\omega$) remain the same as in vacuum, their oscillations in space (represented by wave number $k$) would be modified in a frequency-dependent manner depending on the plasma properties. This leads to a varying wave propagation speed with frequency, known as dispersion. The relationship $\omega = \omega(k)$ is known as the dispersion relation.

The dispersion relation of EM waves propagating in  a non-magnetized, globally neutral plasma can be straightforwardly derived by introducing a space and time variation of all quantities of the form of $\exp i(\vec k \cdot \vec r - \omega t)$ in Maxwell's equations and Newton's second law equation involving the Lorentz force. The final dispersion relation reads \citep[e.g.][]{rybicki79} 
    \begin{equation}
        c^2k^2 = \epsilon \omega^2, ~~{\rm or} ~~ ck = n_r \omega, ~~ {\rm or}~~\omega^2 = \omega_p^2 + k^2 c^2
    \label{eq:dispersion}
    \end{equation} 
where 
    \begin{equation}
        \epsilon \equiv n_r^2 \equiv 1-\frac{4\pi\sigma}{i\omega} = 1 - \left(\frac{\omega_p}{\omega} \right)^2
    \end{equation}
is the dielectric constant, $n_r$ is the index of refraction, $\sigma$ is conductivity defined by ${\vec j} = \sigma {\vec E}$, and 
    \begin{equation}
        \omega_p \equiv \left(\frac{4\pi n_e e^2}{m_e} \right)^{1/2} \simeq (5.63 \times 10^4 \ {\rm s}^{-1}) n_e^{1/2}
     \label{eq:omega_p}
    \end{equation}
is the plasma frequency, where $n_e$ is the plasma density, $e$ and $m_e$ are the charge (absolute value) and mass of the electron, respectively. Noticing that $\epsilon \geq 0$ is required to have a real solution of the dispersion relation $\omega = \omega(k)$, one can see that $\omega_p$ defines a cutoff frequency, below which the EM waves cannot propagate. This is also the oscillation frequency of longitudinal waves (Langmuir waves) in a plasma\footnote{Note that the terms {\em longitudinal} ($\vec k \parallel \vec E$) and {\em transverse} ($\vec k \perp \vec E$) indicate the direction of wave propagation with respect to the electric field $\vec E$. EM waves are transverse waves. On the other hand, the terms {\em parallel} ($\vec k \parallel \vec B$) and {\em perpendicular} ($\vec k \perp \vec B$) indicate the direction of wave propagation with respect to the magnetic field $\vec B$.}. If the FRB frequency (typically $\sim$ GHz) is related to the plasma frequency, one requires $n \simeq (1.2 \times 10^{10} \ {\rm cm}^{-3}) \ \nu_{\rm FRB,9}^2$, where $\nu_{\rm FRB,9} = \nu_{\rm FRB} / (10^9 \ {\rm Hz})$ and throughout the review the convention $Q_n = Q/10^n$ is adopted in cgs units.
    
The dispersion measure (DM) discussed in section (\ref{sec:DM}) is defined through deriving the arrival time difference of a pulse in two different spectral bands. One may start with the 
dispersion relation (\ref{eq:dispersion}), which gives the group velocity of the dispersed wave 
\begin{equation}
v_g(\nu) \equiv \frac{\partial \omega}{\partial k} = c \left(1- \frac{\omega_p^2}{\omega^2}\right)^{1/2}.
\end{equation}
This gives a frequeny-dependent arrival time of radio waves
\begin{equation}
    t(\nu) = \int_0^{D} \frac{dl}{v_g(\nu)} \simeq \int_0^D \frac{dl}{c} \left(1+ \frac{1}{2}\frac{\omega_p^2}{\omega^2}\right),
\end{equation}
where the approximation $\omega \gg \omega_p$ has been adopted. The arrival time difference between two frequencies $\nu_2 > \nu_1$ can be expressed as
\begin{eqnarray}
    \Delta t & = & t(\nu_1) - t(\nu_2) = \frac{e^2}{2\pi m_e c} \left(\frac{1}{\nu_1^2} - \frac{1}{\nu_2^2} \right) \ {\rm DM} \nonumber \\
    & \simeq & (4.15 \ {\rm ms}) \left(\frac{1}{\nu_{\rm 1,GHz}^2} - \frac{1}{\nu_{\rm 2,GHz}^2} \right) \ \frac{\rm DM}{\rm pc \ cm^{-3}} 
\end{eqnarray}
where 
\begin{equation}
{\rm DM} = \int_0^D n_e dl
\label{eq:DM2}
\end{equation}
is defined. 
For a cosmological source, considering that the observed time $t_{\rm obs} = (1+z) t$ and the observed frequency $\nu_{\rm obs} = \nu/(1+z)$, the final expression of DM is Eq.(\ref{eq:DM}) when $t$ and $\nu$ are expressed in terms of the observed values. Defining 
\begin{equation}
{\cal D} \equiv \frac{\Delta t}{\Delta (1/\nu^2)} = \frac{t_{\nu_1} - t_{\nu_2}} {\frac{1}{\nu_1^2}-\frac{1}{\nu_2^2}},
\end{equation}
one can write 
\begin{equation}
    {\rm DM} = K {\cal D},
\end{equation}
where \citep{kulkarni20}
\begin{equation}
    K = 241.0331786(66) \ {\rm GHz^{-2} \ cm^{-3} \ pc \ s^{-1}},
\end{equation}
and DM is in units of $\rm cm^{-3} \ pc$. Notice that many assumptions have entered the above derivation \citep[e.g.][]{kulkarni20}: The motion of ions is neglected, the medium is cold, not moving with respect to the observer, and not magnetized. These factors are not important if the purpose is to give a rough estimate of electron column density along the line of sight but could be essential to perform precise measurements of arrival times and cross check the measurements of the same source by different detectors (e.g. the detection data of FRB 200428 between CHIME and STARE2). 

\subsection{Radio wave propagation in a magnetized plasma}

\subsubsection{General discussion}
When a plasma carries an ordered magnetic field $\vec B$, the dispersion relation is much more complicated. Besides the plasma frequency $\omega_p$, another characteristic frequency, the electron gyration frequency $\omega_{\rm B}$ (also called Larmor frequency $\omega_{\rm L}$), is introduced\footnote{The discussion in this subsection applies to the classical (non-quantum) plasma and wave regime. }. For non-relativistic motion, this frequency depends on $B$ and fundamental constants, i.e.
    \begin{equation}
        \omega_{\rm B} = -\Omega_e \equiv \frac{eB}{m_e c}=(1.76 \times 10^7 \ {\rm s^{-1}}) \ B.
    \label{eq:omega_B}
    \end{equation}
If the FRB frequency is related to $\omega_B$, the required magnetic field strength is $B\simeq (360 \ {\rm G}) \ \nu_{\rm FRB,9}$. Note that $\Omega_e = - eB/m_e c$ is defined as negative to contrast with the positive ion gyration frequency 
\begin{equation}
    \Omega_i = \frac{Z e B}{m_i c} = Z \frac{m_e}{m_i} | \Omega_e |,
\end{equation}
where $m_i$ is the mass of the positive ion and $Z$ is the atomic number of the ion. For an electron-positron ($e^{+}e^{-1}$) pair plasma, one has $\Omega_i = |\Omega_e| = \omega_{\rm B}$. 
    
The existence of $\vec B$ introduces another special direction besides the wave propagation direction 
\begin{equation}
\vec n_r = \frac{c}{\omega}\vec k.
\end{equation}
The dispersion relation becomes angle-dependent. Repeating the exercise of wave expansion for the Maxwell's equations and Lorentz force equation for a global neutral plasma, one gets a dielectric tensor to replace the dielectric constant, which reads \citep[e.g.][]{boyd03,meszaros92,stix92}
    \begin{equation}
        \Vec{\Vec \epsilon}
        \equiv \left(
        \begin{array}{rrr}
         S & -i D & 0 \\
         i D & S & 0 \\
         0 & 0 & P,
        \end{array}
        \right).
    \end{equation}
    This is defined from 
    \begin{equation}
    \vec n_r \times (\vec n_r \times \vec E) = -
    \Vec{\Vec \epsilon} \cdot \vec E
    \label{eq:dispersion1}
    \end{equation}
    (which itself comes from the fourth Maxwell equation, $\vec j = \Vec{\Vec \sigma} \cdot \vec E$ with 
    $\Vec{\Vec \sigma}$ being the conductivity tensor,  $\Vec{\Vec \epsilon}$ is defined as 
    $\Vec{\Vec \epsilon}=\Vec{\vec I}-4\pi
    \Vec{\Vec \sigma} /i\omega$, and $\Vec{\vec I}$ is the unit tensor), where $\vec E$ is the electric field vector of the waves, and the magnetic field direction is defined as the $\hat z$ direction. Here,
\begin{eqnarray}
 S &= &\frac{1}{2}(R+L) = 1 - \frac{\omega_p^2(\omega^2+\Omega_i\Omega_e)}{(\omega^2-\Omega_i^2)(\omega^2-\Omega_e^2)}, \label{eq:S} \\
 D &= &\frac{1}{2}(R-L) = \frac{\omega_p^2\omega(\Omega_i+\Omega_e)} {(\omega^2-\Omega_i^2)(\omega^2-\Omega_e^2)}, \label{eq:D} \\
 R &= &1 - \frac{\omega_p^2}{(\omega+\Omega_i)(\omega+\Omega_e)}, 
     \label{eq:R} \\
 L &= &1 -\frac{\omega_p^2}{(\omega-\Omega_i)(\omega-\Omega_e)},
     \label{eq:L} \\
 P &= & 1-\frac{\omega_p^2}{\omega^2}, \label{eq:P}
\end{eqnarray}    
where $R$, $L$, $P$ denote parameters related to the ``right'', ``left'', and ``plasma'' modes, respectively, and $S$ and $D$ denote ``sum'' and ``difference'', respectively. 

Very generally, $\vec n_r$ and $\vec B$ can have an angle $\theta$. One can write $\vec n_r = (n_r \sin\theta, 0, n_r \cos\theta)$ without loss of generality, so that Equation (\ref{eq:dispersion1}) becomes $(\vec n_r \cdot \vec E) \vec n_r - n_r^2 \vec E + \vec{\vec \epsilon} \cdot \vec E = 0$, or
\begin{equation}
    \left(
    \begin{array}{ccc}
     S - n_r^2 \cos^2\theta & -iD & n_r^2\cos\theta\sin\theta \\
     iD & S-n_r^2 & 0 \\
     n_r^2\cos\theta \sin\theta & 0 & P-n_r^2\sin^2\theta 
     \end{array}
     \right) 
     \left(
     \begin{array}{c}
      E_x \\
      E_y \\
      E_z
     \end{array}
     \right) = 0. 
     \label{eq:general}
\end{equation}
Taking the determinant of the coefficients, the general dispersion relation for waves propagating in a cold, magnetized plasma becomes
\begin{equation}
    A n_r^4 - B n_r^2 + C = 0,
\end{equation}
where
\begin{eqnarray}
 A & = & S \sin^2\theta + P \cos^2\theta, \\
 B & = & RL \sin^2\theta + PS(1+\cos^2\theta), \\
 C & = & PRL.
\end{eqnarray}

In the following, we consider the dispersion relations for a cold, magnetized plasma for different cases of the angle between $\vec k$ (or $\vec n_r$) and $\vec B$: 

\subsubsection{$\vec k \parallel \vec B$} 

When the wave vector is along the magnetic field (e.g. for FRB waves propagating in the open field line region of a magnetosphere), Equation (\ref{eq:general}) becomes  
\begin{equation}
    \left(
    \begin{array}{ccc}
        S-n_r^2 & -iD & 0  \\
        iD & S-n_r^2 & 0 \\
        0 & 0 & P
    \end{array}
    \right) 
     \left(
     \begin{array}{c}
      E_x \\
      E_y \\
      E_z
     \end{array}
     \right) = 0.     
\end{equation}
Besides the $P=0$ plasma mode ($\omega^2 = \omega_p^2$), one has two transverse wave modes, i.e. the R and L modes\footnote{Notice that opposite conventions of R-model and L-mode definitions have been used in different textbooks. For example, \citet{boyd03} defines right(left)-handed with respect to the photon propagation direction while \citet{rybicki79} defines right(left)-handed with respect to the line of sight direction towards the source. We adopt the \citet{boyd03} convention in the following discussion.}:
\begin{eqnarray}
 n_r^2 = R, & ~~{\rm R-mode} \\
 n_r^2 = L, & ~~{\rm L-mode}
\end{eqnarray}
with the dispersion relations
\begin{eqnarray}
\frac{c^2 k^2}{\omega^2} = R \simeq \left\{
\begin{array}{ll}
   1-\frac{\omega_p^2}{\omega(\omega-\omega_B)},  & {\rm ion} \\
   1-\frac{\omega_p^2}{\omega^2-\omega_B^2},  & {\rm pair}
\end{array}, \right. & ~~{\rm R-mode} \label{eq:R-DR}\\
 \frac{c^2 k^2}{\omega^2} = L \simeq \left\{
\begin{array}{ll}
   1-\frac{\omega_p^2}{\omega(\omega+\omega_B)},  & {\rm ion} \\
   1-\frac{\omega_p^2}{\omega^2-\omega_B^2},  & {\rm pair}
\end{array}, \right. & ~~{\rm L-mode} \label{eq:L-DR}
\end{eqnarray}
respectively. Note that hereafter for a pair plasma, the plasma frequency is defined as 
\begin{equation}
 \omega_p \equiv \left(\frac{4\pi n_\pm e^2}{m_e} \right)^{1/2} \simeq (5.63 \times 10^4 \ {\rm s}^{-1}) n_{\pm}^{1/2},
 \label{eq:omega_p2}
\end{equation}
in contrast to Eq.(\ref{eq:omega_p}), where $n_\pm = 2 n_e$ is the pair number density, which is twice of $n_e$ for a neutral pair plasma. If one still uses the electron number density $n_e$ to define $\omega_p$, all the pair-related dispersion relations should have $\omega_p^2$ replaced by $2 \omega_p^2$. This is because in Eqs.(\ref{eq:S})-(\ref{eq:P}), a small term $\omega_{p,i}^2= 4\pi n_e (Ze)^2/m_i$ in parallel to $\omega_p^2$ has been ignored. This terms becomes comparable to $\omega_{p,e}^2$ in the case of pairs. 

Setting $R=0$ and $L=0$ and looking for positive solutions\footnote{Negative frequencies simply mean waves propagating in the opposite direction. So solving positive solutions is complete in solving the propagation problem.}, one can define two cutoff frequencies
    \begin{eqnarray}
        \omega_{\rm R} & \equiv & \left[\omega_p^2+\frac{(\Omega_i-\Omega_e)^2}{4}
        \right]^{1/2}-\frac{(\Omega_i+\Omega_e)}{2} \nonumber \\
        & \simeq & 
        \left\{
         \begin{array}{ll}
        (\omega_p^2+\omega_{\rm B}^2/4)^{1/2}+\omega_{\rm B}/2, & {\rm ion} \\
         (\omega_p^2 + \omega_{\rm B}^2)^{1/2}, & {\rm pair}
         \end{array}
         \right.
    \end{eqnarray}
    and 
    \begin{eqnarray}
        \omega_{\rm L} & \equiv & \left[\omega_p^2+\frac{(\Omega_i-\Omega_e)^2}{4}
        \right]^{1/2}+\frac{(\Omega_i+\Omega_e)}{2} \nonumber \\
        & \simeq & 
        \left\{
         \begin{array}{ll}
        (\omega_p^2+\omega_{\rm B}^2/4)^{1/2}-\omega_{\rm B}/2, & {\rm ion} \\
         (\omega_p^2 + \omega_{\rm B}^2)^{1/2}, & {\rm pair}
         \end{array}
         \right. 
    \end{eqnarray}
    respectively. Here $\Omega_i \ll |\Omega_e|$ and  $\Omega_i = |\Omega_e|$ ($Z=1$) have been adopted for an ion plasma and a pair plasma, respectively. The propagation condition for the R-mode and L-mode waves depends on the sign of the denominators in Equations (\ref{eq:R}) and (\ref{eq:L}), respectively. 
    
 Setting $R\rightarrow \infty$ and $L \rightarrow \infty$, one can define two principle resonances at $\omega_{\rm res,R}=|\Omega_e|=\omega_{\rm B}$ and $\omega_{\rm res,L}=\Omega_i$. The frequency range that radio waves can propagate is defined by $n_r^2 > 0$, which is 
 \begin{eqnarray}
  \omega > \omega_{\rm R}, & ~~{\rm or} \ \omega < \omega_{\rm B}, & ~~{\rm R-mode}, \\
  \omega > \omega_{\rm L}, & ~~{\rm or} \ \omega < \Omega_i, & ~~{\rm L-mode}.
 \end{eqnarray}
    
It is interesting to consider two asymptotic regimes. 
\begin{itemize}
\item In the regions far from the magnetosphere of a neutron star (e.g. in the ISM or IGM), one has $\omega \gg \omega_{\rm B}$ $\omega_p \gg \omega_{\rm B}$ and $|\Omega_e| \gg \Omega_i$. In this case, one has $\omega_{\rm R} \simeq \omega_{\rm L} \simeq \omega_p$. The wave propagation condition is $\omega > \omega_p$ for both R- and L-modes, which is essentially the same as a non-magnetized medium. 
\item In the regions within a neutron star magnetosphere and for a pair plasma, one has $\omega \ll \omega_{\rm B}$, $\omega_p \ll \omega_{\rm B}$ and $|\Omega_e|=\Omega_i=\omega_{\rm B}$. In this case, one has $\omega_{\rm R} \simeq \omega_{\rm L} \simeq (\omega_p^2+\omega_{\rm B}^2)^{1/2} \simeq \omega_{\rm B}$ and the resonances are also $\omega_{\rm B}$. The R-mode and the L-mode become the same and are essentially transparent in all frequencies. 
\end{itemize}

\subsubsection{$\vec k \perp \vec B$}

In another extreme case when the wave vector is perpendicular to the magnetic field (e.g. for FRB waves propagating in the closed field line region of a magnetosphere), Equation (\ref{eq:general}) becomes  
\begin{equation}
    \left(
    \begin{array}{ccc}
        S & -iD & 0  \\
        iD & S-n_r^2 & 0 \\
        0 & 0 & P-n_r^2
    \end{array}
    \right) 
     \left(
     \begin{array}{c}
      E_x \\
      E_y \\
      E_z
     \end{array}
     \right) = 0.     
\end{equation}
One can also define two modes: the ordinary (O-) and the extraordinary (X- or E-) modes, i.e.
\begin{eqnarray}
    n_r^2=P, & ~~{\rm O-mode} \\
    n_r^2 = \frac{RL}{S}, & ~~{\rm X-mode},
\end{eqnarray}
with the O-mode dispersion relation
\begin{equation}
    \frac{c^2k^2}{\omega^2}=P = 1-\frac{\omega_p^2}{\omega^2}, 
\label{eq:O-DR}
\end{equation}
and the X-mode dispersion relation
\begin{equation}
    \frac{c^2k^2}{\omega^2} = \frac{RL}{S} \simeq \left\{
\begin{array}{ll}
   \frac{(\omega^2-\omega_p^2)^2-\omega^2\omega_B^2}{\omega^2(\omega^2-\omega_p^2-\omega_B^2)},  & {\rm ion} \\
   1-\frac{\omega_p^2}{\omega^2-\omega_B^2},  & {\rm pair}
\end{array}. \right. 
\label{eq:X-DR}
\end{equation}
respectively. The O-mode corresponds to the case that the wave electric field vector is parallel to the background magnetic field vector, i.e. $\vec E_w \parallel \vec B$, so that electrons moving in response of $\vec E_w$ oscillations do not feel the existence of the $\vec B$ field. As a result, the dispersion relation is the same as the non-magnetized medium case, and hence, the mode is called ``ordinary''. The X-mode corresponds to the case of $\vec E_w \perp \vec B$. The electrons in response of $\vec E_w$ oscillations would also undergo gyration motion around the background $\vec B$ field, the hence, the mode is called ``extraordinary''. The X-mode has cutoffs ($k \rightarrow 0$) at $\omega_{\rm R}$ ($R=0$) and $\omega_{\rm L}$ ($L=0$), and principle resonances ($k \rightarrow \infty$) at $S=0$, which defines two (upper and lower) hybrid resonance frequencies
\begin{eqnarray}
    \omega_{\rm res,H}^2 & = & \left(\frac{\omega_p^2+\Omega_i^2+\Omega_e^2}{2}\right) \nonumber \\
    & \times & \left[1\pm \left(1+\frac{4\Omega_i\Omega_e(\omega_p^2- \Omega_i \Omega_e)}{(\omega_p^2+\Omega_i^2+\Omega_e^2)^2} \right)^{1/2} \right].
\end{eqnarray}
For an ion plasma, since $\Omega_e^2 \gg \Omega_i^2$, it is interesting to note that the second term in the square root is always $\ll$ 1. One therefore has
\begin{eqnarray}
 \omega_{\rm res,UH}^2 & \simeq & \omega_p^2+\Omega_e^2, \\
 \omega_{\rm res,LH}^2 & \simeq & -\frac{\Omega_i\Omega_e (\omega_p^2-\Omega_i\Omega_e)}{\omega_p^2+\Omega_i^2+\Omega_e^2}.
\end{eqnarray}
For an $e^\pm$ plasma with $\Omega_e^2=\Omega_i^2=\omega_{\rm B}^2$, one has
\begin{eqnarray}
 \omega_{\rm res,UH}^2 & = & \omega_p^2+\omega_{\rm B}^2, \\
 \omega_{\rm res,LH}^2 & = & \omega_{\rm B}^2.
\end{eqnarray}
The frequency range that radio waves can propagate ($n_r^2 > 0$) is
\begin{eqnarray}
 \omega > \omega_p, &  ~~{\rm O-mode}, \\
  \left\{
  \begin{array}{ll}
   &\omega > \omega_{\rm R}, \\
   {\rm or} & \omega_{\rm L} < \omega < \omega_{\rm res,UH}, \\
   {\rm or} & \omega < \omega_{\rm res,LH}.
  \end{array}
  \right.  & ~~{\rm X-mode}
\end{eqnarray}

One can again consider two asymptotic regimes.
\begin{itemize}
    \item In regions far from the magnetosphere of a neutron star (e.g. in the ISM or IGM), one has $\omega \gg \omega_{\rm B}$, $\omega_p \gg \omega_{\rm B}$ and $|\Omega_e| \gg \Omega_i$. In this case, one has $\omega_{\rm R} \simeq \omega_{\rm res,UH}$, $\omega_{\rm L} \simeq \omega_p$ and $\omega_{\rm res,LH} \simeq \omega_{\rm B}$. The wave propagation condition is $\omega > \omega_p$ for both O- and X-modes, which is the same as a non-magnetized medium. 
\item In regions within a neutron star magnetosphere and for a pair plasma, one has $\omega \ll \omega_{\rm B}$, $\omega_p \ll \omega_{\rm B}$ and $|\Omega_e|=\Omega_i=\omega_{\rm B}$. In this case, one has $\omega_{\rm R} \simeq \omega_{\rm res,UH} \simeq \omega_{\rm L} \simeq (\omega_p^2+\omega_{\rm B}^2)^{1/2} \simeq \omega_{\rm B}$ and $\omega_{\rm res,LH} = \omega_{\rm B}$.  So the X-mode is essentially transparent in all frequencies. The O-mode, however, can only propagate when $\omega > \omega_p$. Because of this, when radio waves propagate across the closed field line regions of a neutron star, the X-mode $\vec E_w$ vector would adiabatically rotate to maintain perpendicular to the local $\vec B$ until reaching the radius where $\omega > \omega_p$ is satisfied, at which the polarization vector is frozen out \citep{lu19}. 
\end{itemize}

\subsubsection{Oblique propagation}

When $\left< \vec k, \vec B \right> = \theta$ has an arbitrary angle, the dispersion relation should take Equation (\ref{eq:general}), which is more complicated (not discussed below due the limited space, but see \citet{boyd03,stix92}). Nonetheless, the treatments in the two extreme cases are helpful to discuss the general behavior of the dispersion relations when $\theta$ is small or close to $\pi/2$:
\begin{itemize}
    \item When $\theta \ll 1$, one has the quasi-parallel case. The dispersion relations can be modified from the R- and L-mode relations (Eqs.(\ref{eq:R-DR}) and (\ref{eq:L-DR})) by replacing $\omega_{\rm B}$ by $\omega_{\rm B} \cos\theta$.
    \item When $\theta \rightarrow \pi/2$, one has the quasi-perpendicular case. The O-mode dispersion relation is revised to
    \begin{equation}
        \frac{c^2k^2}{\omega^2} \simeq \frac{\omega^2-\omega_p^2}{\omega^2-\omega_p^2\cos^2\theta},
    \end{equation}
    which can be reduced to Eq.(\ref{eq:O-DR}) when $\theta=\pi/2$. The X-mode dispersion relation can be modified directly from Eq.(\ref{eq:X-DR}) by replacing $\omega_{\rm B}$ by  $\omega_{\rm B} \sin\theta$. 
\end{itemize}

In the literature, for the oblique cases, the X-mode and O-mode are usually defined as the cases when $\vec E_w$ is perpendicular and parallel to the $(\vec k, \vec B)$ plane, respectively. Note that the O-mode defined this way is not completely ``ordinary'', since there is still a $\vec E_w$ component that is perpendicular to $\vec B$. One should be cautious to extend the properties of the O-mode in the $\vec k \perp \vec B$ case to the more general O-mode. For example, the statement that O-mode cannot propagate in a neutron star magnetopshere is only valid in the quasi-perpendicular regime. In the quasi-parallel regime, even the ``O-mode'' is essentially extraordinary, i.e. a significant $\vec E_w$ component is perpendicular to $\vec B$. The waves can therefore also propagate.

\subsection{Faraday rotation}\label{sec:RM2}

Let us take a closer look at the propagation of radio waves in the case of $\vec k \parallel \vec B$ in an ion plasma. Dropping out $\Omega_i$, the R(L)-mode dispersion relations (Equations (\ref{eq:R-DR}) and (\ref{eq:L-DR})) can be generally written as
\begin{equation}
    \omega^2 = k^2 c^2 + \frac{\omega_p^2}{1\mp (\omega_{\rm B}/\omega)} \simeq k^2 c^2 + \omega_p^2 (1 \pm \frac{\omega_{\rm B}}{\omega}),
\end{equation}
where the $\omega_{\rm B} \ll \omega$ approximation has been adopted in the second equation, which is usually valid for the ISM and the IGM. 

Following the same procedure in \S\ref{sec:DM2} and replacing $\omega_p^2$ by $\omega_p^2(1 \pm \omega_{\rm B}/\omega)$ (again valid for $\omega_{\rm B} \ll \omega$), one gets
\begin{equation}
v_g(\nu) = c \left[ 1- \frac{\omega_p^2}{\omega^2}\left(1\pm \frac{\omega_{\rm B}}{\omega}\right) \right]^{1/2}.
\end{equation}
Further requires $\omega_p \ll \omega$, one can derive 
\begin{equation}
    t(\nu) \simeq \int_0^D \frac{dl}{c} \left[ 1+ \frac{1}{2}\frac{\omega_p^2}{\omega^2} \left(1\pm \frac{\omega_{\rm B}}{\omega}\right) \right],
\end{equation}
and
\begin{eqnarray}
    \Delta t & = & t(\nu_1) - t(\nu_2) = \frac{e^2}{2\pi m_e c} \left(\frac{1}{\nu_1^2} - \frac{1}{\nu_2^2} \right) \ {\rm DM} \nonumber \\
    & \pm & \frac{e^3}{(2\pi m_e c)^2} \left(\frac{1}{\nu_1^3} - \frac{1}{\nu_2^3} \right) \ \int_0^D n_e B_\parallel dl \nonumber \\
    & \simeq & (4.15 \ {\rm ms}) \left(\frac{1}{\nu_{\rm 1,GHz}^2} - \frac{1}{\nu_{\rm 2,GHz}^2} \right) \ \frac{\rm DM}{\rm pc \ cm^{-3}} \nonumber \\
    & \pm & (1.16 \times 10^{-11} \ {\rm s}) \left(\frac{1}{\nu_{\rm 1,GHz}^3} - \frac{1}{\nu_{\rm 2,GHz}^3} \right) \frac{\int_0^D n_e B_\parallel dl} {\rm pc \ cm^{-3} \ \mu G}. \nonumber \\
\end{eqnarray}
One can see that the effect of $B$ field in the arrival time has a $\nu^{-3}$ dependence, which is much smaller than the DM term. It depends on $\int_0^D n_e B_\parallel dl$ (a proxy of the rotation measure discussed below), but this term is practically not measurable. 

A measurement of $\int_0^D n_e B_\parallel dl$ is achievable by measuring the rotation of the polarization angle (PA) of linearly polarized waves as a function of frequency known as ``Faraday rotation''. Since linearly polarized waves can be decomposed as the superposition of a right-handed and a left-handed circularly polarized components and since the two modes (R- and L-modes) have different propagation speeds, the PA of the observed waves would display a frequency-dependent variation. Mathematically, this can be denoted as the variation of the phase difference of the circularly polarized waves as a function of frequency. Noticing $k_{\rm R}^2 c^2 = R \omega^2$, $k_{\rm L}^2 c^2 = L \omega^2$, and the phases of the R/L mode waves $\phi_{\rm R,L} = \int_0^D k_{\rm R,L} dl$, the rotation angle can be written as 
\begin{eqnarray}
 \Delta\phi & = & \frac{1}{2} \int_{0}^{D} (k_{\rm L} - k_{\rm R}) dl \nonumber \\
 & \simeq & -\frac{1}{2} \int_0^D \frac{\omega_p^2 \omega_{\rm B}}{c \omega^2} dl \nonumber \\
 & \simeq & -\frac{e^3 \lambda^2}{2\pi m_e^2 c^4} \int_0^D n_e B_\parallel dl \nonumber \\
 & = & \lambda^2 \ {\rm RM},
 \label{eq:Delta-phi}
\end{eqnarray}
where 
\begin{eqnarray}
 {\rm RM } & \equiv & -\frac{e^3}{2\pi m_e^2 c^4} \int_0^D n_e B_\parallel dl \nonumber \\
 & \simeq & (-0.81 \ {\rm rad \ m^{-2}}) \frac{\int_0^D n_e B_\parallel dl} {\rm pc \ cm^{-3} \ \mu G}.
 \label{eq:RM2}
\end{eqnarray}
For cosmological sources, the observed wavelength is $\lambda =(1+z) \lambda_{\rm sr}$, so a more general expression is Equation (\ref{eq:RM}). 

\subsection{Faraday conversion}\label{sec:CM}

More generally, Faraday rotation is a special case of ``Faraday conversion''. In general, a polarized electromagnetic wave can be characterized by four Stokes parameters \citep[e.g.][]{rybicki79}
\begin{eqnarray}
 I & = & {\varepsilon}_0^2, \\
 Q & = & {\varepsilon}_0^2 \cos 2\psi \cos 2\chi, \\
 U & = & {\varepsilon}_0^2 \cos 2\psi \sin 2\chi, \\
 V & = & {\varepsilon}_0^2 \sin 2\psi, 
\end{eqnarray}
where $I=\sqrt{Q^2+U^2+V^2}$ is the total intensity, $L=\sqrt{Q^2+U^2}$ is the intensity of the linear polarization, $V$ is the intensity of the circular polarization, 
$\varepsilon_0 = \sqrt{I}$ is the amplitude of the elliptically polarized EM waves, $\psi = (1/2) \arcsin(V/I)$ is a proxy of the circular polarization degree $\Pi_{\rm o} = V/I$ which is intrinsic to the waves, and $\chi = (1/2) \arctan(U/Q)$ is the angle between the semimajor axis of the ellipse and the $x$-axis defined by the telescope, which is extrinsic to the waves. Notice that  ($I$, $2\psi$ and $2\chi$) are spherical coordinates in a imaginary Poincare sphere, and ($Q$, $U$, $V$) defines a polarization vector $\vec P$ from the center to a point on the sphere in the Cartesian coordinate system, which defines the polarization state of the wave. Faraday rotation is simply the rotation of the $\vec P$ vector around the $V$ axis. When $\vec P$ rotates around axes other than the $V$ axis, there would be conversion between linear polarization $L$ and circular polarization $V$. The waves would undergo Faraday conversion \cite{zheleznyakov64,melrose95}. 

The physics of Faraday conversion can be understood as follows. Any polarization state can be decomposed into superposition of two fundamental modes, either two circular polarization modes (e.g. R- and L-modes) for the quasi-parallel case or two linearly polarization modes (e.g. O- and X-modes) for the quasi-perpendicular case. The different phase velocities of the two eigen modes would make the two modes out of phase and introduce modified polarization behaviors after superposition. For the quasi-parallel case, the different velocities of R- and L-modes introduce rotation of the superposed linear polarization angle, and hence, Faraday rotation. For the quasi-perpendicular case, on the other hand, the difference in the propagation velocities in the O- and X-modes would make the two modes out of phase, making the superposed polarization elliptical. Effectively, part of linear polarization is converted to circular polarization. The amplitude of Faraday conversion is smaller than that of Faraday rotation by a factor of $\omega_{\rm B}/\omega$, which is $\ll 1$ for waves propagating in a medium far outside of the neutron star magnetosphere. 

Mathematically, one may consider that the vector $\vec P$ undergoes rotation around an imaginary vector axis in the direction of
\begin{equation}
\vec \Omega \equiv (g, h, f)
\end{equation}
on the Poincare sphere. The variation of the circular polarization degree can be described by $d \vec{P}/dz = \vec\Omega \times \vec P$, where the $z$-axis is the direction of the $V$ component \cite{gruzinov19}. The three components of $\vec\Omega$ are
\begin{eqnarray}
 f & = & -\frac{1}{c} \frac{\omega_p^2 \omega_{\rm B}}{\omega^2} \hat B_z, \label{eq:f} \\
 h + ig & = & -\frac{1}{c} \frac{\omega_p^2 \omega_{\rm B}^2}{\omega^3} (\hat B_x + i \hat B_y)^2, \label{eq:h&g}
\end{eqnarray}
where $(\hat B_x, \hat B_y, \hat B_z)$ is the unit vector ${\hat B} = \vec B/B$, $f$ denotes the traditional Faraday rotation rate discussed in Equation (\ref{eq:Delta-phi}), and $(h+ig)$ describes the Faraday conversion rate. To order of magnitude, one can see that $h/f \sim g/f \sim \omega_{\rm B}/\omega$, which is $\ll 1$ for waves propagating far outside a neutron star magnetosphere. This means that $\vec \Omega$ is essentially parallel to the $V$ direction and that Faraday conversion is a small-order effect compared with Faraday rotation. 

If one measures oscillations of Stokes parameter $V$, one may define a {\em conversion measure} (CM) as \citep{gruzinov19} 
\begin{equation}
    \left< \Pi_{V} \right> = {\rm CM} \lambda^2,
    \label{eq:CM}
\end{equation}
where $\Pi_V \equiv |V|/I$ and $\left< \Pi_V \right>$ is the rms value of $\Pi_V$. The CM can be related to RM through 
\begin{equation}
    {\rm CM} \simeq \frac{\omega_{\rm B}}{\omega} \ {\rm RM}^{1/2} \sim (10^{-2} {\rm m^{-2}}) \ {\rm RM_m^{1/2}} (B/{\rm G}),
    \label{eq:CM-RM}
\end{equation}
where $B$ is in units of Gauss and ${\rm RM}$ is in units of $\rm rad \ m^{-2}$. This is strictly valid for a small conversion angle $\theta_f$ (the final angle by which the linear-polarization  $Q-U$ plane rotates). For a large $\theta_f$, a more precise expression is \citep{gruzinov19} 
\begin{equation}
    \left< \Pi_{\rm V} \right> = \sqrt{2\left[ e^{-({\rm CM} \ \lambda^2)^2/2}-e^{({\rm CM} \ \lambda^2)^2} \right]}.
\end{equation}
When both CM and RM are measured, one can directly measure $B$ using Equation (\ref{eq:CM-RM}). 

Physically, for a cold plasma Faraday conversion happens when the $B$ field is quasi-perpendicular. Astrophysically, this may be (but is not necessarily) related to the reversal of $B_\parallel$ along the line of sight \citep{melrose10,gruzinov19}. Another possibility of having Faraday conversion is when electrons are no longer ``cold'' but are mildly relativistic with a mean Lorentz factor $\gamma_e > 3$. This is because when considering the response tensor or electrons with a general energy distribution, the expressions of the $h$, $g$, and $f$ parameters depend on the $\gamma_e$ in the medium \citep{huang11}. As $\gamma_e$ increases, $h$ and $g$ increase and $f$ decreases so that conversion becomes progressively more important and rotation becomes less important. The non-detection of Faraday conversion in rFRB 20121102A has been used by \citet{vedantham19} to place an upper limit on the Lorentz factor of the electrons in the medium that generate the conversion, i.e. $\gamma_e < 5$. 

Faraday conversion can be more generally described using the transport equation \citep{huang11,lidz22}
\begin{equation}
  \frac{d \vec S}{ds} =   \left(
     \begin{array}{c}
         \epsilon_I \\
         \epsilon_L \\
         0 \\
         \epsilon_V \\
     \end{array}
    \right) - 
    \left(
     \begin{array}{cccc}
         \eta & \eta_L & 0 & \eta_V \\
         \eta_L & \eta & \rho_V & 0 \\
         0 & -\rho_V & \eta & \rho_L \\
         \eta_V & 0 & -\rho_L & \eta \\
     \end{array}
    \right) \vec S
\end{equation}
for the Stokes vector 
\begin{equation}
    \vec S=\left(
     \begin{array}{c}
         I \\
         Q \\
         U \\
         V \\
     \end{array}
    \right) = \left(
     \begin{array}{c}
         I \\
         L \\
         0 \\
         V \\
     \end{array}
    \right),
\end{equation}
where in the second equation we have replaced $Q$ with $L$ by adopting a coordinate system with $U=0$ without loss of generality. Here $\epsilon$'s are the emission coefficients, $\eta$'s are absorption coefficients, $\rho_V$ (the same as the $f$ parameter in Equation (\ref{eq:f})) is the coefficient for Faraday rotation and $\rho_L$ (essentially the amplitude of $h + ig$ in Equation (\ref{eq:h&g})) is the coefficient for Faraday conversion.  

Apparent oscillations of $L$ and $V$ have been discovered in some bursts from rFRB 20201124A \citep{xuh22}. These features may be interpreted as Faraday conversion or polarization-dependent absorption, which in any case demands a complex magnetized environment around the source \citep{xuh22,lidz22}.

\subsection{Plasma radiation mechanisms}

In classical electrodynamics, charged particles radiate when undergoing acceleration. Below we briefly discuss three well discussed radiation mechanisms involving electron acceleration in electric fields, magnetic fields, and electromagnetic waves, respectively.

\subsubsection{Bremsstrahlung}\label{sec:bremsstrahlung}

An electron in the Coulomb electric field of an ion would radiate through bound-bound (line emission), free-bound (recombination) and {\em free-free} ({\em bremsstrahlung}) processes. The opposite processes give respective absorption processes of the photons. 

For a plasma in thermal equilibrium with temperature $T$, the plasma thermal bremsstrahlung (free-free) emissivity reads \citep{rybicki79}
\begin{eqnarray}
    \epsilon_\nu^{\rm ff} & \equiv & \frac{dE}{dV dt d\nu} \nonumber \\
    &=& \frac{2^5 \pi e^6}{3 m_ec^3}\left(\frac{2\pi}{3k_Bm_eT}\right)^{1/2} Z^2 n_e n_i e^{-h\nu/k_BT} \bar g_{\rm ff} \nonumber \\
    &=& (6.8\times 10^{-38} \ {\rm erg \ cm^{-3} \ s^{-1} \ Hz^{-1}}) \nonumber \\
    &\times& Z^2 n_e n_i T^{-1/2}e^{-h\nu/k_BT} \bar g_{\rm ff},
\end{eqnarray}
where $c$, $k_B$, $e$, and $m_e$ are standard fundamental constants, $T$ is the gas temperature, $n_i$ is the number density of ions, $Z$ is the atomic number of the ions, and $\bar g_{\rm ff}$ is the Gaunt factor. The reason for the factor $n_e n_i$ is that the emissivity of each electron depends on the number density of ions and the total emissivity is proportional to the number density of electrons. Since $Z n_i = n_e$ is needed to maintain charge neutrality, $n_e^2$ enters the problem, so a convenient {\em emission measure} 
\begin{equation}
    {\rm EM} = \int_0^D n_e^2 dl
\end{equation}
can be defined for a radio source, which may be related to the DM of the source through ${\rm EM} L \sim {\rm DM}^2$, where $L$ is the characteristic size of the source. 

The opposite process of bremsstrahlung, i.e. free-free absorption, is relevant to constrain the physical condition to allow the FRBs with the extremely high brightness temperatures to be observed. The absorption coefficient can be expressed as \citep{rybicki79}
\begin{eqnarray}
    \alpha_\nu^{\rm ff} &=& \frac{4 e^6}{3 m_e h c}\left(\frac{2\pi}{3k_Bm_eT}\right)^{1/2} Z^2 n_e n_i \nu^{-3} (1-e^{-h\nu/k_BT}) \bar g_{\rm ff} \nonumber \\
    &=& (3.7\times 10^{8} \ {\rm cm^{-1}}) Z^2 n_e n_i T^{-1/2} \nu^{-3} (1-e^{-h\nu/k_BT}) \bar g_{\rm ff}, \nonumber \\
\end{eqnarray}
or, in the Rayleigh-Jeans regime
\begin{eqnarray}
    \alpha_\nu^{\rm ff} &=& \frac{4 e^6}{3 m_e k c}\left(\frac{2\pi}{3k_Bm_e}\right)^{1/2} T^{-3/2} Z^2 n_e n_i \nu^{-2} \bar g_{\rm ff} \nonumber \\
    &=& (0.0018 \ {\rm cm^{-1}}) T^{-3/2} Z^2 n_e n_i  \nu^{-2} \bar g_{\rm ff}.
    \label{eq:alpha_nu^ff}
\end{eqnarray}
Integrating over distance, one gets the optical depth \citep{cordes02}
\begin{eqnarray}
    \tau_{\nu}^{\rm ff} & = & \int_0^D \alpha_\nu^{\rm ff} dl \nonumber \\
    & = & (5.47 \times 10^{-8}) T_4^{-3/2} Z^2 \nu_9^{-2} \bar g_{\rm ff} \frac{\rm EM}{\rm pc \ cm^{-6}}.
\end{eqnarray}
An FRB is transparent only if $\tau_{\nu}^{\rm ff} < 1$ is satisfied in the emission region and also in the local environment surrounding the FRB source. 

For a relativistically hot plasma, the emissivity and absorption coefficient should be multiplied by a correction factor. The frequency-integrated correction factor is $(1+AT)$, where $A=4.4 \times 10^{-10} \ {\rm K}^{-1}$ \citep{rybicki79}. 

\subsubsection{Cyclotron, synchrotron and curvature radiation mechanisms}

Electrons gyrate in magnetic fields and radiate. For non-relativistic electrons, the emitted {\em cyclotron radiation} spectrum is line-like, with the main power at the Larmor frequency $\omega_{\rm B}$ and progressively lower powers at its higher harmonics. 

A relativistic electron with Lorentz factor $\gamma_e$ radiates {\em synchrotron radiation}  with a characteristic radiation frequency \citep{rybicki79}
    \begin{equation}
     \omega_{\rm SR} \simeq \frac{3}{2} \gamma_e^2 \frac{eB }{m_e c}\sin\alpha,
     \label{eq:omega_SR}
    \end{equation}
    where $\alpha$ is the pitch angle between the electron velocity and magnetic field.
    The power 2 for $\gamma_e$ is due to the following three factors: (1) the relativistic mass is larger by a factor of $\gamma_e$; (2) the fraction of the orbital time with radiation beamed towards an observer is smaller by a factor of $2/\gamma_e$ due to the relativistic beaming effect; and (3) the observed timescale is shorter than the emission timescale by roughly a factor of $(1-\beta) \sim 1/(2\gamma_e^2)$, where $\beta$ is the dimensionless speed of the electron. If synchrotron radiation is responsible for the FRB emission (e.g. within the framework of the synchrotron maser model), the required condition is $\gamma_e^2 B \sin\alpha \simeq (360 \ {\rm G}) \ \nu_{\rm FRB,9}$.
    
    The relativistic beaming effect for synchrotron radiation is valid under the vacuum approximation. In a plasma, with the refraction index $n_r \equiv \sqrt{\epsilon} < 1$, the beaming angle $\theta_b$ becomes $\sqrt{1-n_r^2\beta^2}$ rather than $\sqrt{1-\beta^2}$. If $n_r$ deviates from unity much more than $\beta$, one has $\theta_b \sim \sqrt{1-n_r^2} = \omega_p/\omega$ and synchrotron radiation is suppressed \citep{rybicki79}. One may define the Razin frequency by equating $\theta_b$ and $1/\gamma_e$, which gives
    \begin{equation}
        \omega_{\rm Razin} = \gamma_e \omega_p.
    \label{eq:omega_Razin}
    \end{equation}
    Synchrotron radiation is suppressed when $\omega < \omega_{\rm Razin}$. Matching the Razin frequency with GHz, the condition is $\gamma_e^2 n_e \simeq (1.2 \times 10^{10} \ {\rm cm}^{-3}) \ \nu_{\rm FRB,9}^2$.
    
    In a strong magnetic field environment such as the magnetosphere of a pulsar or a magnetar, the synchrotron cooling timescale, $t_{\rm c,SR} \sim \gamma_e m_e c^2 / [(4/3) \gamma_e^2 \beta_e^2 c \sigma_{\rm T} (B^2/8\pi) ] \sim (8\times 10^{-20} ~{\rm s}) \ B_{12}^{-2}\gamma_{e,2}^{-2}$, is extremely short. As a result, charged particles stay at the lowest Landau level and essentially slide along magnetic field lines in the local inertial (co-rotating) frame. Since the field lines are usually curved, particles will radiate when they accelerate in the curved trajectory. The characteristic frequency of such {\em curvature radiation} can be calculated by replacing the electron gyration radius in the synchrotron radiation formula by the curvature radius $\rho$ of the field lines so that
    \begin{equation}
        \omega_{\rm CR} = \frac{3}{2} \gamma_e^3 \frac{c}{\rho}.
    \label{eq:CR}
    \end{equation}
    The origin of $\gamma_e^3$ is similar to synchrotron radiation, except that there is no $\gamma_e m_e$ suppression in gyration frequency for synchrotron radiation (the mass does not enter the problem since the curvature radius of the field line does not depend on mass). To match the GHz emission, the  parameters should satisfy $\gamma_{e,2}^3 \rho_7^{-1} \simeq 1.4 \ \nu_{\rm FRB,9}$.

\subsubsection{Compton and inverse Compton scattering}

An initially at-rest electron oscillates in electromagnetic waves and emits essentially isotropically at the same incident frequency if $\hbar \omega_{i} \ll m_e c^2$ with cross section equals the Thomson scattering cross section $\sigma_{T} = (8\pi/3) (e^2/m_e c^2)^2 \simeq 6.65\times 10^{-25} \ {\rm cm}^2$. When the electromagnetic waves have an extremely large amplitude so that the electron reaches a relativistic speed (relevant to FRBs near the FRB generation site), the electron motion trajectory becomes complicated and cross section much enhanced (e.g. \citet{yang20a}, see Section \ref{sec:large-a} for details). The existence of a strong background magnetic field further complicates the picture \citep{beloborodov21,qu22c}.

When an electron moves relativistically and interacts with electromagnetic waves with angular frequency $\omega_i$, it would inverse Compton scatter the waves to a higher frequency
\begin{equation}
    \omega_s \sim \gamma_e^2 (1-\beta\cos\theta_i) \omega_i.
\end{equation}
Such a process could be relevant to FRB radiation (Section \ref{sec:ICS}).

\section{General constraints on the models}\label{sec:general}

In order to interpret FRBs, a competent model needs to invoke a radiation mechanism model to address individual burst properties (brightness temperature, polarization properties, spectral down-drifting, radio efficiency, high-energy emission, etc)  and a source model that accounts for the global properties of the bursts (energetics, burst rate, luminosity/energy function, redshift distribution, host galaxy properties, etc.). Before discussing these in detail in Sections \ref{sec:radiation}) and \ref{sec:source}), one may place some generic, essentially model-independent constraints on the models based on some basic observational facts and physical principles.

\subsection{Burst duration (width) and engine size}\label{sec:size}

After correcting for the convolution effects from scattering and instrumental effects (\S\ref{sec:temporal}), the intrinsic duration $W_i$ of an FRB defines a length scale 
\begin{equation}
    R_i = c W_i = (3\times 10^7 \ {\rm cm}) \ {W}_{-3},
    \label{eq:Ri}
\end{equation}
where $W_{-3}$ is the intrinsic duration in units of milliseconds. 
The size of the FRB central engine $R_0$ should satisfy $R_0 \lesssim R_i$. This is straightforward if the FRB emitter does not move with a relativistic speed. The reason is that if $R_0 > R_i$, even if the emission region is lit up simultaneously everywhere, the duration of the event should be $R_0 / c > W_i$ due to the light propagation delay between the front end and the rear end of the emission region with respect to the observer. 

If the FRB emitter is moving towards the observer with a relativistic speed (which is likely the case as discussed in \S\ref{sec:variability} and \S\ref{sec:induced} below), the situation is more complicated but the conclusion of $R_0 \lesssim R_i$ remains valid. Most generally, let us assume that the emitter travels with a bulk Lorentz factor $\Gamma$ in a direction with an angle $\theta$ with respect to the line of sight. In the lab frame, let us consider that the central engine sends off two light signals towards the relativistic emitter (the fastest causal connection is through propagation of photons), and the emitter promptly reacts to the two signals and send off two signals to the observer immediately after receiving the two central engine signals. Approximating the emitter as a point source and ignoring cosmic expansion, one can write the following relation between the three intervals \cite{zhang18}:
\begin{equation}
   \frac{1-\beta\cos\theta}{1-\beta} \Delta t_{\rm eng} = (1-\beta\cos\theta) \Delta t_e = \Delta t_{\rm obs},
\end{equation}
where $\Delta t_{\rm eng}$ is the time interval for the engine to emit two signals; $\Delta t_e$ is the time interval for the relativistic emitter to receive the two signals from the engine and also the time interval for the emitter to send off two signals; and $\Delta t_{\rm obs}$ is the time interval for the observer to detect the two signals. Here $\beta$ is the dimensionless speed of the emitter, $\theta$ is the angle between the direction of motion and line of sight, and the factor $(1-\beta\cos\theta)$ (which $\simeq 1/2 \Gamma^2$ for $\theta=0$) is a factor accounting for the propagation effect, and $\Gamma=(1-\beta^2)^{-1/2}$ is the Lorentz factor of the emitter. One can see that even though the emitter timescale is stretched due to its motion, the observed timescale $\Delta t_{\rm obs}$ still track the central engine timescale $\Delta t_{\rm eng}$ ($t_{\rm obs}= t_{\rm eng}$ for $\theta=0$)\footnote{If the line of sight is outside the emission beam, $(1-\beta\cos\theta) \sim (1-\cos\theta)$ which becomes  $\gg (1-\beta)$, so $\Delta t_{\rm obs}$ becomes $> \Delta t_{\rm eng}$. One can see a longer burst with a lower flux. Such off-axis FRBs also know as ``slow radio bursts'' may be detectable from Galactic magnetars or other FRB-emitting sources \citep{zhang21}. }. As a result, $W_i$ can be used to constrain the size of the central engine in any case. 

Equation (\ref{eq:Ri}) immediately suggests that the most compact, stellar-mass objects in the universe, i.e. a neutron star or a stellar-mass black hole, are the most likely candidates for FRB engine.  Larger objects (e.g. white dwarfs, stars, and even supermassive black holes) have been invoked to interpret FRBs in some models, but these models must invoke contrived conditions to allow only a small enough region to power an FRB. 

\subsection{Variability timescale and emission radius}\label{sec:variability}

The rapid variability timescale, in particular the $\sim 60$ ns timescale observed in rFRB 20200120E from the M81 globular cluster, can be used to further constrain the emission radius of FRBs \citep{beniamini20,lu22}. For an on-beam FRB (i.e. $\theta \sim 0$ for a point source, or $\theta < \theta_j$ for a conical jet with an opening angle $\theta_j$), a natural variability timescale\footnote{Scintillation (see \S\ref{sec:scattering} for more discussion) can introduce modulations in shorter timescales but with small amplitudes. Distinct pulses in an FRB lightcurve should be intrinsically related to the size of the source or emission region.} is
\begin{equation}
    \delta t \simeq \frac{R_{\rm FRB}}{2c \Gamma^2}.
    \label{eq:t_var}
\end{equation}
This timescale defines both the observed time for the emitter to travel to the emission radius $R$ in the rising phase, and also the angular spreading timescale due to the propagation delay of a spherical jet front in the decaying phase. In principle, if one is allowed to arbitrarily increase the Lorentz factor of the emitter, any small $\delta t$ can be reproduced for any $R$. So Eq.(\ref{eq:t_var}) alone is not constraining. Interesting constraints can be posed when the duration of the burst $W$ is considered. For certain models, for example, the synchrotron maser model invoking the external shock \citep{metzger19}, the emission radius can be estimated as $R_{\rm FRB} \sim \Gamma^2 c W$. This immediately suggests that $\delta t$ cannot be significantly shorter than $W$. The 60-ns variability from the ms-duration bursts of the M81 globular cluster FRB \cite{nimmo22} therefore disfavors the external shock model of FRBs \citep{lu22}. The synchotron maser internal shock model \cite{beloborodov20} is still allowed. However, it suffers from other drawbacks. For example, the frequency down-drifting feature, which the external shock model interprets as the shock propagating to progressively larger radii \cite{metzger19}, is no longer straightforwardly interpreted within the internal shock models. On the other hand, rapid variability of FRBs is not a challenge to the magnetospheric models, as a 0.4-nanosecond pulse has been observed from the magnetosphere of the 33-ms Crab pulsar \cite{hankins07} (even though FRBs are more energetic than nano-shots).

\subsection{Periodicity}\label{sec:periodicity2}

The special source, FRB 20191221A, was detected to have a 0.2168 s periodic separation during a 3 s duration \cite{chime-period}. Since known sources of a sub-second period are all rotating neutron stars (pulsars), this source offers a definite clue that at least some FRBs originate from pulsar-like objects. Further arguments can be made that the FRB radiation region (at least for this source) is the magnetosphere of an underlying pulsar or magnetar \citep{chime-period,beniamini22b}. This is because models invoking emission regions outside the magnetosphere do not have well-defined geometric windows to maintain a strict periodic window. 

The lack of periodicity from active repeaters such as rFRB 20121102A \cite{zhangy18b,lid21,hewitt22} and rFRB 20201124A \cite{xuh22,niujr22}, on the other hand, places less constraints on the models. \citet{katz20} argued that this suggests a black hole rather than a neutron star origin of repeating FRBs. This argument is not strong because unlike pulsar emission, FRB radiation pressure is so strong that the magnetospheric structure is likely significantly distorted so that a well-defined magnetospheric window (the conventional open field line region) likely does not exist and it is entirely possible that an FRB emitting neutron star emit bursts at random phases. The radio bursts from the magnetar SGR J1935+2154 seem to be emitted from a much wider phase window than the narrow window for pulsed emission \cite{zhu22}. With the burst data alone, it appears that the source does not have a strict periodicity, even though the magnetar has a strict 3.24-s period.  

So far, only rFRB 20180916B has been confirmed to possess a long-term 16-day periodicity \cite{chime-periodic,paster-marazuela21,pleunis21}. Its origin is subject to debate. The most natural interpretation would be to attribute this to the orbital period of a binary system, with the emission from the FRB emitter only reaching the observer in a particular phase window \cite{ioka20,lyutikov20,wada21}. Other interpretations to the 16-d period of rFRB 20180916B include magnetar precession \cite{levin20,yangzou20}, slowly rotating magnetars \cite{beniamini20b}, and even precession of a black hole accretion disk \cite{katz22c}. None of these scenarios for the 16-d periodicity were theoretically predicted before the discovery of the rFRB 20180916B. So, it would be uncomfortable, at least to me, if such long-term periodicty is a common feature of active repeaters because that would require such periodicity being at the heart of FRB generation mechanisms \cite{zhang20d}. It is now clearer that such a long-term periodicity is not commonly observed among active repeaters (the case of rFRB 20121102A is to be confirmed, see \S\ref{sec:repetition}). Whatever mechanism that is operating in rFRB 20180916B likely applies in rare cases and is probably due to a chance coincidence. 

\subsection{Energetics, radio emission efficiency, and beaming}\label{sec:beaming}

The derived isotropic energies of individual bursts and the energy-dependent burst rates for repeaters can be used to place interesting constraints on the average luminosity and total energy budget of the underlying FRB source, which may be used to constrain FRB source models. 

For one-off FRBs, the true peak luminosity and energy of the burst are
\begin{eqnarray}
 L_p & = & L_{\rm p,iso} f_b \eta_r^{-1}, \nonumber \\
 E & = & E_{\rm iso}  f_b \eta_r^{-1},
\end{eqnarray}
where $L_{\rm p,iso}$ (Eq.(\ref{eq:Liso})) and $E_{\rm iso}$ (Eq.(\ref{eq:Eiso})) are the measured isotropic radio peak luminosity and energy directly from observations, $\eta_r$ is the radio emission efficiency, and 
\begin{equation}
    f_b \equiv \frac{\delta \Omega}{4\pi}
\end{equation}
is the beaming factor of an individual burst, with $\delta \Omega$ being the solid angle of the burst. Note that $f_b$ reduces and $\eta_r^{-1}$ increases the energy budget of the source so that the effects of the two factors tend to cancel out each other. Both factors are not well constrained from observations. The X-ray burst associated with FRB 200428 was more than $10^4$ more energetic than the radio burst itself \citep{HXMT-SGR,Integral-SGR,Konus-SGR}, so for this particular event, the upper limit of $\eta_r$ is $\sim (10^{-4}-10^{-5})$. Various X-ray flux upper limits for extragalactic FRBs places a lower limit on $\eta_r$, which is of this order or even smaller \cite{piro21,laha22a,laha22b}.

\begin{figure}
\includegraphics
[width=\columnwidth]{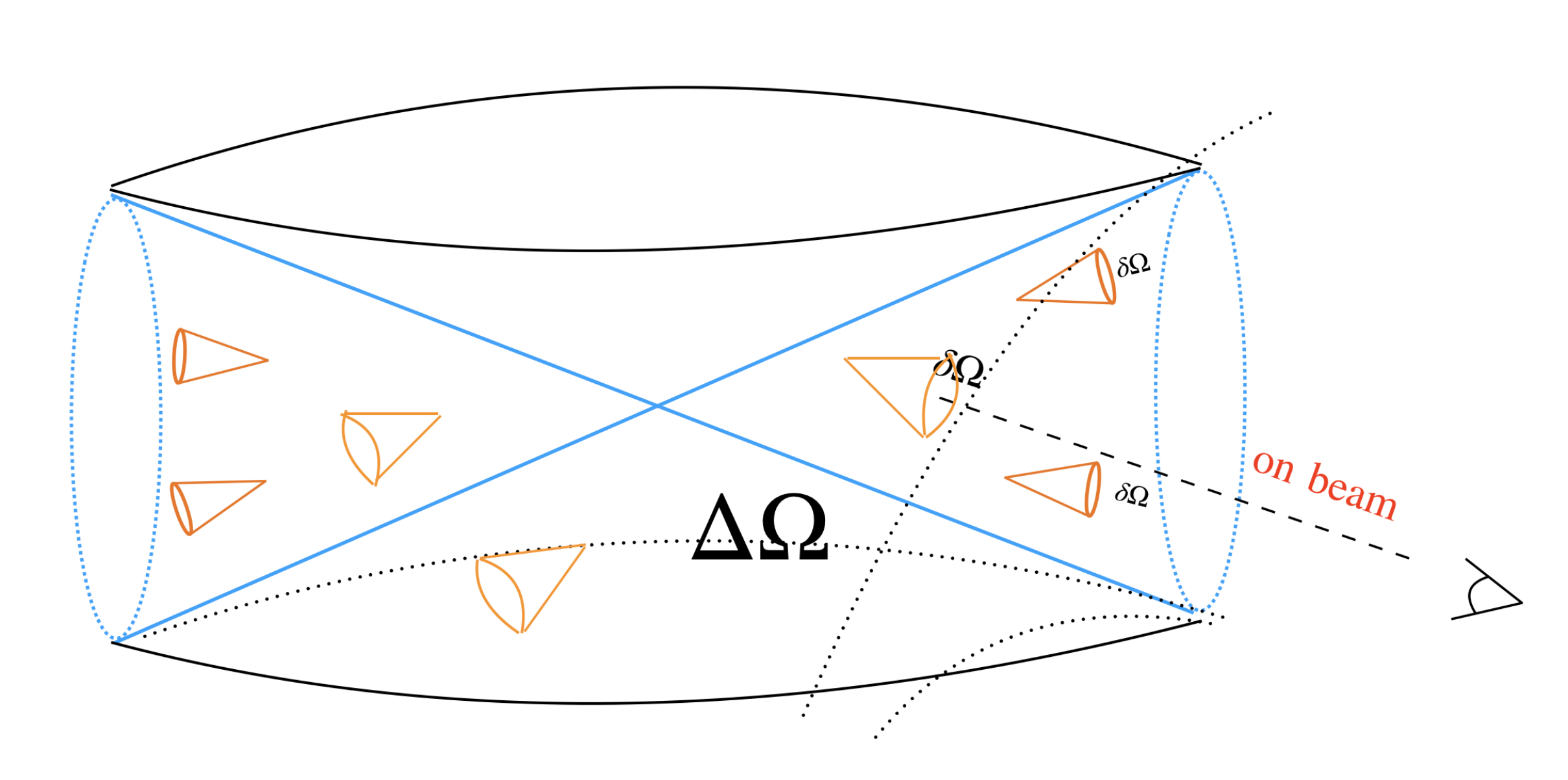}
\caption{A cartoon picture of the beaming factor of individual bursts (with a solid angle $\delta \Omega$) and global beaming (with a solid angle of $\Delta\Omega$). A  fan-beam from a magnetospheric rotator is illustrated as an example for global beaming, but a more general geometry is possible.}
\label{fig:beaming}
\end{figure}

For repeating sources, one should consider the average energy-dependent bursting rate $dN/dt dE_{\rm iso}$ during the active phase and the observational duty cycle of the active phase $\zeta$ (e.g. for the rFRB 20121102A observing campaign with FAST \cite{lid21}, the observational duty cycle is about 60 hours out of 47 days). One should also introduce a global beaming factor
\begin{equation}
    F_b \equiv \frac{\Delta \Omega}{4\pi},
\end{equation}
which can be larger than $f_b$ of the individual bursts. This is because the global emission beam can have a larger solid angle $\Delta\Omega$, inside which each burst could have a narrower beam (See Fig.\ref{fig:beaming}). The average source luminosity that is used to make FRBs would be
\begin{equation}
    L_{\rm src}= \int_{E_{\rm iso,m}}^{E_{\rm iso,M}} \left(\frac{dN}{dt dE_{\rm iso}} \right) E_{\rm iso} (F_b \eta_r^{-1}) d E_{\rm iso},
    \label{eq:L_src}
\end{equation}
where $E_{\rm iso,m}$ and $E_{\rm iso,M}$ are the minimum and maximum isotropic FRB energy from the source. The reason that $F_b$ rather than $f_b$ is adopted is the following. Even though each burst has a beaming factor $f_b$, there are altogether $\Delta\Omega / \delta\Omega$ such bursts on average most of which are not detected but are added to the energy budget of the source. The final beaming factor is therefore $(\Delta\Omega / \delta\Omega) f_b = F_b$. 

Assume that the repeater source has a lifetime $\tau$ and the activity level remains unchanged during lifetime. The total energy budget of the source over a duration $\tau$ can be estimated as
\begin{equation}
    E_{\rm src}= \int_0^{\tau} \int_{E_{\rm iso,m}}^{E_{\rm iso,M}} \left(\frac{dN}{dt dE_{\rm iso}} \right) E_{\rm iso} (F_b \eta_r^{-1} \zeta^{-1}) d E_{\rm iso} dt,
    \label{eq:E_src}
\end{equation}
where $\zeta=\tau_{\rm obs}/\tau$ is the observational duty cycle. 

Interesting constraints on the energy budget of repeaters have been made. For rFRB 20121102A, \citet{lid21} reported that the 1652 bursts detected in $\sim 60$ hours during a 47-day observational campaign. The total emitted radio energy (corresponding to the integral in Eq.(\ref{eq:E_src}) without the $(F_b \eta_r^{-1}\zeta^{-1}$ factor) is  $\sim 3.4\times 10^{41}$ erg. Considering $\zeta=60/(47 \cdot 24) =0.053$, $\eta_r = 10^{-4} \eta_{r,-4}$, and $F_b = 0.1 F_{b,-1}$, one can derive that the total source energy used to make FRBs is $E_{\rm src} = (6.4\times 10^{45} \ {\rm erg}) F_{b,-1} \eta_{r,-4}^{-1} (\zeta/0.053)^{-1}$. This is already $\sim 4\%$ of the total dipolar magnetic energy ($E_B \sim (1/6) B^2 R^3 \sim (1.7\times 10^{47} \ {\rm erg}) B_{15}^2 R_6^3$) of a magnetar. Since rFRB 20121102A already existed more than a decade, this observation already poses a significant energy budget issue for the magnetar model, unless $\eta_r$ is larger or $F_b$ is smaller. The magnetospheric models satisfy these constraints\footnote{Because of unidentified coherent mechanism of FRBs, the radio efficiency in the magnetospheric models cannot be predicted. However, the radio emission efficiency of radio pulsars is observed to range from $10^{-8}$ to close to unity \cite{szary14}. }, but the synchrotron maser shock model is already severely constrained by the data (see \S\ref{sec:magnetars} for details). Note that before the FAST observations, \citet{margalit20b} already posted a tight energy budget constraint on rFRB 20121102A based on the previous data within the framework of the magnetar synchrotron maser model. The many more bursts detected with FAST \cite{lid21} only tightened the constraints for the source. Even more stringent constraints on the magnetar synchrotron maser model have been established for another active repeater rFRB 20201124A \citep{xuh22,zhangyk22}.

\subsection{Brightness temperature and coherent radiation}\label{sec:Tb}

Astronomical objects emit four levels of electromagnetic radiation with increasing complexity (Table II): 
blackbody radiation, thermal radiation, incoherent non-thermal radiation, and coherent non-thermal radiation. Blackbody and thermal radiations require that the emitting particles are in thermal equilibrium (defined by the gas temperature $T$), with the blackbody radiation having an additional requirement that photons have a large enough optical depth to reach thermal equilibrium as well \citep{rybicki79}. Thermal radiation includes blackbody radiation, but also allows photons not to achieve thermal equilibrium. One example is thermal bremsstrahlung which has a different spectral shape from blackbody with the cutoff energy defined by the gas temperature $T$. If particles are accelerated to deviate from thermal equilibrium, say, in shocks or magnetic reconnection regions,  the radiation becomes non-thermal. As non-thermal particles typically have a (one segment or multi-segment) power-law distribution,  non-thermal radiation spectra are typically (broken) power-laws. In particular, in the low-frequency regime, non-thermal radiation is usually subject to self-absorption from the opposite process of the emission mechanism if particle radiation is incoherent. 

\begin{table*}
\caption{Astrophysical radiation mechanisms}
\begin{ruledtabular}
\begin{tabular}{cccc}
Mechanisms & Particles & Photons & Examples \\
\hline
blackbody & thermal equilibrium & thermal equilibrium & CMB, stars \\
thermal & thermal equilibrium & may/may not be in thermal equilibrium & disks, intracluster medium  \\
incoherent non-thermal & non-thermal & subject to self-absorption limit & SNRs, GRBs,  blazars\\
coherent non-thermal & non-thermal & not subject to self-absorption limit & radio pulsars, FRBs
\end{tabular}
\end{ruledtabular}
\end{table*}\label{tab:radiation}

Coherent non-thermal radiation may be defined in different ways, but the most straightforward way is through its ability of overcoming the self-absorption limit. The self-absorption-defined specific luminosity limit at a particular frequency is the blackbody specific luminosity at that frequency for a gas with the maximum temperature and source size allowed by the emitter. For electrons with a characteristic Lorentz factor $\gamma_e$ in the comoving frame, the comoving-frame effective temperature would be $kT' \sim \gamma_e m_e c^2$. For a synchrotron radiation source, $\gamma_e = {\rm max} (\gamma_m, \gamma_a)$ is usually the maximum of the following two: the minimum injection Lorentz factor $\gamma_m$ and the corresponding Lorentz factor for self-absorption $\gamma_a$ \citep[e.g.][]{kumar15}. As a result, 
\begin{equation}
    k T' = \gamma_e m_e c^2 = {\rm max}(\gamma_m, \gamma_a) m_e c^2,
\end{equation}
or 
\begin{equation}
T' =(5.9\times 10^{11} \ {\rm K}) \gamma_{e,2}
\end{equation}
defines the maximum incoherent brightness temperature in the comoving frame, where $\gamma_e \sim 100$ has been adopted. For radio galaxies, \citet{kellermann69} showed that the observations had a maximum brightness temperature of about $T_{\rm b,max}\sim 10^{12}$ K, which corresponds to a typical electron Lorentz factor $\gamma_e \sim (10^2-10^3)$. They argued that this limit is physically related to the requirement that the second-order Compton scattering power does not exceed that of the first-order (synchrotron self-Compton) through self-regulated Compton cooling. For systems like GRBs or blazar jets, $\gamma_e$ is related to the bulk Lorentz factor of the jet, which directly defines the internal energy density (and hence, the effective temperature) in the emission region. 

For a relativistic emitter beaming towards earth, the allowed maximum radio specific flux at a frequency is larger than the comoving value by a factor of ${\cal D}$ for an extended source or ${\cal D}^3$ for a point source \citep{zhang18}, where
\begin{equation}
    {\cal D} \equiv \frac{1}{\Gamma (1-\beta\cos\theta)} \simeq \Gamma
\end{equation}
is the Doppler factor,  $\Gamma$ and $\theta$ carry the same meaning defined earlier, and the final approximation applies to the regime $\theta \leq 1/\Gamma$. Compared with the observationally defined brightness temperature (Eq.(\ref{eq:Tb})), one may perform either of the following two approaches. One is to derive brightness temperature in the comoving frame ($T'_b$) and compare with the maximum $T'$; the other is to derive the maximum allowed $T$ in the observer frame and compare with observationally-defined $T_b$ (Eq.(\ref{eq:Tb})). We adopt the more straightforward latter approach and derive the condition that coherence is required by the data if
\begin{equation}
    T_b \geq {\cal D} \gamma_e m_e c^2/k \simeq (5.9\times 10^{13} {\rm K}) \Gamma_2 \gamma_{e,2}.
    \label{eq:Tb-condition}
\end{equation}
This result is consistent with \citet{lyubarsky21} who adopted the opposite approach. Since FRBs have the observed $T_b$ much greater than this value, their radiation mechanisms must be coherent. We will discuss various coherent mechanisms in \S\ref{sec:radiation}. 

\subsection{Attenuation processes}\label{sec:attenuation}

In order to have high-$T_b$ radio pulses detectable from Earth, the radio waves must overcome various absorption or scattering processes along the propagation paths. The three important processes to attenuate the radio emission flux are induced Compton scattering, free-free absorption, and synchrotron absorption, which we discuss below in turn.

\subsubsection{Induced Compton scattering and Lorentz factor lower limit}\label{sec:induced}

With the existence of free electrons, photons with a particular frequency can be scattered out of the state and other photons with different frequencies can be scattered into the state. The Thomson scattering optical depth can be estimated as $\tau_{\rm T} \sim n_e \sigma_{\rm T} R$, where $n_e$ is the electron number density, $\sigma_{\rm T} = (8\pi/3) (e^2/m_e c^2)^2 \simeq (6.65 \times 10^{-25} \ {\rm cm}^2)$ is the Thomson cross section, and $R$ is the size of the emission region. This optical depth is relevant for scattering of high-frequency photons (for very high-energy photons, the Klein-Nishina correction is needed), but in the low-frequency regime, scattering can be enhanced significantly by induced Compton scattering if $T_b$ is high enough \citep{kompaneets57,wilson78,thompson94,lyubarsky08}. The essence of this mechanism can be summarized as follows \citep{wilson78}. Consider two photon states (not electron states) $a$ and $b$ (defined by both the energies and directions of the photons) with photon occupation numbers $n_a$ and $n_b$, respectively. The spontaneous change in $n_a$ because of scattering from $a$ to $b$ is $dn_a/dt \propto -n_a$. However, since photons are bosons that satisfy the Bose-Einstein statistics, the existence of photons at $b$ actually boost the scattering rate from $a$ to $b$ by a factor of $(n_b+1)$, i.e. $dn_a/dt = -n_a(1+n_b)$. Similarly, the scattering rate from $b$ to $a$ is $d n_a/dt \propto (n_a+1) n_b$. The net change at level $a$ is $dn_a/dt \propto [(n_a+1) n_b - n_a(1+n_b)]$ which essentially cancels out but leaves a small term related to the recoil frequency shift due to Compton scattering, i.e. $\Delta\nu /\nu = (h\nu/m_ec^2) (1-\cos\theta)$, where $\theta$ is the angle between the directions of $a$ and $b$. It is found that for cold electrons without bulk motion, induced Compton scattering becomes important when $(k_BT_b / m_ec^2) \Omega^2 > 1$ (where $\Omega$ is the solid angle of the uniform beam) \citep{wilson78}\footnote{The factor $(k_BT_b / m_ec^2)$ is the product of the photon occupation number $k_BT_b/ h\nu$ and the fractional change of energy $\sim h\nu / m_ec^2$. }. As a result, the optical depth due to induced Compton scattering is enhanced with respect to Thomson scattering by the same factor, i.e.\footnote{A coefficient $3/8\pi^2$ is added with precise calculations  (e.g. W. Lu, 2021, unpublished notes). Note that this expression makes use of the assumption a uniform electron number density. It has been suggested that the medium may be subject to filamentation due to the propagation of FRBs, which would significantly reduce the optical depth for induced Compton scattering \citep{sobacchi22b}.} 
\begin{equation}
    \tau_{\rm C} \simeq \frac{3}{8\pi^2} \left(\frac{k_BT_b}{m_ec^2} \Omega^2\right) \tau_{\rm T} \simeq (6.4 \times 10^{24}) \Omega^2 T_{b,36}\tau_{\rm T} .
\label{eq:tauC1}
\end{equation}
The detailed expression depends on the explicit problems one is addressing. For example, if one considers the induced Compton scattering constraint in an emitting source, the expression can be written as \cite{lyubarsky08}
\begin{equation}
    \tau_{\rm C} \simeq \frac{3\sigma_{\rm T}}{8\pi} \frac{c n_e {\cal S}_{\nu}^{\rm obs}}{m_e \nu^2} \left( \frac{D_{\rm L}}{r_0}\right)^2 Z,
\end{equation}
where ${\cal S}_{\nu}^{\rm obs}$ is the observed specific flux of the FRB, $\nu$ is the FRB frequency, $D_{\rm L}$ is the luminosity distance of the source, $r_0$ is the radius of the launching point, and $Z$ is an integral that has the dimension of $r_0/c$ and carries the information of $\Omega$. For another example, if one considers the induced Compton scattering by a medium as an FRB from a separate source passes through it, the expression becomes \cite{ioka20}
\begin{equation}
    \tau_{\rm C} \simeq \frac{3\sigma_{\rm T}}{32\pi^2} \frac{L_\nu n_e c \Delta t}{r^2 m_e \nu^2},
\end{equation}
where $r$ is the distance between the FRB source and the scatterer. Note that the above discussion applies to an unmagnetized plasma with $\omega_{\rm B} \ll \omega_p$. In a highly magnetized environment such as the magnetosphere of a neutron star, charged particles are confined in strong magnetic fields, so that the particles required to have the right directions and energies for induced Compton scattering are not available. As a result, there is no need to consider the induced Compton scattering constraint in the emission region if FRBs are emitted from the magnetosphere of a central engine. 

When the emitter is moving relativistically with a bulk Lorentz factor $\Gamma$, the induced Compton scattering optical depth drops significantly \citep{lyubarsky08}. From Eq.(\ref{eq:tauC1}), noticing  $\Omega \propto \Gamma^{-2}$ and $T'_b = T_b / {\cal D} \simeq T_b / \Gamma $, one gets
\begin{equation}
    \tau_{\rm C} \sim \frac{k_B T_b}{m_e c^2} \frac{\tau_{\rm T}}{\Gamma^5},
\label{eq:tauC2}
\end{equation}
which is significantly smaller than the case without bulk motion. 

The induced Compton scattering optical depth also drops if the electron gas is relativistically hot. \citet{lu18} suggested that for a narrow Gaussian-like spectrum with a characteristic electron energy $\gamma_e$, the approximated optical depth is 
\begin{equation}
    \tau_{\rm C} \sim \frac{k T_b}{m_e c^2} \frac{\tau_{\rm T}}{\gamma_e^5}.
\label{eq:tauC3}
\end{equation}
For a power law photon spectrum, the results depend on the spectral index $p$ (convention $I_\nu \propto \nu^p$) but the suppression factor is shallower than $\gamma_e^{-5}$. 

Some FRB emission models invoke relativistic shocks as the emission site \citep[e.g.][]{lyubarsky14,beloborodov17,metzger19,beloborodov20,plotnikov19}. The emission region would be also relativistically hot in the comoving frame. Combining Equations (\ref{eq:tauC1}), (\ref{eq:tauC2}) and (\ref{eq:tauC3}) and noticing $\Omega^2$ is already included in the $1/\Gamma^5$ suppression factor, one may derive that the condition of $\tau_{\rm C} < 10$ is\footnote{Notice that induced Compton scattering mainly modifies the shape of the spectrum rather than exponentially attenuate photon flux. As a result, a larger optical depth than unity, e.g. $\tau_{\rm C} = 10$, is adopted as the transition point where the effect becomes important.}
\begin{equation}
    \Gamma \gamma_e \gtrsim (5.8 \times 10^4) T_{b,36}^{1/5} \tau_{\rm T}^{1/5}
\end{equation}
Since $\gamma_e \propto \Gamma$ is generally expected\footnote{This is straightforwardly expected for external shocks. For internal shocks, $\gamma_e$ is more related to the relative Lorentz factor between the colliding shells, which may also scale with $\Gamma$. }, one may place a lower limit of $\Gamma$ as
\begin{equation}
    \Gamma \gtrsim 240 \xi_e^{1/2} T_{b,36}^{1/10} \tau_{\rm T}^{1/10},
\end{equation}
where $\gamma_e = \xi_e \Gamma$ has been assumed. Similar constraints have been derived by \citet{lyubarsky08,murase16}. Note that within the relativistic shock models, a Lorentz factor of this order is also required by the duration and variability constraint (Eq.(\ref{eq:t_var})), so induced Compton scattering constraint is usually satisfied in the shock model without introducing an additional condition \citep[e.g.][]{metzger19,beloborodov20}.

\subsubsection{Free-free absorption}

Radio emission can be also attenuated via free-free absorption, the inverse process of free-free emission or bremsstrahlung. The importance of free-free absorption for FRBs has been discussed by various authors \cite{luan14,murase16,metzger17,yangzhang17,kumar17,kundu21}. 

The free-free absorption coefficient (Eq.(\ref{eq:alpha_nu^ff}) together with the relativistic correction factor $(1+A T)$ (\S\ref{sec:bremsstrahlung}) can be used to estimate the optical depth against free-free absorption. An FRB is transparent if the optical depth is below unity.

Free-free absorption is important when the density of the emitter or environment medium is high. Therefore, the free-free absorption constraint was adopted \citep{luan14} to disfavor an early FRB model invoking flaring stars \citep{loeb14}. For repeating FRB models invoking a young magnetar born from a supernova explosion, free-free absorption was  used to place a lower limit on the age of the supernova remnant before which the remnant shell is too dense to allow FRBs to escape freely \cite{metzger17,yangzhang17} (see \S\ref{sec:environment} for details). For FRB systems invoking relativisic shocks, either as the site of FRB emission or as a screen in front of FRB produced at an inner radius, free-free absorption in the hot shocked plasma could be important if the total kinetic energy exceeds $\sim 10^{44}$ erg, which may account for the frequency down-drifting feature observed in some FRBs \citep{kundu21}. 

\subsubsection{External synchrotron absorption}

For active repeaters surrounded by a persistent radio source (PRS) \citep{chatterjee17,niu22}, coherent FRB emission needs to pass through the PRS, which is likely powered by synchrotron radiation. Under certain conditions, FRBs could be  absorbed by the PRS via synchrotron absorption and the PRS source could be subsequently heated up by the absorbed FRBs \citep{yang16}.  

Assuming that the nebula electrons have an initial differential number density spectrum $N(\gamma_e,0) = K \gamma_e^{-p}$, one can estimate the synchrotron optical depth as
\begin{equation}
    \tau_{\rm \nu,SR} = \frac{e^2 K R}{4 m_e c} \frac{1}{\nu_B} \left(\frac{\nu}{\nu_B} \right)^{-\frac{p+4}{2}} f(p),
\end{equation}
where $\nu_B = eB/(2\pi m_e c)$, $R$ is the radius of electron acceleration region, and $f(p)$ is a function of order unity. Solving $\tau_{\rm \nu,SR} = 1$, one can derive synchrotron absorption frequency \citep{yang16}
\begin{equation}
    \nu_a = \nu_B \left[ \frac{\pi}{2} \frac{eRK}{B} f(p) \right]^{\frac{2}{p+4}}.
\end{equation}
The spectrum of the nebula needs to be solved numerically by including electron injection, synchrotron cooling, as well as heating by FRBs, which would give rise to complicated spectra for both electrons and photons. The predicted spectra \citep{yang16} turn out to share the general shape of the later observed PRS spectrum of rFRB 20121102A \cite{chatterjee17,marcote17}, as shown in \citet{liqc20}. The small nebula size and not too high a synchrotron self-absorption frequency constrain the parameter space for such models in general \cite{metzger17}. Mode-dependent synchrotron absorption may change the polarization mode and enhance linear polarization \cite{quzhang23}. 

\subsection{Ordered magnetic fields and strengths}

The fact that FRB emission is linearly polarized with a high polarization degree poses a generic constraint, namely, there must exist ordered magnetic fields in the FRB emission region. Indeed, current leading models to interpret FRBs invoke either  magnetospheres of magnetized central engines or relativistic shocks with ordered magnetic fields. See \citet{quzhang23} for a survey of various emission mechanisms to produce high polarization in FRBs. 

Further constraints on the strength of magnetic fields have been discussed in the literature \citep{kumar17,lyutikov17}. The argument is that the electromagnetic wave energy density in the emission region should not exceed the magnetic energy density of the emitter in the same region before the FRB is emitted. Such a constraint can be placed if the FRB emission originates from dissipation of magnetic fields, or the magnetic field in the emission region confines the generated FRB emission. Note that such a condition in general is not always necessary for producing intense electromagnetic radiation. For example, the fireball model for GRBs does not require to abide by such a condition, with the electromagnetic energy of radiation generated from the thermal energy or the dissipated kinetic energy in the fireball \cite{zhang18}. In the case of coherent radiation, on the other hand, many models require that ordered $B$ fields should remain ordered during the emission processes. As a result, such a condition is quite relevant. 

The electromagnetic wave energy density, independent of the emission frequency, may be estimated as $L_{\rm iso} / (4\pi R_{\rm FRB}^2 c)$, where $R_{\rm FRB}$ is the radius where FRB emission is radiated. The condition
\begin{equation}
    \frac{L_{\rm iso}}{4\pi R_{\rm FRB}^2 c} < \frac{B^2}{8\pi}
    \label{eq:B-condition}
\end{equation}
gives 
\begin{equation}
    B > \sqrt{\frac{2L_{\rm iso}}{ c}} \frac{1}{R_{\rm FRB}}\simeq (8.2 \times 10^{15} \ {\rm G}) L_{\rm iso,42}^{1/2} R_{\rm FRB}^{-1}.
    \label{eq:B}
\end{equation}
The key is how to estimate $R_{\rm FRB}$. If one assumes $R_{\rm FRB} = c W_i = (3\times 10^7 \ {\rm cm}) \ (W_{\rm ms})$, one obtains $B > (2.7 \times 10^8 \ {\rm G}) L_{\rm iso,42} (W_{\rm ms})^{-1}$, which leads to the conclusion that the emission region has to be within the magnetosphere of a neutron star \citep{lyutikov17}. This argument, however, is flawed, since $R_{\rm FRB}$ cannot be always simply estimated as $cW_i$. If the emitter is moving relativistically with a bulk Lorentz factor $\Gamma$, as is envisaged in the synchrotron maser models, one has $R_{\rm FRB} = \Gamma^2 c W_i = (3\times 10^{13} \ {\rm cm}) \Gamma_3^2 \ W_{-3}$. The $B$-field constraint becomes 
\begin{equation}
    B > \sqrt{\frac{2L_{\rm iso}}{c}}\frac{1}{\Gamma^2 c W_i} \simeq (2.7 \times 10^{2} \ {\rm G}) L_{\rm iso,42}^{1/2} \Gamma_3^{-2} W_{-3}^{-1}.
    \label{eq:B2}
\end{equation}
Note that the magnetic field strength at the light cylinder of a magnetar is $B_{\rm lc} \simeq B_* (cP/2\pi R_*)^{-3} = (9.2\times 10^3 \ {\rm G}) B_{*,15} P^{-3} R_{*,6}^3$. So this estimate allows the emission region to be outside of a neutron star magnetosphere.

\subsection{Afterglow}

A generic constraint can be placed on the brightness of the multi-wavelength afterglows of FRBs. Afterglow observations for GRBs have been essential in identifying their multi-wavelength counterparts and host galaxies as well as measuring their redshifts. In the case of FRBs, the isotropic energy is typically more than 10 orders of magnitude smaller than GRBs ($E_{\rm iso,FRB} \sim 10^{39}$ erg vs. $E_{\rm iso,GRB} \sim 10^{52}$ erg). The expected FRB afterglow emission is expected to be much fainter \cite{yi14}. One possible way of enhancing afterglow emission is to assume that the FRB radiative efficiency $\eta_r$ is very low so that the afterglow kinetic energy can be boosted by a factor of $\eta_r^{-1}$. According to the standard GRB afterglow model \citep{meszarosrees97,sari98,zhang18}, the characteristic synchrotron frequency of injected minimum-energy electrons and the peak synchrotron specific flux for a relativistic jet being decelerated by a constant-density medium read
\begin{eqnarray}
    \nu_m & = & (3.3\times 10^{8} \ {\rm Hz}) (1+z)^{1/2} t_d^{-3/2}  \epsilon_{B,-2}^{1/2}  \nonumber \\
    &\times& [\epsilon_{e,-1} (p-1)/(p-2)]^2 (E_{\rm FRB,38}/\eta_{r,-6})^{1/2} \\
    F_{\rm \nu,max} & = & (1.6 \times 10^{-8} \ {\rm mJy}) (1+z) \epsilon_{B,-2}^{1/2} \nonumber \\
    &\times & (E_{\rm FRB,38}/\eta_{r,-6}) n^{-1} D_{\rm L,28}^{-2},
\end{eqnarray}
where the blastwave kinetic energy in normalized to $10^{44}$ erg (which assumes $\eta_r = 10^{-6}$ for $E_{\rm FRB} = 10^{38}$ erg), $\epsilon_e$ and $\epsilon_B$ are shock equipartition parameters for electrons and magnetic fields, respectively, $p$ is the power law index of the injected electrons, $n$ is the medium density, $t_d$ is the observing time in units of day, and $D_{\rm L,28}$ is the luminosity distance of the source in units of $10^{28}$ cm. One can see that the afterglow emission peaks in the radio band and is extremely faint. Detailed calculations \cite{yi14} suggest that a detection is possible only if the source is extremely nearby and the FRB is extremely energetic (i.e. the radio efficiency is very low), e.g. $E=E_{\rm FRB}/\eta_r = 10^{47} E_{\rm FRB,40} \eta_{r,-7}$ erg. For a relativistic, mildly magnetized jet, the reverse shock emission could be brighter than the forward shock emission, which would ease the detection of the afterglow \cite{yi14}. 

No confirmed FRB afterglow has been detected so far (even for the Galactic FRB 200428). This is consistent with the theory and suggests that $\eta_r$ is not extremely low. It is worth noting that in the synchrotron maser model invoking external shocks \cite{metzger19}, the multi-wavelength counterpart associated with the FRB could be regarded as its own ``afterglow'', even though the electron energy distribution is assumed to be thermal rather than a power law. No Fermi acceleration of particles is envisaged, which could be a problem theoretically. The two hard spikes observed in the X-ray counterpart \cite{HXMT-SGR,Integral-SGR} of FRB 200428 \cite{bochenek21,CHIME-SGR} can be interpreted within this model as the external shock emission \cite{margalit20}, even though it is more naturally interpreted as emission within the magnetar magnetosphere (\citet{lu20}, \citet{yangzhang21}, \citet{ioka20b}). 

\section{Coherent radiation mechanisms}\label{sec:radiation}

Coherent radiation mechanisms invoke fundamental plasma physics, which could be shared among different source models. For example, coherent curvature radiation by bunches has been discussed in many different contexts involving magnetospheres, such as radio emission from the inner magnetospheres of pulsars and magnetars \cite{ruderman75,katz14,kumar17,yangzhang18}, from ejected magnetospheres from ``blitzars'' \cite{falcke14,zhang14}, from kinetic-energy ``combed'' magnetospheres \cite{zhang17}, from magnetopsheres during asteroid-NS collisions \cite{geng15,dai16,dai20}, as well as from the global magnetospheres formed by merging charged objects \cite{zhang16a}. The synchrotron maser mechanism in relativsitic shocks, on the other hand, has been invoked in the magnetar internal \cite{beloborodov17,beloborodov20} or external \cite{lyubarsky14,metzger19} shock models, shocks from low-$B$ compact objects \cite{waxman17,long18}, and even black hole accreting systems \cite{sridhar21}. Therefore, it is reasonable to detach radiation models from source models and discuss the general physics behind each radiation model. This is the task of this section. 

\subsection{Coherent radio emission overview}

Following the discussion in \S\ref{sec:Tb}, we can summarize two fundamental properties of a coherent radiation mechanism: (1) the observed luminosity, $L_{\rm obs}$, exceeds the sum of the emitted power $P_e$ for individual particles, i.e. $L_{\rm obs} > N_e P_e$, where $N_e$ is the total number of electrons; and (2) the observed luminosity is not subject to self-absorption, so that Equation (\ref{eq:Tb-condition}) is satisfied. 

There are several ways to classify coherent radiation mechanisms. Based on differences in general physics, one may classify the mechanisms in the following three types \cite[e.g.][]{melrose78}. Each mechanism has its emission properties and back-reaction mechanisms. 
\begin{itemize}
    \item Coherent emission by bunches (or the ``antenna'' mechanism): In this mechanism, emitting particles are physically clustered in six-dimensional phase space, i.e. in both 3-D position space and 3-D momentum space. This is how coherent emission is emitted from antennae in radio stations. Within this mechanism, microscopic particles (e.g. electrons) are physically bunched together to radiation as a global particle with a total charge $N_{e,b} e$, where $N_{e,b}$ is the number of charges in each bunch, typically distributed within a unit volume defined by the wavelength of the radio waves ($N_{e,b} \sim n_e \gamma^2 \lambda^3$, where $n_e$ is the charge number density, $\lambda$ is the wavelength, and $\gamma$ is the bulk Lorentz factor of the bunch). The emission power of the bunch, depending on the degree of coherence, can reach a maximum of $N_{e,b}^2 P_e$ \cite[e.g.][]{yangzhang18}. The total luminosity of the system would be $\sim N_{e,b}^2 N_b P_e$, where $N_b \sim N_e / N_{e,b}$ is the number of bunches in the emission region. The back-reaction effects of such bunched emission are two folds: due to internal Coulomb repulsion, bunches tend to disperse in space. Radiation reaction may also make the particles disperse in the momentum space \cite{melrose78}. 
    \item Hydrodynamic instabilities (or ``plasma masers''): In this mechanism, some oscillation modes in a plasma exponentially grow with time, with macroscopic particles clustering in the momentum space. The MHD waves eventually escape in the form of electromagnetic waves in the radio band. The back-reaction effect is that as the mode grows, dispersion in the momentum space occurs and the instability would suppress itself. 
    \item Kinetic instabilities (or ``vacuum masers''): In this mechanism, electromagnetic waves detached from the plasma fluid would undergo negative absorption in an energy-population-inverted medium so that the amplitude of emission grows with distance, reaching a high brightness temperature. The effect of back-reaction is that masers tend to reduce population inversion so that the instability also suppresses itself. 
\end{itemize}

Only a few types of objects are observed to emit coherent radio emission, e.g. Sun, Jupiter, astronomical maser sources, pulsars, and FRBs. The mechanisms operating in different types of objects can achieve different degrees of coherence (i.e. different values of $T_b$). \citet{melrose17} reviewed the mechanisms of coherent emission in different types of objects and suggested that they have different origins: (1) Plasma emission at the plasma frequency $\omega_p$, which invokes Langmuir plasma waves (longitudinal oscillations) through a streaming instability as the trigger mechanism, likely applies to solar radio bursts; (2) Electron cyclotron maser emission at the cyclotron frequency $\omega_B$, which invokes a cyclotron plasma instability, likely applies to Jupiter and Earth Aurora;  (3) pulsar coherent emission must have a different mechanism, which has at least four possibilities: curvature emission by bunches, linear acceleration emission, relativistic plasma emission, and anomalous Doppler emission. However, all four mechanisms encounter difficulties and the pulsar coherent mechanism remains an enigma after more than half a century of study.

The prospect of understanding FRB coherent emission is not bright, either, since it involves more extreme processes to produce coherent emission. In any case, many mechanisms have been discussed in the literature and some have been briefly reviewed in \citet{zhang20}, \citet{lyubarsky21} and \citet{xiao21}. In the following, we will present a critical review on various FRB coherent radiation models, which are generally grouped into two types based on the emission region: those involving magnetospheres (also called ``close-in'' or ``pulsar-like'' models) and those invoking relativistic shocks far outside of the magnetopsheres (also called ``far-away'' or ``GRB-like'' models). 
 
\subsection{Magnetospheric models}\label{sec:magnetosphere}

All the pulsar-like mechanisms proposed for FRBs, as expected, have been proposed to interpret pulsar radio emission. In the following, we will discuss these mechanisms in turn, each with a brief introduction within the pulsar context, and then with a critical evaluation on its motivations and issues to account for FRB emission. Some pulsar mechanisms that have not been reinvented for FRBs are discussed in the end. 

\subsubsection{Pulsar magnetosphere basics}

Before going over detailed pulsar-like models, it is informative to review the basic physics of pulsar magnetospheres. 

Consider a pulsar that carries a plasma-loaded magnetic field and rotates with an angular velocity $\vec\Omega$. Let us make two idealized assumptions: 1. The plasma has infinite conductivity so that the net force received by each particle is zero, i.e. $e(\vec E + (1/c) ((\vec \Omega \times \vec r) \times \vec B)) = 0$ (the ideal MHD condition, which is also the force-free condition as explained below); 2. The rotating magnetosphere is in a steady state so that the $\partial / \partial t$ terms in Maxwell equations are zero (strictly applies to a uniformly rotating, $\vec\Omega \times \hat \mu_B=0$ rotator, where $\hat \mu_B$ is the direction of the magnetic axis, which is either parallel or anti-parallel to the direction of the spin axis $\vec \Omega$). From Maxwell equations and with some basic vector calculus, one can derive that everywhere in the magnetosphere within the light cylinder radius
\begin{equation}
    R_{\rm LC} = \frac{c}{\Omega} = \frac{cP}{2\pi} = (4.8\times 10^9 \ {\rm cm}) (P/1 \ {\rm s}),
\end{equation}
the net charge density as observed in the inertial frame of an observer who watches the star rotates is the Goldreich-Julian density \cite{goldreich69}
\begin{equation}
    \rho_e = \rho_{\rm GJ} \equiv -\frac{\vec \Omega \cdot \vec B}{2\pi c} \frac{1}{1-\left( \frac{\vec \Omega \times \vec r}{c}\right)^2} \simeq -\frac{\vec \Omega \cdot \vec B}{2\pi c},
\end{equation}
where $\vec B$ is the local magnetic field at a location in the magnetosphere, and for a dipolar field, its strength falls with radius $r$ as $B \simeq B_s (r/R)^{-3}$, where $R$ is neutron star radius and $B_s$ is the surface magnetic field strength. The last approximation applies to the region well within the light cylinder. This corresponds to a net charge number density
\begin{equation}
     n_{\rm GJ} = \rho_{\rm GJ} / e  \sim (6.9\times 10^{10} \ {\rm cm}^{-3}) \ B_{12}P^{-1}.
\end{equation}
By definition, with such a density there is no $\vec E$ component parallel to the local $\vec B$ vector (i.e. $E_\parallel = 0$), and $(\vec E \times \vec B)$ drift velocity is just the velocity $\vec v$ to allow particles to be frozen in the magnetic fields and co-rotate with the star, i.e. $(\vec E \times \vec B)/B^2 = \vec v/c$. The local current density can be simply denoted as $\vec j = \rho_{e} \vec v$, so the ideal MHD condition $\vec E + (1/c) (\vec v \times \vec B) = 0$ condition can be also translated to the ``force-free'' condition $\rho_e \vec E + (1/c) (\vec j \times \vec B) = 0$ \citep[e.g.][]{contopoulos99,timokhin06}. For an oblique rotator ($\vec\Omega \times \hat \mu_B \neq 0$), the $\partial/\partial t=0$ assumption is no longer satisfied, but particle-in-cell (PIC) simulations show that the GJ density is still an excellent description of the local charge density in a force-free magnetosphere \citep[e.g.][]{spitkovsky06}. Note that the Goldreich-Julian density does not depend on the specific assumption regarding the magnetic field configuration.

A force-free magnetosphere is boring, with no particle acceleration and emission. In reality, however, maintaining a force-free magnetosphere is not easy. One needs to have abundant electron-positron pairs with a number density $n_\pm = \xi n_{\rm GJ}$ and a multiplication factor $\xi \gg 1$, in order to maintain a net charge density matching the GJ density everywhere in the magnetosphere. Without copious pair production, deviation from the GJ density would be quickly built up even if initially a GJ magnetosphere is realized. This is because the centrifugal force drives particles away due to the rapid spin of the star. As a result, various charge deficit regions, or ``gaps'', where $|\rho| < |\rho_{\rm GJ}|$ is satisfied, would form in the magnetosphere \citep{ruderman75,arons79,cheng86,muslimov92}. In these gaps, $E_\parallel$ no longer vanishes. Charged particles are accelerated and radiate curvature radiation or inverse Compton scattering, producing $e^{\pm}$ pairs via either the $\gamma B$ or $\gamma\gamma$ QED processes \citep{daugherty96,zhangharding00,hibschman01}. The pairs subsequently redistribute in the $E_\parallel$, forming an opposite $E_\parallel$ field and eventually ``screen'' the original $E_\parallel$. The magnetosphere then again approaches the GJ force-free configuration. Such processes are likely unsteady, driving refreshed generation of pairs. Production of pairs has long been regarded as the necessary condition to power pulsar radio emission, with the radio pulsar ``deathline'' defined such that pair production conditions fail \citep{ruderman75,zhang00}. 

Another way of modifying the GJ magnetosphere is to introduce a global current $\vec J$ in the magnetosphere \citep{thompson02,beloborodov09}. In this case, the net charge density as observed by a lab-frame observer becomes \cite{thompson02}
\begin{equation}
    \rho_e = \rho_{\rm GJ} + \rho_{\rm twist},
\end{equation}
where
\begin{equation}
    \rho_{\rm twist} = \frac{1}{4\pi c} \vec\Omega \cdot [\vec r \times (\nabla \times \vec B)] \simeq \frac{1}{c^2} \vec\Omega \cdot (\vec r \times \vec J)
\end{equation}
describes a new charge density component to induce a twisted magnetic field component around the current (Ampere's law). A twisted magnetosphere can be still force-free, but is not in a steady state and would gradually untwist via dissipation within the twist-supported current with a non-zero potential \citep{beloborodov09}. \citet{chen17} showed from PIC simulations that an electric ``gap'' with unscreened parallel electric field can form in a twisted magnetar magnetosphere, which continuously accelerate particles and main pair production. Twisted magnetospheres are usually discussed within the context of the magnetars after X-ray flares, which undergo secular untwisting in an extended period of time. 

Recent PIC simulations revealed that besides charge-depleted gaps for pair starved magnetospheres, another promising energy dissipation and particle acceleration site for a pair rich magnetosphere is the equatorial current sheet region outside the light cylinder \cite{kala14,philippov18,kala18}. This region is regarded as a possible new site for high-energy emission from pulsars. 

Phenomenological studies and geometric modeling of pulsar radio emission suggest that there are potentially three types of pulsar radio emission:
\begin{itemize}
    \item Inner magnetospheric radio emission: Radio emission from old, slowly rotating pulsars is consistent with emission from the inner magnetosphere in the open field line regions. The double-peak pulse profile and its ``radius-to-frequency mapping'' (wider separations at a lower frequencies) as observed in a large sample of pulsars strongly support this geometric configuration. Modeling suggests that the radius of the emission is about 10s of stellar radii \citep{rankin93}.  
    \item Outer magnetospheric radio emission: Young pulsars such as the Crab pulsar have a pair of pulses that clearly align with the high-energy ($\gamma$-ray, and X-ray) pulses \citep{hankins07}. Since the latter has to be emitted from the outer magnetosphere (the predicted high-energy cutoff due to $\gamma B$ pair production from inner magnetosphere models for $\gamma$-ray emission was not detected), this radio component must be generated from the outer magnetosphere or even in the current sheet region outside the magnetosphere. 
    \item Magnetar radio emission: Magnetars are poor radio emitters and usually do not emit radio pulses during the quiescent state. However, they can become transient radio pulsars after bursting activities. When they emit, the radio pulses sometimes show a broad pulse profile and a flat or even rising spectrum, in apparent contrast to the pulses from normal pulsars \citep{camilo07}. SGR J1935+2154 was detected by FAST to show a pulsar phase five months after FRB 200418, with 795 pulses detected in 16.5 hours over 13 days \cite{zhu22}. Unlike the radio pulses of radio pulsars, these pulses have an opposite phase with respect to the X-ray pulses from the magnetar. It is unclear whether magnetar radio emission shares the same origin as one of the two mechanisms operating in normal pulsars or has its distinct origin. 
\end{itemize}

FRB emission has a typical luminosity  $\sim 10$ orders of magnitude higher than pulsar radio emission. It is unclear whether any of the three above mentioned mechanisms can apply to FRBs. 

\subsubsection{Coherent curvature radiation by bunches}

This mechanism has been widely discussed in both the pulsar and FRB fields. Within the pulsar context, \citet{ruderman75} suggested that unsteady vacuum gap discharges release ``sparks'' composed of secondary electron-positron pairs, which collide at a distance of 10s of neutron star radius. Two stream instabilities drive the formation of bunches \citep{usov87,melikidze00}, which radiate coherently in curved magnetic field lines to produce pulsar radio emission from the inner magnetosphere. The mechanism was found ``user-friendly'' to account for the phenomenology of pulsar radio emission, including the characteristic frequency, radius-to-frequency mapping, polarization properties, etc. \citep{ruderman75}. The formation and maintenance of the bunches were regarded as the main drawbacks for such a mechanism \cite{melrose78}, but various suggestions to overcome these criticisms have been discussed in the literature \citep[e.g.][]{melikidze00}. 

The application of this mechanism to FRBs has been discussed by several authors \citep{katz14,kumar17,yangzhang18,lu18,katz18b,katz20c,lu20,yang20c,wanglai20,cooper21,wangwy22,wangwy22b}. Because of the extremely high $T_b$ of FRB emission, some novel aspects of the mechanism have been noticed. The key ingredients of such a mechanism can be summarized as follows:
\begin{itemize}
    \item Characteristic frequency: According to Eq.(\ref{eq:CR}), the frequency of curvature radiation is $\nu_{\rm CR} \sim 0.72 \ {\rm GHz} \ \gamma_{e,2}^3 \rho_7^{-1}$. For 1 GHz radiation, the required electron Lorentz factor is
    \begin{equation}
      \gamma_e \simeq 110 \nu_9^{1/3} \rho_7^{1/3},
      \label{eq:gamma_e}
    \end{equation} 
    which is in the range of $10^2-10^3$ for a wide range of curvature radius $\rho$, from $\sim 10^7$ cm ($10 R_{\rm NS}$) to $\sim 10^{10}$ cm (around the light cylinder radius).
    \item Emission power of a bunch: The emission power of an individual electron is
    \begin{eqnarray}
        P_e & = & \frac{2}{3} \frac{\gamma_e^4 e^2 c}{\rho^2} \nonumber \\
        & \simeq & (4.6\times 10^{-15} \ {\rm erg \ s^{-1}}) \ \gamma_{e,2}^4 \rho_7^{-2} \nonumber \\
        & \simeq & (7.2\times 10^{-15} \ {\rm erg \ s^{-1}}) \ \nu_9^{4/3} \rho_7^{-2/3}. 
    \end{eqnarray}
    A bunch of $N_{e,b}$ electrons would emit with a power $\sim N_{e,b}^2 P_e$ (Strictly, this is the maximum value,  \citet{yangzhang18}.). The number $N_{e,b}$ in a bunch can be estimated as
    \begin{equation}
     N_{e,b} = A_b \lambda n_e \simeq A_b \lambda \zeta n_{\rm GJ} \simeq 3\times 10^{21} \ A_{b,9}\nu_9^{-1}\zeta_1 n_{\rm GJ,10},
    \end{equation} 
    where $\zeta$ is the net-charge multiplicity with respect to the Goldreich-Julian density, and $A_b$ is the cross section of the bunch, whose radial size is fixed roughly as the wavelength $\lambda$ of the emission (Fig.\ref{fig:bunch}). The most conservative estimate gives $A_{\rm b,min} \sim \pi (\gamma_e \lambda)^2$, which requires that the transverse coherence region covers the wavelength in the electron comoving frame \citep{kumar17,wanglai20}. The bunch cross section can be in principle much larger, up to the radius whose projection in the direction of line-of-sight is $\lambda$, i.e. $r_{\perp,1} \sim \sqrt{r_0 \lambda}$ (the Fresnel length, Fig.\ref{fig:bunch}); but is limited to the casually connected region size $r_{\perp,2} \sim \rho/\gamma$. So one may write 
    \begin{eqnarray}
        & A_{\rm b,max}  \simeq  {\rm min} \left[\pi r_0 \lambda, \pi (\rho/\gamma)^2 \right] \nonumber \\
        & \simeq {\rm min} (9.4\times 10^8 \ {\rm cm} \ r_{0,7} \nu_9^{-1}, 3.1\times 10^{10} \ {\rm cm} \ \rho_7^2 \nu_9^{-2}). \nonumber \\
        &
    \end{eqnarray}
    Here $r_0$ is the distance between the FRB emission region and the effective origin of the field line tangents\footnote{The introduction of $r_0$ is purely for a geometric purpose to estimate the maximally allowed the size of the bunch. Physically, particles are ejected from the inner magnetosphere of the neutron star and whether there is transverse coherence up to $\pi r_0 \lambda$ depends on the detailed particle injection and bunch formation processes.}. This is especially the case when field lines are nearly parallel in the outer magnetospheres. 
\begin{figure}
\includegraphics
[width=\columnwidth]{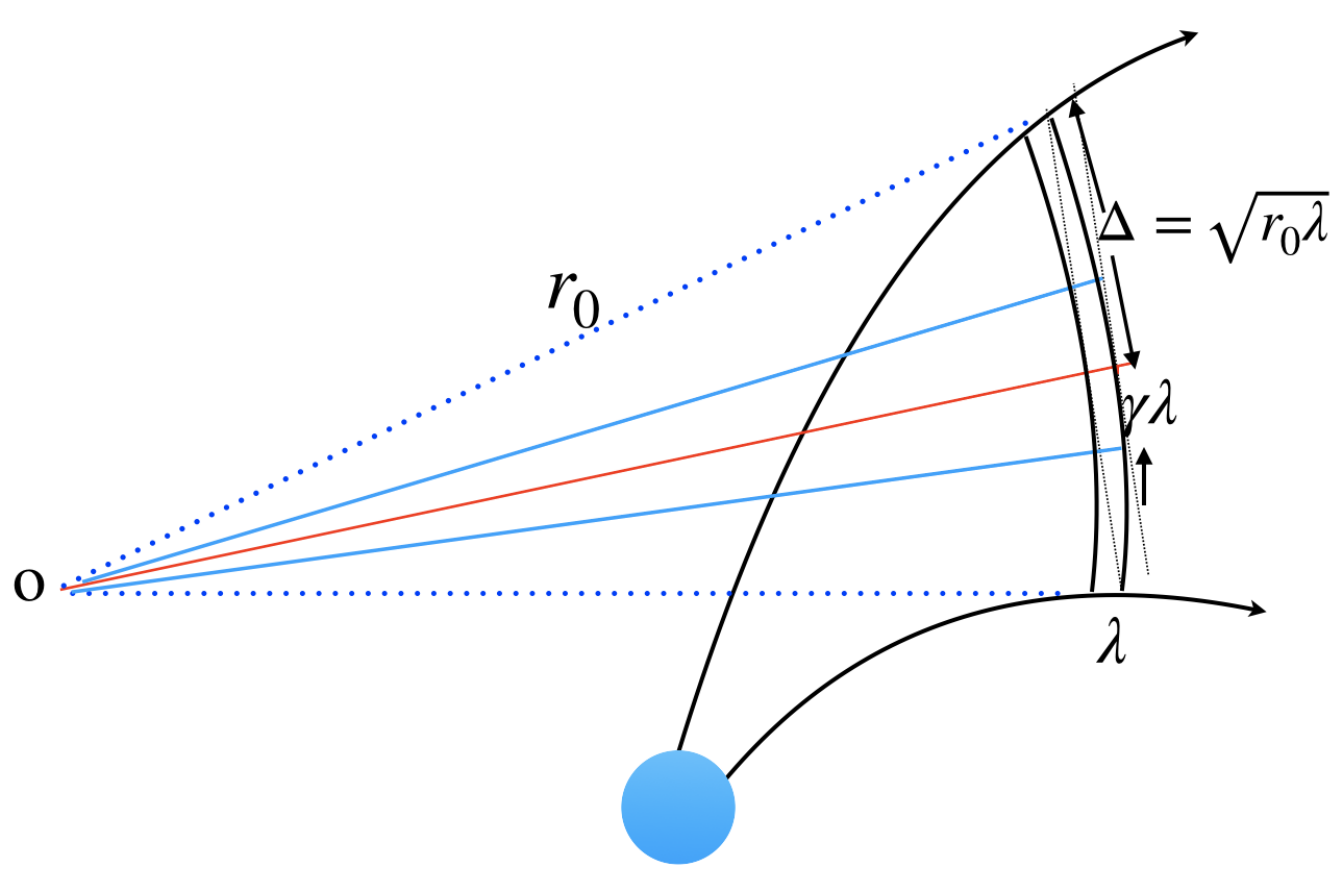}
\caption{An illustration of the shape of bunch. The radial size is approximately limited by the wavelength, i.e. $\sim \lambda$. The maximum transverse size is at least $\sim \gamma_e\lambda$, but can be as large as the Fresnel length $\sim \sqrt{r_0 \lambda}$. Note that in order to show the geometry clearly the bunch size is greatly exaggerated. In reality, the distances from the two edges of the bunch to the NS as well as $r_0$ are similar to each other. }
\label{fig:bunch}
\end{figure}

    \item Observed luminosity: Because of the light propagation effect as discussed in \S\ref{sec:size}, the observed power of an individual emitting electron is greater than its emitted power by a factor of $\sim (1-\beta_e \cos\theta)^{-1} \sim \gamma_e^2$ (when $\theta \leq 1/\gamma_e)$. Considering the possible existence of $N_b$ independent bunches that contribute to the observed luminosity at an epoch, one may write the total true luminosity (not isotropic equivalent) as
    \begin{eqnarray}
      L & \simeq & N_b N_{e,b}^2 P_e \gamma_e^2 \nonumber \\
      & \simeq & (7.8\times 10^{38} \ {\rm erg \ s^{-1}}) \ N_{b,6} A_{b,9}^2 \zeta_1^2 n_{\rm GJ,10}^2,
      \label{eq:L_bunch}
    \end{eqnarray}
    where $N_b$, $A$ and $n_{\rm GJ}$ are normalized to their respective typical values ($N_b$ may be estimated as $\Delta r/\lambda = (3.3\times 10^5 \ {\rm cm}) \Delta r_7 \nu_9$, where $\Delta r$ is the depth of the field line that contributes to instantaneous radiation). Interestingly, $\nu_9$ and $\rho_7$ are apparently cancelled out in Equation (\ref{eq:L_bunch}). This ``true'' luminosity from the model may be compared with the beaming corrected luminosity derived from the observed isotropic luminosity as discussed in \S\ref{sec:beaming}, i.e.
    \begin{equation}
        L \simeq L_{\rm iso} {\rm max} (\theta_j^2/4, (4 \gamma_e^{2})^{-1}),
    \end{equation}
    where the solid angle of an individual FRB $\delta\Omega$ is written as $\pi \theta_j^2$ where $\theta_j$ is the half opening angle of the min-jet. Note that when $\theta_j \leq \gamma_e^{-1}$, our treatment is consistent with \citet{kumar17}, who used a $\gamma_e^4$ parameter to make a connection between the emitted power and the observed isotropic luminosity. Our treatment is more general that includes the $\theta_j > \gamma_e^{-1}$ regime. One can see that for plausible parameters, the observed FRB isotropic-equivalent luminosity can be reproduced. 
    \item Cooling time and the required $E_\parallel$: \citet{kumar17} first pointed out that for models invoking curvature radiation by bunches, a steady $E_\parallel$ is needed in the magnetosphere to continuously inject energy to the bunches to maintain the observed luminosity for the typical FRB duration. Consider a bunch of $N_{e,b}$ electrons radiating coherently. The total energy is $E_b = N_{e,b} \gamma_e m_e c^2$ and the total emission power is $N_{e,b}^2 P_e$. So the cooling timescale is
    \begin{equation}
        t_c = \frac{\gamma_e m_e c^2}{N_{e,b} P_e} \simeq (4.5\times 10^{-12} \ {\rm s}) A_{b,9}^{-1} \rho_7 \zeta_1^{-1} n_{\rm GJ,10}^{-1}. 
    \end{equation}
    This is much shorter than the typical FRB duration. In order to maintain FRB emission power, the electrons need to continuously gain energy from an electric field $E_\parallel$ such that $(N_{e,b} e) E_\parallel c = N_{e,b}^2 P_e$. This gives
    \begin{equation}
        E_\parallel = \frac{N_{e,b} P_e}{ec} \simeq (1.4 \times 10^6 \ {\rm esu}) \nu_9^{1/3} \rho_7^{-2/3} A_{b,9} \zeta_1 n_{\rm GJ,10}.
    \end{equation}
    The existence of such a field is required to apply the coherent curvature radiation by bunches to explain FRB emission. \citet{kumar20a} proposed that such an $E_\parallel$ may be provided by the propagation of Alf\'ven waves to a charge starved region at an altitude of 10s of neutron star radii. \citet{lu20} argued that such a mechanism can account for FRBs with a wide range of luminosities. \citet{cooper21} investigated the maximum luminosity of this mechanism by considering the effect of the induced current of the emitting bunches. They confirmed that the mechanism can generate emission with FRB luminosities. \citet{qu23} showed that the existence of such an $E_\parallel$ is essential to overcome the plasma suppression effect for bunched coherent emission \cite{gil04,lyubarsky21}.
    \item Spectrum: The radiation spectrum of coherent curvature radiation by three-dimensional bunches in a realistic pulsar magnetosphere was calculated in detail by \citet{yangzhang18} and \citet{wangwy22}. The spectrum is found to be in the form of a broken power law separated by a few characteristic frequencies defined by the length and the opening angle of the bunch. The spectral indices of different segments depend on the relative ordering among the characteristic frequencies, and the high-frequency spectral index depends on the power law index of the emitting electrons $p$. The possible self-absorption effect by other bunches was studied by \citet{ghisellini18}. If charges are spatially separated, the shape of the coherent spectrum would have a narrower peak than the regular case  \citep{yang20c}. In general, to achieve narrow spectra for the bunch models, one needs to invoke convolution of the intrinsically broad spectrum of individual bunched charges and their spatial distribution \cite{katz18b}. 
    \item Polarization: Both O-mode and X-mode polarized waves can be generated with curvature radiation \citep{wang12,kumar17}. Nearly 100\% linear polarization degree is expected if the observer is within the $1/\gamma_e$ cone of the electron emission beam, but circular polarization can develop outside the emission cone \cite{wangwy22,wangwy22b,tong22,quzhang23}. Depending on the location of the emission region, the polarization angle can either display a swing (for an inner magnetospheric location and/or a rapid rotation of the magnetosphere), as seen in radio pulsars within the framework of the rotating vector model \citep{rad69}, or stay nearly flat (for an outer magnetospheric location and/or a slow rotation of the magnetosphere). 
    \item Radius to frequency mapping and frequency downdrifting: \citet{wang19} showed that there is a simple interpretation to the frequency downdrifting feature observed in some FRBs. Since charged bunches need to be radiation-reaction limited within this model (balance between $E_\parallel$ accceleration and curvature cooling), $\gamma_e$ may maintain a roughly constant value along field lines. Since the curvature radius continuously increases as the bunches move away from the magnetosphere, the curvature radiation frequency continuously decreases with the increasing height. Suppose several bunches in adjacent field lines were launched simultaneously from the base, as the magnetosphere rotates, the line of sight always catches emission from lower altitudes (and hence, with a higher frequency) first and the emission from higher altitudes (and hence, with a lower frequency) later, so that frequency downdrifting should be commonly expected\footnote{\citet{lyutikov20b} later proposed a similar idea to interpret frequency downdrifting using radius-to-frequency mapping, even though the radiation mechanism was not specified.}. Allowing that the bunches can be ejected at somewhat different times, \citet{wangwy20b} showed that occasionally a frequency updrifting FRB may be observed, but the downdrifting pattern should prevail. This is consistent with the observations \citep{zhou22}. 
\end{itemize}

Despite the success of this simple model to interpret a broad range of pulsar and FRB phenomenology, the mechanism has been criticized by several authors:
\begin{itemize}
    \item \citet{melrose78} pointed out that coherent curvature radiation by bunches suffers from the difficulties of bunch formation and maintenance. The bunch formation mechanisms have been explored extensively in the pulsar field. The common ingredient of the models is a two-stream instability \citep[e.g.][]{melikidze00}, which is likely realized in the violent event that powers an FRB. The maintenance of bunches is more difficult to realize. Strong repulsion force within the bunches tend to disperse the bunch spatially and radiation reaction tends to disperse the bunch in the momentum space \cite{katz18b,katz20c}. However, since the FRB duration is short, the maintenance mechanism only needs to apply for a millisecond duration. 
    \item \citet{lyubarsky21} emphasized that the plasma effect, which tends to limit brightness temperature and is moderately severe for pulsar radio emission \citep{gil04}, becomes substantial in suppressing coherent emission from FRBs. If the bunch moving with $\gamma_e$ is surrounded by a plasma moving with $\gamma_p$, \citet{lyubarsky21} suggests that the emission power of the bunch $N_{e,b}^2 P_e$ is suppressed by a huge factor of the order of $10^{-10}$.
    \citet{qu23} revisited the arguments of \cite{gil04,lyubarsky21} and found that the plasma suppression effect is not important in FRB problems. If a strong $E_\parallel$ exists in the emission region, as expected in the realistic FRB models \citep{kumar17}, there is essentially no suppression if the bunch is in the radiation-reaction-limited regime for coherent curvature radiation. 
    \item The curvature radiation spectrum, similar to the synchrotron spectrum, might be too broad to interpret the narrow-band spectrum as observed in some FRBs, especially the repeaters. Charge separation can alleviate this criticism \citep{yang20c,katz18b}.
\end{itemize}

\subsubsection{Coherent ICS emission by bunches, free electron laser and linear acceleration emission}\label{sec:ICS}

Besides curvature radiation, there are another family of models that invoke vacuum-like coherent mechanisms that do not depend intrinsically on the dispersive properties of the plasma. Within these models, bunched particles resonate coherently in some low-frequency waves, either of electromagnetic or electrostatic types, and inverse Compton scatter the waves to higher frequencies to make FRBs. A relatively simple model is to invoke low frequency electromagnetic waves, which might be excited near the neutron star surface by crustal oscillations \citep{zhang22}. Usually it is believed that crustal oscillations would excite Alfv\'en waves, and indeed the bulk of the energy that eventually powers an FRB is likely carried by Alfv\'en waves. On the other hand, if a small amount of oscillation energy would be converted to electromagnetic waves by coherently oscillating charges in the near-surface magnetosphere, then the waves (all modes for a quasi-parallel configuration and X-mode only for a quasi-perpendicular configuration) would penetrate through the magnetosphere unimpeded. Suppose that there are relativistic bunched charges moving with a bulk Lorentz factor $\gamma$, the low-frequency electromagnetic waves with angular frequency $\omega_0$ and frequency $\nu_0 \sim 10^4$ Hz would be upscattered to a frequency
\begin{eqnarray}
    \omega & = & \gamma^2 \omega_0 (1-\beta\cos\theta_i), \label{eq:omegaICS} \\
    \nu &= & (1 \ {\rm GHz}) \ \gamma_{2.5}^2 \nu_{0,4} (1-\beta\cos\theta_i), \label{eq:nuICS}
\end{eqnarray}
where $\theta_i$ is the incident angle. Such an inverse Compton scattering (ICS) model has been considered to interpret pulsar radio emission \citep{qiao98,qiao01,xu00} and its promise to interpret FRB emission has been discussed in detail in \citet{zhang22}.

The advantage of the ICS mechanism is that the emission power of a single particle, $P_e^{\rm ICS} \sim (1.6\times 10^{-7} \ {\rm erg \ s^{-1}}) (\delta B_{0,6})^2 r_8^{-2}$ (where $\delta B_0$ is the oscillation amplitude of magnetic field of the electromagnetic waves and $r$ is radius of the emission region) is much greater than that of curvature radiation, $P_e^{\rm CR} \sim (4.6\times 10^{-15} \ {\rm erg \ s^{-1}}) \gamma_{2.5}^4 \rho_8^{-2}$. As a result, the required degree of coherence to interpret the FRB high brightness temperature is greatly reduced. Indeed, even if one adopts the most conservative bunch cross section $A_{\rm b,min} = \pi (\gamma \lambda)^2$, the required $N_{e,b}$ is so low that a charge number density of the order of $n_{\rm GJ}$ is already enough to account for the FRB luminosity. As a result, the bunches do not need to have a large plasma density and the criticism of bunch emission suppression due to the plasma effect \citep{lyubarsky21} is greatly alleviated \citep{qu23}. Even a small fluctuation in charge number density with respect to the background Goldreich-Julian density \citep{yangzhang18} would be adequate to produce bunched ICS radiation, so that the criticisms of bunch formation and maintenance \citep{melrose78} are also alleviated. The frequency of the scattered waves (Eq.(\ref{eq:nuICS})) depends on $\omega_0$, $\gamma$ and $\theta_i$. All could be nearly constant within an FRB ($\gamma$ is radiation-reaction limited), so the  bunched coherent ICS mechanism has the advantage of generating narrow-band spectra than curvature radiation, consistent with observations \citep{zhou22}. The frequency down-drifting feature may be also produced via a radius-to-frequency mapping feature or the intrinsic damping of the low frequency waves toward longer wavelengths \citep{zhang22}. This mechanism can also produce intrinsic circular polarization with a proper viewing geometry \cite{quzhang23}.

In the case that vacuum-like electromagnetic waves are not excited from the near surface region, coherent radio emission may be still excited by relativistic particles scattering off an oscillating electric field along the magnetic field line via the amplified linear acceleration emission (ALAE) or off Alfv\'en waves via the free-electron laser (FEL) mechanism. 

ALAE was introduced in early years by \citet{melrose78} as a mechanism to replace bunched curvature radiation to interpret pulsar radio emission. It was further developed by \citet{rowe95}. It invokes an oscillating $E_\parallel$ along the direction of particle motion, with particles radiating coherently in such an accelerating field. This mechanism has not been investigated in detail within the FRB context. Since the emission occurs nearly the neutron star surface, it is unclear whether the mechanism can produce the observed high $T_b$ of FRBs without being absorbed or scattered within the inner magnetosphere \citep[e.g.][]{ioka20b,beloborodov21}. 

The FEL mechanism invokes the interaction between an Alfv\'en wave disturbance (also called wiggler) and a relativistically moving bunch. This mechanism has been studied intensively in the laboratories \citep[e.g.][]{benford92} and was discussed by \citet{fung04} within the pulsar radio emission context.  It was investigated in detail by \citet{lyutikov21} within the FRB context. The characteristic angular frequency of emission is defined by 
\begin{equation}
    \omega \simeq 4\gamma^2 (c k_w),
    \label{eq:omegaFEL}
\end{equation}
where $k_w$ is the wavenumber of the low-frequency wiggler waves, similar to the direct ICS case Eq.(\ref{eq:omegaICS}) where $c k_w$ is replaced by $\omega_0$. The trajectory of bunched relativistic electrons in  wiggling Alfv\'en waves can be solved. The resulting emission spectrum has a narrow band, which can interpret the spectral feature of the Crab pulsar and narrow spectra of repeating FRBs \citep{lyutikov21}. The mechanism is also more powerful than curvature radiation, and hence, easier to satisfy the brightness temperature constraint. In general, the FEL mechanism \citep{lyutikov21} and coherent ICS mechanism by bunches \citep{zhang22} are intrinsically similar mechanisms and share several common features and advantages in interpreting FRB coherent radio emission. 

\subsubsection{Magnetospheric maser mechanisms}

The bunching mechanisms discussed above do not invoke negative absorption or growth of plasma modes that depend on the dispersive properties of the plasma. In this section, we discuss several magnetospheric maser mechanisms for coherent radio emission. These mechanisms include ``vacuum masers'' that invoke negative absorption of electromagnetic radiation as if in an vacuum, and ``plasma masers'' that invoke growth of plasma modes. In order for the latter models to work, several requirements are needed: (1) The plasma should support modes whose frequency falls into the observed frequency band (e.g. GHz); (2) there should be an unstable particle distribution in the relevant frequency band, which should resonate the plasma mode; (3) the mode should grow rapid enough to reach the desired amplitude to account for the high brightness temperature; and (4) the plasma mode should eventually escape the region as electromagnetic waves. To date, none of such magnetospheric models have been found suitable to interpret FRB radio emission. As a result, even though the following mechanisms have been introduced to interpret coherent radio emission from other astronomical objects, so far they have not been found successful in interpreting FRB emission (see \citet{lu18} for a critical study of maser mechanisms within the context of FRBs). Nonetheless, they are listed below for completeness. 

\begin{itemize}
    \item Relativistic plasma emission: Plasma emission is a multi-stage process \citep{melrose17}. The first step is to drive a longitudinal Langmuir waves through a streaming instability between a fast beam and a background plasma. The subsequent stages include amplification of the Langmuir turbulence and the conversion of the plasma mode to electromagnetic waves that could eventually escape the magnetosphere. The characteristic emission frequency should be the plasma frequency $\omega_p$ or the boosted plasma frequency $\Gamma \omega_p$ in the relativistic version. Even though plasma emission has been identified as the main mechanism producing coherent solar radio bursts, it has not been successful in interpreting pulsar radio emission. The main reason is that the instability growth rate is too small due to the limitation of the relatively small plasma property gradients. 
    \item Electron cyclotron maser emission: This mechanism involves plasma maser emission at the non-relativistic cyclotron frequency $\omega_{\rm B}$ and its harmonics $s \omega_{\rm B}$, with decreasing amplitudes at higher $s$ values. The mechanism was found to be responsible to the decametric radio emission of Jupiter and the auroral kilometric radiation of Earth \citep{melrose17}. However, within a neutron star magnetosphere (especially for a magnetar), $\omega_{\rm B}$ is usually much higher than the radio frequency so that this mechanism is usually not relevant for magnetospheric models\footnote{For slow rotating non-magnetar pulsars, the condition may be satisfied. However, the energetics of the neutron star would not be large enough to power FRBs.}.
    \item Curvature radiation maser: Curvature radiation maser is not possible for a rotating dipole because there is no solution for negative absorption. However, if the pulsar magnetosphere is distorted, under certain conditions negative absorption would be possible for curvature radiation \citep{luo95}. Such a model is not attractive to interpret FRBs since it is unclear how the specific magnetospheric configuration might be realized in an FRB emitting source. 
    \item Anomalous cyclotron-Cherenkov and Cherenkov drift resonances: Instabilities occur when a dispersion relation has a term whose denominator approaches zero, termed as a resonance. In a pulsar magnetosphere, maser-type plasma instabilities can operate at the anomalous cyclotron-Cherenkov resonance $\omega - k_\parallel v_\parallel + \omega_{\rm B}/\gamma = 0$ ($\gamma$ is the plasma Lorentz factor in the pulsar frame) and the Cherenkov drift resonance $\omega - k_\parallel v_\parallel - k_\perp u_d = 0$ ($u_d$ is the drift velocity). Even though this mechanism is a plausible candidate to account for pulsar radio emission \citep{lyutikov99}, they are not favored in interpreting FRBs because the conditions for the maser mechanisms to operate either cannot be realized or demand unreasonable parameters \citep{lu18}. 
\end{itemize}

\subsubsection{Other magnetospheric mechanisms}

Two more magnetospheric coherent mechanisms have been proposed to interpret FRB emission, which deserve special discussion. 

The first model was proposed by \citet{lyubarsky20}. This model invokes a large scale magnetic perturbation to form a magnetic pulse, which strongly compresses the magnetospheric plasma and pushes it away. The pulse propagates from the flare site within the magnetosphere outwards and eventually reaches the current sheet that separates the oppositely oriented magnetic fields beyond the light cylinder. The FRB is powered by the enhanced magnetic reconnection in the current sheet region. It is conjectured that coalescence of magnetic islands in the reconnection current sheet produces magnetosonic waves, which propagate on top of the magnetic pulse and eventually escape as electromagnetic waves. The characteristic frequency is defined by the dimension of the magnetic islands $\xi a'$ where $a'$ is the width of the current sheet, so that 
\begin{equation}
    \omega = \delta \omega' = \delta \frac{c}{\xi a'},
\end{equation}
where the primed quantities are measured in the rest frame of the magnetic pulse, and $\delta$ is the Doppler factor of the pulse. In order to match the observed FRB frequency, $\xi \sim (10-100)$ is required. The emission is polarized along the rotation axis of the magnetar. The advantage of this model is that the emission site is slightly beyond the light cylinder, which avoids the criticisms regarding FRB propagation within the magnetosphere \citep{beloborodov21}. PIC numerical simulations of such a scenario have been carried out, which show the excitement of narrow-band GHz emission \cite{mahlmann22}. An alternative radiation mechanism within this scenario is bunched ICS of relativistic electrons accelerated from the current sheet off the low frequency waves generated from the inner magnetosphere \citep{zhang22}. 

The second model is the direct electromagnetic wave generation from non-uniform pair production across different field lines. Within the framework of radio pulsars, \citet{timokhin10,timokhin13} showed that unsteady pair production is norm near pulsar polar caps regardless of whether there is strong binding of particles from the pulsar surface. Through 1D simulations, they showed the existence of broad-band superluminal electrostatic waves in the unsteady pair-screening region, which they suspect as a candidate pulsar radio emission mechanism.  \citet{philippov20} showed from large-scale two-dimensional kinetic plasma simulations that such non-steady pair production and screening of electric fields along the magnetic field lines by freshly produced pairs can naturally generate electromagnetic waves that can escape the magnetosphere. They propose that such a mechanism could be responsible for the coherent radio emission of radio pulsars. Within the context of FRBs, \citet{wadiasingh20} speculated that such a mechanism may power FRBs from a magnetar magnetosphere. \citet{yangzhang21} developed an analytical toy model for this process and showed that the mechanism can indeed apply to FRBs given that non-steady, non-uniform pair production could be realized in an FRB environment. They argued that crustal oscillations of a magnetar could be the engine for such non-steady, non-uniform pair production processes. 

\subsubsection{Transparency of FRBs from magnetospheres}

One criticism to the magnetospheric mechanism of FRBs is that the FRB waves may undergo strong scattering by the magnetospheric plasma so that the high brightness temperature would not be achievable \citep{beloborodov21}. Due to their high intensities, FRB waves have a large oscillation amplitude for the wave $\vec E_w$  field (also wave $\vec B_w$ field) such that the dimensionless amplitude parameter \citep[e.g.][]{luan14}
\begin{equation}
    a \equiv \frac{eE_w}{m_e c \omega} \simeq \frac{eB_w}{m_e c \omega}= \frac{\omega_{\rm B,w}}{\omega} \gg 1
\end{equation}
in the magnetosphere (see \S\ref{sec:large-a} for more discussion on the large-amplitude wave effect). This amplitude factor $a$ denotes how fast an electron moves in response to the waves and when $a \gg 1$, the electron speed approaches the speed of light. Without an external magnetic field, the electron would move under both the oscillating $E_w$ field and the Lorentz force due to $B_w$, making a ``8'' shape trajectory \citep{sarachik70,yang20a}. The electron is accelerated to a Lorentz factor of the order of $a$. The scattering cross section can be generally defined as $\sigma = P / S$, where $P$ is the emitting power and $S$ is the received photon flux. Because of the relativistic motion of the electron, the emitted power is enhanced by a factor of $\sim a^2$ with respect to Thomson scattering, so that \citep{sarachik70,yang20a}
\begin{equation}
    \sigma \sim a^2 \sigma_{\rm T}.
    \label{eq:sigma}
\end{equation}

In a neutron star magnetosphere with background $B$, the situation is more complicated. When $B \gg B_w$, the electron motion is dictated by $B$ rather than $B_w$, so that the enhancement of $\sigma$ (Eq.(\ref{eq:sigma})) would not occur. In a magnetosphere, since $B \propto r^{-3}$ for a dipolar configuration and since $B_w   = \sqrt{L/c r^2}\propto r^{-1}$ for an EM wave, there will be a point where $B$ drops below $B_w$. Recall $\omega_B = eB/m_e c$, this condition can be also written as 
\begin{equation}
    a > \frac{\omega_B}{\omega}.
\end{equation}
The scattering cross section is greatly increased and the optical depth is greatly enhanced:
\begin{equation}
    \tau_{es} \sim n \sigma r_c \simeq 0.4 \tilde\sigma L_{42}^2 \xi B_{s,15}^{-1}\nu_9^{-2} P^{-1} R_6^{-4}. 
\end{equation}
Here $n=\xi n_{\rm GJ}$, $\tilde\sigma = \sigma/( [a(r_c)]^2 \sigma_{\rm T})$ is the cross section normalized to $a^2\sigma_{\rm T}$, and $r_c = (B_s R^3)^{1/2} (c/L)^{1/4} \simeq (4.2 \times 10^8 \ {\rm cm}) L_{42}^{-1/4} B_{s,15}^{1/2} R_6^{1/2}$, which is the critical radius at which $B_w = B$. This estimate suggests that the FRB waves would indeed become opaque to Thomson scattering in a magnetar magnetosphere, if $L$ and $\xi$ are large \citep{beloborodov21}. The situation worsens since the relativistic motion of electrons in complicated trajectories would radiate $\gamma$-rays, which may produce additional pairs to increase the opacity.

However, the above arguments are based on two assumptions: (1) the magnetospheric plasma is essentially at rest and the angle between wave propagation and the local $B$ field, $\theta_{kB}$ is nearly $90^{\rm o}$, i.e. the FRB is trying to penetrate through the closed field line region. \citet{qu22c} argued that both assumptions are likely invalid in  realistic magnetospheric emission models for FRBs. Various mechanisms (the standard pulsar mechanism, Alfv\'en wave propagation, and ponderomtive force acceleration) likely drive a relativistically moving plasma in the open field line region of a magnetar magnetosphere. The propagation of the intense FRB waves also tend to align the $\vec k$ and $\vec B$ vectors so that $\theta_{kB}$ is likely $\ll 1$. Both effects would reduce the scattering optical depth significantly and it is shown that FRBs are transparent in a magnetar magnetosphere even for high-luminosity FRBs with a large pair multiplicity, if the plasma Lorentz factor $\gamma_p > 10^2$ \citep{qu22c}.

FRBs are likely associated with X-ray and $\gamma$-ray photons emitted from a magnetar magnetosphere. The transparency of FRBs depends on the competition between the FRB and X-ray luminosities. \citet{ioka20b} showed that FRB photons can break out of the pair-rich magnetopshere with radiation pressure if the FRB emission radius is larger than a few tens of NS radii.  As long as the work done by the FRB waves on the $e^\pm$ is small compared with the initial FRB energy, the FRB can successfully break out the magnetosphere. \citet{ioka20b} showed that the breakout condition is satisfied in the high $L_{\rm FRB}$ low $L_X$ regime. According to this result, SGR giant flares may not be associated with successful FRBs since the bright X-ray emission would likely choke the FRB jet (see  \citet{katz16} for discussion of alternative possibilities of non-detection of an FRB associated with SGR 1806-20 giant flare). This is consistent with the radio luminosity upper limit of the SGR 1806-20 giant flare \cite{tendulkar16}. 

\subsection{Relativistic shock models}\label{sec:shocks}

The second general type of models invoke relativistic shocks to generate coherent radio emission. The term ``synchrotron maser'' has been adopted to describe several very different scenarios. We discuss the three versions of the model below, with the decreasing order of their relevance to ``synchrotron maser'', which incidentally, is also the reverse order of popularity. 

\subsubsection{Vacuum synchrotron maser}

The first model is literally ``synchrotron maser''. For a synchrotron emitting source, the synchrotron absorption coefficient can be written in the form of \citep{rybicki79,ghisellini17,waxman17,lu18}
\begin{equation}
    \alpha_\nu = -\frac{1}{2 m_e \nu^2} \int_1^\infty\gamma^2 j_\nu(\gamma,\psi) \frac{\partial}{\partial \gamma} \left(\frac{dN/d\gamma}{\gamma^2}\right)  d\gamma ,
\end{equation}
where $j_\nu(\nu, \psi)$ is the viewing-angle-dependent emissivity for a single electron (in $\rm erg \ s^{-1} \ Hz^{-1} \ sr^{-1}$, different from the volume emissivity commonly defined). One may also write the net absorption cross section per particle \citep{ghisellini91,ghisellini17,lu18} 
\begin{equation}
    \sigma_{a,\nu} \simeq \frac{1}{2 m_e \nu^2} \frac{1}{\gamma^2} \frac{\partial}{\partial \gamma} [\gamma^2 j_\nu(\gamma,\psi)].
\end{equation}
A vacuum maser is possible when either $\sigma_{a,\nu}$ is negative ($\gamma^2 j_\nu (\gamma,\psi)$ is a decreasing function of $\gamma$), or $\alpha_\nu$ is positive ($dN / d\gamma$ distribution is steeper than $\gamma^{-2}$, i.e. population inversion).

\citet{ghisellini17} found that if the emission region has an extremely ordered magnetic field and if the emitting electrons have a very narrow distribution for both pitch angle and energy, $\sigma_{a,\nu} <0$ is possible in a certain range of the viewing angle $\psi > 1/\gamma_e$, where $\gamma_e$ is the electron Lorentz factor. Even though relativistic shocks are not specified in the model discussed by \citet{ghisellini17}, the required magnetic field strength $B$ and electron energy $\gamma_e$ for the characteristic synchrotron frequency to fall into the FRB band are consistent with the typical values for shock models. The plasma effect is not important in this model, so the emission is the ``vacuum'' type.

Even though the mechanism is clean and straightforward, the difficulties of the model include how to maintain extremely ordered $B$ field (within $1/\gamma_e$ angle), how to accelerate particles to maintain a narrow pitch angle distribution (again within $1/\gamma_e$ angle) and how to accelerate particles to maintain a narrow energy distribution. Known astrophysical particle acceleration mechanisms, e.g. relativistic shocks and magnetic reconnection, usually accelerate particles to a power law energy distribution and the accelerated relativistic electrons typically have a wide angular distribution with respect to the local $B$ field. Perturbations usually introduce wiggles of magnetic field lines. As a result, this mechanism may not be realized in nature due to the contrived physical conditions required. 

\subsubsection{Plasma synchrotron maser in non-magnetized relativistic shocks}

Accelerated particles in relativistic shocks usually have an energy distribution $dN/d\gamma \propto \gamma^{p}$ with $p \sim -2$ above the minimum Lorentz factor $\gamma_m$. Maser emission is therefore impossible for the frequency range defined by $\gamma > \gamma_m$, since $\alpha_\nu$ is positive. Nonetheless, population inversion ($p > 2$) may be possible at $\gamma < \gamma_m$ \citep{sagiv02,waxman17}. In the extreme case, a sharp cutoff of $\gamma$-distribution below $\gamma_m$ mimics a $\delta$-function, which is much steeper than $\gamma^2$. In the frequency space, maser emission occurs at \citep{sagiv02}
\begin{equation}
    \nu < \nu_{R^*} = {\rm min} [\gamma_m, (\nu_p / \nu_B)^{1/2}] \nu_p,
\end{equation}
where $\nu_{R^*}$ is the modified Razin frequency below which the plasma effect becomes dominant, $\nu_p = \omega_p/2\pi$, and $\nu_B = \omega_B/2\pi$. The relativistic beaming effect for synchrotron radiation is suppressed because of the role played by the refractive index $n_r$ \citep{rybicki79}. The traditional synchrotron radiation is suppressed, but the possibility for maser emission is opened. This mechanism is a plasma version of synchrotron maser, and it applies to a weakly magnetized plasma with $\omega_B \ll \omega_p$. In a hydrodynamical shock, one usually defines microscopic parameters $\epsilon_e$ and $\epsilon_B$ as the fraction of shock internal energy that are distributed in electrons and magnetic fields. Observations of GRBs show that typically $\epsilon_B \ll \epsilon_e \ll 1$ (\citet{kumar15} and references therein). Since $\nu_p / \nu_B = \omega_p / \omega_B \sim  (\epsilon_e / \epsilon_B)^{1/2}$, the condition for plasma synchrotron maser emission is satisfied. 

For this model to work, a weakly magnetized central engine is preferred. Demanding model parameters to satisfy FRB observational constraints, \citet{long18} showed that neutron stars with surface magnetic fields $B_* \leq 10^{11}$ G is preferred. This is at odds with the observational constraint that magnetars are responsible to at least some FRBs. Also, since the emission region is weakly magnetized, such a model does not predict an extremely high polarization degree as is observed in the majority of FRBs. As a result, this mechanism, if relevant, would not be responsible for the majority of FRBs. 

\subsubsection{Bunched coherent cyclotron/synchrotron radiation in highly magnetized relativistic shocks}

Another version of the relativistic shock models invokes a highly magnetized upstream. The upstream magnetic field lines are highly ordered. As the shock propagates into the magnetized medium, magnetic fields are amplified and particles coherently gyrate around these field lines, forming a ``ring'' in the momentum space, even though they can spread in a wide position space. They then radiate coherently as a global bunch at the gyration frequency $\sim \omega'_B = e B' / m_e c$, where $B'$ is the downstream magnetic field strength in the comoving frame of the fluid. The observed frequency is Doppler boosted by a factor of the bulk Lorentz factor $\Gamma$ if the shock moves towards the observer relativistically. Such a mechanism, even still termed as ``synchrotron maser'', is in fact more analogous to bunched coherent cyclotron/synchrotron radiation mechanism (the electron Lorentz factor is typically a few), even though bunching occurs in the momentum space. The mechanism was introduced to the FRB field by \citet{lyubarsky14} and studied by various teams to interpret FRB observations \citep{beloborodov17,beloborodov20,metzger19,margalit20,margalit20b,lu20,yu21}. The physical process of this mechanism has been verified via particle-in-cell numerical simulations \citep{plotnikov19,babul20,sironi21}. This mechanism is physically robust (with some requirements such as ordered $B$ field and cold plasma) and user friendly in interpreting observations. As a result, it is the most competitive mechanism within the relativistic shock model category. 

The features, strengths and weaknesses of this model can be summarized as follows.

\begin{itemize}
    \item The most important condition for such a mechanism to operate is the existence of ordered magnetic fields in the upstream. Such a feature allows electrons to gyrate coherently in momentum space so that their cyclotron/synchrotron radiation power could be coherently added. A commonly suggested scenario is that an FRB magnetized pulse collides with a magnetized magnetar wind that carries a global ordered $B$ field. The FRB is emitted in the forward shock region. There are two versions of this model: the external shock type in which the upstream is an electron-ion wind produced from a previous magnetar flare \citep{metzger19}, and the internal shock type in which the upstream is a relativistic rotationally-powered electron-positron pair wind \citep{beloborodov20}. In any case, because of the highly ordered magnetic field, very high linear polarization degree is expected \cite{quzhang23}. The linear polarization angle is expected to stay constant during each burst, as has been observed in some repeaters \citep[e.g.][]{michilli18,jiangjc22}. Because of the same reason, the condition for this maser mechanism to operate is also demanding. Irregularities in the field configuration would greatly suppress coherent emission. Also a rapid swing of linear polarization angle across individual bursts \citep[e.g.][]{luo20} poses a great challenge to such a model.
    \item Another condition for such a mechanism to operate is that the upstream media should remain ``cold''. Random motion of electrons in a hot plasma would smear up or even destroy the ``ring'' in the momentum space, leading to suppression of coherent emission \citep{babul20}. As a result, this feature poses a constraint on the waiting time of successive FRBs within the external shock model. Shortly after a collision, both the shocked wind and the shocked FRB ejecta would be hot. The magnetic field configurations may be also distorted due to the irregularities introduced during the collision. If another FRB pulse collides into this remnant of previous collision, strong coherent emission would be likely suppressed. A long waiting time of the order of $\sim 100$ s would be reasonable \citep{metzger19}, which is consistent with the second peak of the waiting time distribution of active repeaters \citep{lid21,xuh22}. However, active repeaters also have another peak in the waiting time distribution, which is of the order of milliseconds \citep{lid21,xuh22}. These closely connected bursts, also known as ``burst storms'' \citep{hewitt22} or ``burst clusters'' \cite{zhou22}, pose a challenge to the external shock version of this model. This is not an issue for magnetospheric models, since different pulses are related to different emission regions in a rotating magnetosphere as they sweep across the line of sight.
    \item The magnetization parameter
    \begin{equation}
        \sigma \equiv \frac{B^2}{4\pi \Gamma \rho c^2} = \frac{{B'}^2}{4\pi \rho' c^2}
    \end{equation}
    is defined as the Poynting-flux-to-kinetic-flux ratio in the lab frame or the magnetic internal energy density (magnetic energy density plus magnetic pressure) over mass density in the comoving frame. For an electron-positron plasma, one also has $\sigma = {{\omega'_B}^2}/{{\omega'_p}^2} = {\omega_B^2}/{\omega_p^2}$ where $\omega_B=eB/\Gamma m_e c$ and $\omega_p = (4\pi n e^2/\Gamma m_e)^{1/2}$. For this mechanism to operate efficiently, the upstream $\sigma$ value should be in the Goldilocks zone with a value $\sigma \sim 1$. At smaller $\sigma$ values, since magnetic energy is not dominant, global magnetic fields are likely subject to turbulent perturbation so that the field lines tend to be more tangled. The coherent mechanism cannot operate efficiently. At higher $\sigma$ values, the fraction of energy carried by particles reduces (most energy is still in magnetic fields) so the efficiency of making coherent emission also drops. PIC simulations suggest that the maser efficiency scales as $\eta \sim 10^{-3} \sigma^{-1}$ \citep{sironi21}\footnote{\citet{plotnikov19} suggested $\eta = 7 \times 10^{-4} \sigma^{-2}$ from an earlier 1D simulation. The results of 3D simulations by \citet{sironi21} are generally consistent with the 1D results. The difference in $\sigma$-dependence is different frames used. The $\sigma^{-2}$-dependence applies to the shock frame, while the $\sigma^{-1}$-dependence applies to the downtream frame, which is more relevant in estimating the maser efficiency (L. Sironi, 2021, private communication).}. One interesting question regarding these models is to address why at the FRB emission radius, $\sigma \sim 1$ is by chance achieved. Pulsar wind theories suggest that a pulsar wind with an initial magnetization $\sigma_0 \gg 1$ tends to reach $\sigma \sim \sigma_0^{2/3}$ at the sonic point where the wind speed is as high as the fast sonic wave speed so that the magnetic ``piston'' losses pressure to accelerate the outflow \citep[e.g.][]{li92}. Beyond this radius, magnetic acceleration is rather slow (unless there exists an external pressure confinement to maintain a significant magnetic pressure gradient) so that it is difficult to reduce $\sigma$ further down to unity. In general, this mechanism predicts a relatively low radio emission efficiency $\eta \ll 10^{-3}$, suggesting that FRBs should be accompanied by bright high-energy emission in X-rays \citep{margalit20} or optical \citep{beloborodov20}. It also suggests that the total energy budget required in the shock models is generally higher than that required in the magnetospheric models. It turned out that the Galactic FRB 200428 has an X-ray-to-radio luminosity ratio of the order of $\sim 10^{4}$, which can be accounted for from both models \citep{lu20,margalit20}. However, active repeaters rFRB 20121102A and rFRB 20201124A already have a very high total energy budget in the radio band during their active bursting periods. This demands that the radio efficiency cannot be much smaller than $10^{-3}$ in order to satisfy the total energy budget of magnetars \citep{lid21,xuh22,zhangyk22}. 
    \item In the downstream comoving frame the characteristic frequency for maser emission is $\omega'_B$ \citep{sironi21}, which is defined by the strength of upstream magnetic field $B$ at the emission radius $R_{\rm FRB}$, bulk Lorentz factor $\Gamma$, and the central engine parameters (e.g. surface magnetic field $B_s$, spin period $P$ of the magnetar). Demanding the observed frequency $\Gamma \omega'_B$ to be in the $\sim$ GHz regime and combining with other constraints (e.g. $R_{\rm FRB}$ as the deceleration radius defined by FRB energy, ambient density, and $\Gamma$, and the duration of the FRB defined by $w=R_{\rm FRB}/c\Gamma^2$), one can place some interesting constraints on model parameters. This has been done for the Galactic FRB 20200428 \citep{lu20,margalit20,yu21}. The general conclusion is that the observations can be reproduced, even though some special physical conditions have to be satisfied. To overcome such a fine-tuning issue, \citet{metzger19} argued that the peak of the FRB spectrum sweeps across a wide frequency range as it decelerates and the observer only sees them when the peak is in the radio band. This idea is also used to interpret the spectral down-drifting observed in repeating FRBs \cite{metzger22}. On the other hand, observationally there is no systematic peak-frequency time evolution among adjacent bursts or a duration-spectral width correlation to suppose this speculation \citep{zhou22}. 
\end{itemize}

\subsection{Summary}

The discussion in this section can be summarized as follows:
\begin{itemize}
    \item There are many coherent radio emission models proposed in the literature to interpret FRB emission, which generally fall into two categories: magnetospheric (closer-in, pulsar-like) models and relativistic shock (far-out, GRB-like) models. Some models (e.g. the bunched curvature radiation model and magnetized synchrotron maser model) have been extensively studied and demonstrated ability of interpreting certain FRB data. Some other models (e.g. magnetospheric maser models and two other versions of shock maser models) suffer from some significant criticisms so may not be strong candidates to power FRBs. Some  other models (e.g. bunched ICS and FEL mechanisms, reconnection in current sheet, non-steady pair production induced radiation) deserve closer investigations and confrontation with the data. Current observations cannot pin down exactly which mechanism is at play to power FRBs.
    \item Purely from the theoretical perspective, none of the proposed models are free of issues or difficulties. Within the magnetospheric models, the bunching coherent curvature radiation or ICS models demand an $E_\parallel$ to continuously inject energy into the bunches to satisfy the energy budget constraint. The origin of the $E_\parallel$ is not well identified. Various particle - low frequency wave interaction models beg the existence of these low frequency waves, whose existence can be justified for a neutron star model (e.g. through star quakes or glitches) but may not be justified all types of central engine models (e.g. black hole engines). Magnetospheric models in general need to address the opacity of high-luminosity bursts, which demands a relativistically moving plasma in the magnetopshere \citep{qu22c}. The magnetized synchrotron maser model in relativistic shocks need to address the origin of the demanding requirements including very ordered, cold upstream plasma, the Goldilocks $\sigma$ value, as well as special model parameters required from the data. 
    \item From the observational perspective, data can be used to differentiate among some models. In particular, the following four criteria \citep{zhang20b} would be helpful: 
    \begin{enumerate}
        \item Polarization angle swings: even though a flat PA curve can be accounted for by both shock and magnetosphericc models, a significant PA swing is consistent with magnetospheric models but poses a great difficulty to the shock models; 
        \item Radio efficiency: a high/low radio emission efficiency may offer support to the magnetospheric/shock models. The constraints on efficiency may be based on the energy of the high-energy counterpart of the FRB (e.g. the X-ray burst associated with FRB 200428) or theoretically derived total energy budget; 
        \item Beaming angle: Magnetospheric models predict a narrower emission beam than the shock model. Therefore, the identification of narrow beaming for certain FRBs may offer a support to the magnetospheric models. Evidence in support of narrow beaming may include the lack of FRBs associated with most X-ray bursts from SGR 1935+2154 \citep{lin20}, possible detection of off-beam FRBs, or ``slow radio bursts'' \citep{zhang21,chenzhang23}, and frequency-dependent periodic window of rFRB 20180916B \citep{lidz21}. 
        \item Rapid variability: Since $\delta t \sim R/c\Gamma^2$ (Eq.(\ref{eq:t_var})), a very small $\delta t$ would point toward a small $R$ (if $\Gamma$ is constrained) \citep{beniamini20}. The 60-ns variability \cite{nimmo21} observed in rFRB 20200120E from a globular cluster in M81 disfavors a shock origin of the FRB \citep{lu22}.
    \end{enumerate}    
\end{itemize}

Looking ahead, upcoming abundant FRB data may shed light on the radiation mechanism of FRBs. It is optimistic that data may provide clues on the location of the FRB emission (magnetospheres vs. shocks), but the identification of the very coherent mechanism(s) may not be easy, as the experience in understanding pulsar radio emission mechanism speaks itself. The current available data seem to support the magnetospheric origin of at least some FRBs. It is possible that both magnetospheric and shock models operate (e.g. the latter works for the most energetic bursts while the former works for less energetic ones), but the current data of burst properties have not demanded an  explanation involving dichotomy yet. 

\section{Source models}\label{sec:source}

In this section, we discuss various source models for FRBs. Since repeaters seem to be common and since there is no proof that intrinsically one-off FRBs (those associated with cataclysmic events) exist (but see \citet{moroianu22}), all the subsections except the last one in this section discuss sources for repeating FRBs. Different from the previous theory review \citep{platts19} that lists models in a stamp-collecting manner (see also the FRB theory Wiki page\footnote{https://frbtheorycat.org}), we attempt to provide critical comments on these models. \S \ref{sec:magnetars}-\ref{sec:interactingNSs} discuss the neutron star models, which are the most likely models. This is followed by other non-neutron-star astrophysical models (\S \ref{sec:nonNS}) and more exotic models (\S \ref{sec:exotic}) for repeating FRBs. Finally, cataclysmic models are discussed in \S\ref{sec:cataclysmic}. 

\subsection{Magnetars}\label{sec:magnetars}

The leading source model for FRBs is the magnetar model. Magnetars (\citet{duncan92,thompson95,thompson96}, see also \citet{katz82}) may be generally defined as neutron stars with dipolar surface magnetic fields exceeding $\sim 10^{14}$ G, but there is no clear separation line between magnetars and high-$B$ pulsars. Observationally they appear as soft $\gamma$-ray repeaters (SGRs) and anomalous X-ray pulsars (AXPs), both having quiescent X-ray luminosities exceeding their spin-down luminosities, with the former displaying repeated soft-$\gamma$-ray/hard-X-ray bursts \cite{kaspi17}. Later observations suggest that some neutron stars emit SGR-like bursts but have surface dipolar magnetic fields below $10^{14}$ G \citep{rea10}. These sources may have strong multi-polar magnetic fields near the surface and are also included in the magnetar population. There are 30 magnetars currently known\footnote{http://www.physics.mcgill.ca/$\sim$pulsar/magnetar/main.html}, including 16 SGRs and 14 AXPs. In another research front in transient astrophysics, a type of millisecond magnetar has been hypothesized \citep{usov92}, which has been widely discussed as the central engine of gamma-ray bursts (GRBs) and superluminous supernovae (SLSNe) \citep[e.g.][]{zhang01,metzger11,woosley10,kasen10}.

The connection between FRBs and magnetars has been discussed by many authors within different contexts. The earliest suggestion was by \citet{popov10} who interpreted the Lorimer burst \citep{lorimer07} as SGR hyperflares. In the paper by \citet{thornton13} who reported four additional FRBs, the authors discussed several possibilities and pointed out that the inferred FRB rate is consistent with the rate of SGR flares. The SGR-like model was later further discussed by \citet{kulkarni14} and \citet{katz16}. Interactions between magnetar flares and an ambient wind were introduced by \citet{lyubarsky14} as a mechanism to generate FRBs. Prompted by the discovery of the active repeater rFRB 20121102A that resides in a dwarf star forming galaxy similar to the hosts of long GRBs and SLSNe, \citet{metzger17} suggested that millisecond magnetars could be the engine of active repeating FRBs. This model was further developed by \citet{beloborodov17,margalit18,metzger19,beloborodov20} within the framework of the synchrotron maser model. \citet{kumar17,yangzhang18}, on the other hand, consider the requirement of producing FRB emission from neutron star magnetospheres, and drew the conclusion that the isolated neutron stars that can power FRBs are likely magnetars. \citet{wadiasingh19} proposed that magnetars with a low-twist of magnetic fields would initially not have enough pairs to screen $E_\parallel$ so that a pair cascade may be triggered to eventually power an FRB. \citet{wadiasingh20} discussed the line of death of FRB emission from magnetars and suggested that FRB emission is favored in magnetars with long periods. \citet{lyubarsky20} proposed that enhanced magnetic reconnection in the current sheet region of a magnetar could power FRBs. Recent developments in magnetar FRB models include coherent inverse Compton scattering model \cite{zhang22}, the free electron laser model \cite{lyutikov21}, and a direct emission model from a magnetized shock \cite{thompson22}.  Prompted by the discovery of FRB 200428 associated with SGR J1935+2154 \citep{CHIME-SGR,STARE2-SGR}, many studies have been carried out to investigate how the magnetar model can produce FRBs within the magnetophere \citep{lu20,yang20c,yangzhang21} or in relativistic shocks \cite{margalit20b,yu21}. The 0.286-s period of FRB 20192112A offers a strong support to the magnetospheric magnetar models \cite{chime-period,beniamini22b} at least for this special source.

Various versions of the magnetar models have the following common ingredients:
\begin{itemize}
    \item Energy budget: These models  make use of two energy reservoirs: either the rotation energy of the magnetar
    \begin{equation}
        E_r = \frac{1}{2} I \Omega^2 \simeq (2.0 \times 10^{46} \ {\rm erg}) I_{45} P^{-2}
    \end{equation}
    or the magnetic energy of the magnetar\footnote{This estimate includes the dipolar magnetic field only. Magnetars may store a toroidal magnetic field component, which may be stronger than the poloidal component. Thus, this estimate is a lower limit.}
    \begin{equation}
        E_{\rm B} \lesssim \frac{1}{6} B_s^2 R^3 \simeq (1.7 \times 10^{47} \ {\rm erg}) B_{s,15}^2 R_6^3,
    \end{equation}
    where $I$ is the moment of inertia, $P$ is the spin period, $B_s$ is the surface dipolar magnetic field at the pole, and $R$ is the radius of the neutron star. One can immediately see that the rotation energy reservoir becomes smaller than the magnetic energy reservoir when $P > (0.34 \ {\rm s}) I_{45}^{1/2} B_{s,15}^{-1} R_6^{-3/2}$ is satisfied. The total energy of bursts for repeating FRBs should be bound by these limits. For example, rFRB 20121102A emitted a total amount of energy $\sim 3.4\times 10^{41}$ erg in the radio band assuming isotropic emission from 1652 bursts detected in 59.5 hours in a 47-day time span \citep{lid21}. Correcting the observational duty cycle, the total energy emitted would exceed $(6.4\times 10^{45} \ {\rm erg}) F_{b,-1} \eta_{r,-4}^{-1}$ assuming a radio efficiency $\eta_r \sim 10^{-4}$ and a global beaming factor $F_b = 0.1$. This is a substantial fraction of the magnetic energy available from a magnetar. The magnetar models involving a wide beaming angle and a low radiative efficiency (e.g. the synchrotron maser model) are greatly constrained by the data. Magnetospheric models invoke a smaller global beaming factor and a higher $\eta_r$, which are favored \citep{zhang20b}. 
    \item Energy loss/dissipation rate: The average FRB emission luminosity should be bound by the average energy loss/dissipation rate of the magnetar. The energy loss rate due to magnetic dipole spindown is
    \begin{equation}
        \dot E_r  = \frac{B_s^2 R^6 \Omega^4}{6c^3} \simeq (10^{37} \ {\rm erg \ s^{-1}}) B_{s,15}^2 P^{-4} R_6^6,
    \end{equation}
    and the average energy dissipation rate of the magnetic energy may be estimated as 
    \begin{equation}
        \dot E_{\rm B} = \frac{E_{\rm B}}{\tau_d} \simeq (3.2\times 10^{35} \ {\rm erg \ s^{-1}}) E_{\rm B,47} \tau_{d,4}^{-1}
    \end{equation}
    where $\tau_d = (10^4 \ {\rm yr}) \tau_{d,4}$ is the characteristic decay time scale of magnetic fields \citep[e.g.][]{colpi00}. Note that since FRB emission has a very low duty cycle (even for very active repeaters), the luminosities of individual bursts are {\em not} subject to these average energy loss/dissipation rate bounds as long as the average FRB energy emission rate is below this. For example, the active episode of rFRB 20121102A occurred in 2019 emitted of the order of $10^{41} \ {\rm erg}$ energy in radio band in 47 days. Consider that the source has a $\sim 160$-d period \citep{rajwade20} and that the source is not active in some of the projected cycles, one may roughly estimate the average radio-band energy emission rate as $\sim 10^{41} \ {\rm erg} / 2 \ {\rm yr} \sim (1.6 \times 10^{33} \ {\rm erg \ s^{-1}})$. This is smaller than both $\dot E_r$ and $\dot E_{\rm B}$. However, the requirement on $\eta_r$ is tight, i.e. \begin{equation}
        \eta_r > \left\{ \begin{array}{ll}
            (5\times 10^{-3}) E_{\rm B,47}^{-1} \tau_{d,4}, & E_{\rm B}~{\rm budget} \\
            (1.6\times 10^{-4}) B_{s,15}^{-2}P^4 R_6^{-6}, & E_{\rm r}~{\rm budget}. 
        \end{array}\right.
    \end{equation}
    Again models with a low $\eta_r$ are disfavored, unless $E_B$ is much larger or $\tau_d$ much shorter\footnote{\citet{beloborodov17} and \citet{margalit19} argued that magnetars different from the Milky Way known population with a larger core magnetic field and a shorter magnetic decay timescale may exit in other galaxies to power active repeaters.}.
    \item Triggering mechanism: All magnetar FRB models, regardless of how the radio waves are emitted, rely on some common trigger mechanisms. One commonly discussed trigger mechanism is crust cracking at the neutron star surface \citep[e.g.][]{thompson01,beloborodov07,wangwy18,wadiasingh19,dehman20,yangzhang21}, even though some authors \citep{levin12} suggested that crust cracking may not proceed in an abrupt way. Alternative trigger mechanisms include elastic deformation and magnetar oscillations without exceeding the yield strain of the crust \cite{wadiasingh20b} or fast ambipolar diffusion in the core \cite{beloborodov17}. In any case, oscillations of the crust would send Alfv\'en waves to the magnetosphere, triggering various processes that might be related to FRB production (e.g. bunched curvature radiation \citep{kumar17,yangzhang18,cooper21} or inverse Compton scattering \citep{zhang22}, direct electromagnetic wave generation due to non-uniform pair production \citep{philippov20,yangzhang21}, enhanced reconnection in the current sheet region outside the magnetosphere \citep{lyubarsky20}, as well as ejecting magnetic pulses outside the magnetosphere to produce FRBs via magnetized relativistic shocks \citep{yuan20}. Alternative trigger mechanisms include sudden magnetic reconnection events in the magnetosphere \citep{popov10}, sudden discharge of vaccum gaps \citep{katz17}, or sudden triggers from an external event \citep{zhang17,dai20}.
\end{itemize}

The proposed magnetar models also differ in several aspects:
\begin{itemize}
    \item Emission site: From small to large distance from the neutron star surface, there are four versions of magnetar models: 1. models invoking FRB emission region inside the magnetosphere, typically 10s to 100s of the neutron star radii \citep[e.g.][]{kumar17,yangzhang18,wadiasingh19,kumar20a,lu20,yangzhang21,lyutikov21,zhang22}; 2. models invoking the current sheet region outside the light cylinder as the FRB emission site \cite{lyubarsky20,mahlmann22}; 3. models invoking internal shocks due to collisions between magnetic blobs \citep{beloborodov17,beloborodov20}; and 4. models invoking external shocks\footnote{Under certain conditions, the emission radius of the \citet{metzger19} model can be smaller than that of the internal shock model \citep{beloborodov20}.} formed when magnetic shells are decelerated by the magnetar wind \citep{metzger19,margalit20b,thompson22}. 
    \item Radiation mechanism: Many mechanisms discussed in \S\ref{sec:radiation} have been proposed for various versions of the magnetar models. The magnetospheric models invoke bunched curvature radiation \citep{kumar17,yangzhang18,lu20,cooper21}, bunched inverse Compton scattering \citep{zhang22}, free electron laser \citep{lyutikov21}, or direct EM generation due to non-uniform pair production \citep{philippov20,yangzhang21} as radiation mechanisms. In the current sheet region, magnetosonic waves excited by coalescence of magnetic islands \citep{lyubarsky20} or coherent inverse Compton scattering \citep{zhang22} are invoked to produce FRB emission. In both the magnetic internal and external shock regions, the specific version of the synchrotron maser (bunched coherent cyclotron/synchrotron radiation) mechanism is invoked to produce FRB emission \citep{plotnikov19,sironi21}.
\end{itemize}

There are many open questions regarding the magnetar models for FRBs. Besides the question regarding trigger mechanism, emission site, and radiation mechanism discussed above, the following are some other examples of open questions;
\begin{itemize}
    \item {\em Can magnetars produce all FRBs in the universe?} Shortly after the discovery of FRB 200428 in association with the Galactic magnetar SGR 1935+2154, the enthusiasm and confidence of interpreting all FRBs in the universe as being generated by magnetars have grown tremendously \citep[e.g.][]{lu20,margalit20}. The assumption is that all FRBs are intrinsic repeaters. Regular magnetars such as those observed in Milky Way may be responsible for the apparently non-repeating FRBs and those repeaters with a low repetition rate, while young magnetars may be responsible for active repeaters observed in cosmological distances. The fact that there is no active repeating FRB sources from the Milky Way is interpreted as the lack of very young magnetars in the Galaxy (or if there is any, the FRB emission beam does not point towards Earth). The repeaters in association with the old population such as globular clusters \citep{kirsten22,nimmo22} were interpreted as a new population of young magnetars born from accretion induced collapse or mergers of binary neutron stars \citep{margalit19,wang20}, binary white dwarfs \citep{kremer21}, or NS-WD binaries \citep{zhongdai20}. However, growing evidence suggests that this most conservative, ``magnetars make them all'' suggestion for the FRB origin may not be adequate to account for all the FRB observational data. For example, the rFRB 20200120E-like sources may be very common. However, none of the known Galactic magnetars are associated with globular clusters. The `magnetars make them all'' scenario likely runs into the event rate issue. The general delay with respect to star formation rate required for the inferred FRB redshift distribution \cite{zhangzhang22,qiang22,hashimoto22} also raises a flag to this simple scenario. 
    \item {\em Does FRB emission favor young or old magnetars?} Active repeaters are widely interpreted as being produced by new-born magnetars. The arguments in support of this idea include the association of a dwarf star-forming galaxy \citep{tendulkar17} with rFRB 20121102A, the associations of a persistent radio source with rFRB 20121102A \citep{chatterjee17,marcote17} and rFRB 20190520B \citep{niu22}, as well as a larger energy reservoir (both magnetic and spin energies) and probably a faster decaying rate \citep{metzger17,beloborodov17} in young magnetars. The issues of having very young magnetars as prolific FRB emitters include significant free-free absorption and induced Compton scattering in a dense environment associated with supernova remnants or pulsar wind nebulae around the new-born magnetars. On the other hand, charge-starvation seems to be favorable for magnetars to make FRBs within their magnetopsheres. Older magnetars tend to more easily reach charge starvation because of the reduced pair production due to slow spin and low twist \citep[e.g.][]{wadiasingh19,wadiasingh20,beniamini20b}. 
    \item {\em What is the mechanism for $E_\parallel$ in magnetar magnetospheres?} A charge starved region in a magnetar magnetosphere is where $E_\parallel$ is developed and particles accelerated. Magnetospheric FRB models require the existence of an $E_\parallel$ to continuously supply energy to otherwise rapidly cooling particle bunches \citep{kumar17,zhang22}. The exact mechanism to generate $E_\parallel$ in the FRB emission region is not identified. One possibility is that $E_\parallel$ can be developed as Alfv\'en waves propagate to the outer magnetosphere where $e^\pm$ density is not sufficient to supply the current required to sustain the Alfv\'en waves \citep{kumar20a}. Another mechanism is the traditional pulsar mechanism that opens various types of gaps in the pulsar magnetosphere \citep{ruderman75,arons79,cheng86,zhang97,harding98,muslimov04}. The energetics of these gaps, on the other hand, are limited by the spindown power of the magnetars, which is not large enough to power FRBs for slow rotators. Sudden crust cracking may excite global readjustment of the magnetospheric configuration, leading to temporarily enhanced gaps with large $E_\parallel$, which could be another mechanism to power FRBs. 
    \item {\em What is the role of Alfv\'en waves?} Various FRB models invoke Alfv\'en waves as an important ingredient. The role of Alfv\'en waves varies in different models. \citet{kumar20a,lu20} invoked Alfv\'en waves as the agent to produce $E_\parallel$ at a large enough radius to accelerate bunched particles to power FRB emission. \citet{chen22b} questioned this possibility by a numerical simulation that shows that particles are advected without forming a significant charge starved region.  \citet{kumar22b} performed simulations in a longer duration and found that an $E_\parallel$ of the order of a few percent of the Alfv\'en wave amplitude can be indeed generated. \citet{yuan20} showed that low-amplitude AlfvÃ©n waves from a magnetar quake propagate to the outer magnetosphere and convert to ``plasmoids'' (closed magnetic loops). The plasmoids are accelerated from the star, driving blast waves into the magnetar wind. \citet{lyubarsky20} invokes Alfv\'en waves to significantly compress the current sheet region outside the light cylinder to enhance relativistic magnetic reconnection, which may facilitate the generation of FRBs in the reconnection region  \citep{lyubarsky20,mahlmann22,zhang22}.
\end{itemize}

\subsection{Other isolated neutron star models}\label{sec:NSs}

Besides magnetars, other types of isolated neutron stars have been discussed as the source of FRBs.

\begin{itemize}
    \item {\em Giant pulses from young pulsars.} Giant radio pulses have been observed from some young pulsars, such as the Crab pulsar. The brightest giant pulse (GP) observed so far has a peak amplitude $S_{\rm \nu.max} = 2.2$ MJy at 1 GHz and a pulse width $<0.4$ ns, corresponding to a brightness temperature $T_b \gtrsim 10^{41.3}$ K (thanks to its very short duration) \citep{hankins07}. An immediate inference is that similar GPs from nearby galaxies would be detected as FRB-like events by Earth observers  \citep{connor16,cordes16}. Unlike magnetar-powered FRBs that possibly consume magnetic energy of the parent star, these GP-like FRBs likely consume spin energy of the parent star. Placing GP-emitting pulsars to larger distances suggests that they could be detected up to $\sim 100$ Mpc, but not to larger cosmological distances as suggested by the DM excess of most FRBs. So FRBs in these models are also called ``ERBs'' -- extragalactic radio bursts \citep{cordes16}. Now it has been confirmed that most FRBs originate from cosmological distances greater than 100 Mpc \citep[e.g.][]{tendulkar17,macquart20,bhandari21}. The simplest version of this model is incapable of interpreting the data unless much brighter GPs are invoked.
    \item {\em Pulsar lightening}. \citet{katz17} argued that the FRB phenomonology is similar to atmosphere lightening and conjectured that FRBs are produced when vacuum gaps in pulsar magnetospheres break down to suddenly drive currents in the magnetosphere. The FRB energetics in this model is also limited by spindown power of the underlying pulsar.
\end{itemize}
 
\subsection{Interacting neutron star models}\label{sec:interactingNSs}

A number of FRB models invoke neutron stars interacting with an external agent. These interacting neutron star models come in different flavors depending on the energy budget that is invoked to explain FRB emission. In the extreme versions of the interaction models, the ultimate energy comes externally from the gravitational energy of a falling object, or the kinetic energy of an external moving fluid. In milder versions of the models, the ultimate energy still comes from the neutron star itself (e.g. the spin or magnetic energy), but the external agent may play a role of triggering FRB emission or shaping the detectability of FRBs. We now discuss several of such models in the literature.
\begin{itemize}
    \item {\em Comet/asteroid interaction models:} One suggested way of making FRBs is through interactions between comets or asteroids and a neutron star. The direct impact model \citep{geng15,dai16,bagchi17,dai20,dai20b,smallwood19} invokes a gravitational energy budget
    \begin{eqnarray}
        E_{g} & = & \frac{G M m}{R} \simeq  (1.9\times 10^{40} \ {\rm erg}) \nonumber \\
        & \times & \left(\frac{M}{1.4 M_\odot}\right) \left(\frac{m}{10^{20} \ {\rm g}}\right) R_6^{-1}
    \end{eqnarray}
    to power FRBs, where $M$ and $R$ are the mass and radius of the neutron star, and $m$ is the mass of the small body. One can see that the Galactic FRB 20200428 from SGR 1935+2154, which has an radio luminosity/energy smaller by orders of magnitude than cosmological FRBs, already requires a small body mass\footnote{Beaming correction is usually not considered in these models because the emission solid angle is expected to be large.} $m\sim 10^{20}$ g. Scaling the required mass up based on the luminosities to cosmological FRBs, one finds that the demanded comet/asteroid mass is immense. Take the 2019 active episode of rFRB 20121102A as an example \citep{lid21}. The total energy emitted in radio in 1652 bursts detected in 59.5 hours during 47 days is $\sim 3.4\times 10^{41}$ erg. Counting on the missed FRBs outside of the FAST observing window, the total radio energy would be $\sim 6.4\times 10^{42}$ erg.  Recall that the Galactic FRB 20200428 with a radio energy of a few $10^{35}$ erg \citep{STARE2-SGR,CHIME-SGR}. The total small body mass to power the rFRB 20121102A for that emission episode is already a few times of $10^{27}$ g, which is of the order of the Earth mass. So the comet/asteroid collision model is a ``very expensive'' mechanism which consumes a lot of mass. The total mass in the Kuiper belt of our own solar system is about 2\% of Earth mass \citep{pitjeva18}. Furthermore, a significant fraction of comets/asteroids are dynamically ejected when a neutron star enters the comet/asteroid belt \citep{smallwood19}. The huge mass budget and the very short waiting times (as short as several miliseconds) between some bursts \citep{lid21,xuh22,zhangyk22} essentially rule out the direct impact model at least for rFRB 20121102A.
    
    Another version of the comet/asteroid interaction model does not invoke direct impact. \citet{mottez14} suggested that small bodies orbiting a pulsar at low orbits could periodically interact with the pulsar winds to drive two stationary Alfv\'enic structures called Alfv\'en wings. The destablisation of the plasma by the Alfv\'en wing's current may excite coherent radiation and make FRBs. \citet{mottez14} estimated that a multi-Jy level radio burst may be generated if the source is at a distance of $D=1$ Mpc and if the small body is $r=1$ AU from the pulsar (flux depends on $r^{-2}$). Interpreting cosmological FRBs within this model require narrow beaming and low orbits. In general, this model is energetically much more efficient than the direct impact models, since it does not require the small body being destroyed. Nonetheless, since the ultimate emission power comes from the spin energy of the pulsar, the same energy requirements for single neutron stars also applies to this model. Observationally, it may be difficult to distinguish this model from some isolated neutron star models.
    
    \item {\em Cosmic comb model:} \citet{zhang17} suggested that a sudden interaction between a fluid flow (also called an astronomical stream) from a nearby source of an otherwise isolated neutron star can make coehrent radio emission. An FRB is observed by an Earth observer when the ``combed'' magnetosphere sweeps across the line of sight. The sources of the stream could be energetic events such as supernovae, gamma-ray bursts, tidal disruption events, or more moderate events such as AGN flares or even erratic outflows from a companion. As a result, an FRB may or may not be associated with bright counterparts depending on the source of astronomical stream. Note that the specific version of this model invoking interaction between a supernova and a neutron star was proposed earlier \citep{egorov09} which was overlooked by \citet{zhang17}. rFRB 20121102A was interpreted as a repeatedly combed regular neutron star near a massive black hole \citep{zhang18b}. The ultimate energy power of this model comes from the kinetic energy of the astronomical stream. The kinetic luminosity received by the neutron star may be estimated as 
    \begin{equation}
        L_{\rm kin} \sim \frac{L}{4\pi r^2} \pi \left(\frac{c P}{2\pi}\right)^2,
    \end{equation}
    where $L$ is the luminosity of the source of the astronomical stream, $r$ the distance of the neutron star from that source, and $P$ is the spin period of the neutron star ($cP/2\pi$ is the light cylinder radius). The condition for a cosmic comb event to happen is that the ram pressure exceeds the magnetic pressure at the light cylinder, i.e.
    \begin{equation}
        \rho v^2 > \frac{B_s^2}{8\pi} \left(\frac{2\pi R}{cP} \right)^6,
    \end{equation}
    where $\rho$ and $v$ are the density and velocity of the stream at the interaction radius, $B_s$ and $R$ are the surface magnetic field and radius of the neutron star, respectively. 
    
    The 16-day period of rFRB 20180916B \cite{chime-periodic} may be interpreted as the orbital period of a binary system containing an FRB pulsar (or magnetar) and a massive star or neutron star companion \citep{ioka20,lyutikov20}. For a total mass $M_{\rm tot}=10 M_\odot$ in the binary, the separation between the two stars is $\sim 4\times 10^{12}$ cm. The kinetic luminosity received by the FRB pulsar is $L_k = (3.6\times 10^{32} \ {\rm erg \ s^{-1}}) L_{39} P^2$, where $L=(10^{39} \ {\rm erg \ s^{-1}}) L_{39}$ is the companion's kinetic luminosity normalized to its Eddington luminosity. Such a luminosity is too small to interpret the repeated FRBs from rFRB 20180916B unless an extremely narrow beam or a much greater luminosity than the Eddington value are assumed. As a result, the original version of the cosmic comb model is not adequate to interpret the observations of at least rFRB 20180916B. 
    
    \item {\em Binary comb models:} \citet{ioka20} proposed the binary comb model for periodically repeating FRBs. The role of the companion is no longer to directly provide the power of FRBs. Rather, the interaction between the companion wind and the FRB pulsar magnetosphere defines a funnel from which FRBs, intrinsically produced by the FRB pulsar itself, can escape and be detected from Earth. The similar scenario was independently proposed by \citet{lyutikov20}, who also displayed the companion-wind-defined funnels through numerical simulations. \citet{wada21} expanded on the funnel mode of \citet{ioka20} and identified two more modes ($\tau$-crossing mode and inverse funnel mode) to define the FRB escaping window for periodic FRBs. Even though within the binary comb model the companion wind only plays a passive role of defining the detectability of the bursts, it was nonetheless speculated \citep{ioka20} that the so-called aurora particles entering the magnetopshere of the FRB pulsar may play an active role in driving coherent radio emission and powering FRBs.

    \item {\em Magnetospheric interaction models:} It is possible that direct interaction between the magnetospheres of two neutron stars may make FRBs. Possible FRB-like electromagnetic field signals have been discussed within the context of binary neutron star mergers shortly before the merger \citep[e.g.][]{hansen01,lai12,piro12,wang16,wada20}. The commonly discussed energy release mechanism is the unipolar effect as a neutron star with a weak magnetic field travels in the magnetosphere of another neutron star with a stronger magnetic field. \citet{gourgouliatos19} studied several configurations between the two magnetospheres of the two neutron stars in the pre-merger phase and discussed possible energy dissipation. They mentioned the possibility of connecting these interactions with repeating FRBs. \citet{zhang20} showed that for typical parameters similar to the double pulsar system \citep{kramer08}, strong magnetosphere interactions between the two inspiring neutron stars occur decades to centuries before the merger. He argued that such systems could be ideal candidates for producing repeating FRBs through magnetic reconnection with the expense of the magnetic energy (and ultimately the spin energy) of the two neutron stars. Invoking the beaming effect (which is expected from magnetospheric interaction induced events), he argued that the energy budget in the system is more than enough to power rFRB 20121102A-like active repeaters. The model has several predictions \citep{zhang20}: 1. The activity level elevates with time as the two neutron stars get closer with time; 2. Active repeaters could be mHz gravitational wave sources detectable by future space-born GW observatories such as LISA \cite{LISA}, Taiji \cite{Taiji} and TianQin \cite{Tianqin}; 3. There could be quasi-periodic signals at the orbital period, which is typically 100s of seconds. The environment of the globular cluster rFRB 20200120E \citep{kirsten22,nimmo22} is consistent with that of a binary neutron star merger, even though models invoking BNSs require that the source lasts for a much longer duration, e.g. $10^6$ yr \citep{kremer21,lu22}.
    \item {\em White-dwarf-fed neutron star model:} \citet{gu16,gu20} delineated a scenario that invokes a compact NS-WD binary in which the WD already fills its Roche lobe so that matter from the WD can be channelled towards the NS. The authors speculated that magnetic reconnection may be triggered by episodic accretion of WD materials approaching the NS surface. Curvature radiation is then envisaged to happen as relativistic particles stream out along the magnetic field lines. Such a scenario is speculative since known neutron star accreting systems (e.g. X-ray pulsars) tend to produce thermal emission in the accretion column. In general, it is difficult to produce delicate magnetospheric coherent radio emission in an accreting system. 
\end{itemize}

\subsection{Non-neutron-star astrophysical models}\label{sec:nonNS}

\begin{itemize}
    \item {\em Stellar-mass black hole sources.} Besides neutron stars, the only other kind of objects whose sizes are small enough to accommodate millisecond durations of FRBs are stellar mass black holes. The difference between a black hole engine and a neutron star (e.g. magnetar) engine is that the former may or may not have a clean magnetosphere as the latter does because of the dirty accretion environment, so that the magnetospheric radiation mechanisms associated with neutron star models may not be straightforwardly applied. Nonetheless, \citet{katz17b,katz22c} speculated that the accretion disk of a black hole may collimate a ``funnel'' from which jet-like emission may be released. In order to account for the short duration of FRBs, he further speculated that the jet may be rapidly wandering and that the duration of the FRB corresponds to the duration when the very narrow FRB jet sweeps across the line of sight. \citet{katz20} further argued that the lack of periodicity in repeating FRBs at the typical neutron star spin period favors the black hole origin of FRBs (but see Section \ref{sec:periodicity2} for counter arguments). These papers did not specify the FRB emission site and the coherent radiation mechanism. \citet{lilb18} suggested that the accretion system involving a black hole and a white dwarf with Roche lobe overflow may launch magnetic blobs and produce FRBs via the synchrotron maser mechanism. \citet{sridhar21} proposed a detailed model for periodic FRBs invoking an accreting black hole binary similar to the BH ultraluminous X-ray (ULX) sources. The FRB mechanism is hypothesized as the synchrotron maser mechanism in relativistic, magnetized shocks, similar to GRB-like models for magnetars. Such a model makes some specific predictions (e.g. some known ULX sources will produce FRBs someday). Since these models rely on the synchrotron maser models as the radiation mechanism, the general theoretical and observational caveats discussed in \S\ref{sec:shocks} also apply to these models.
    \item {\em Supermassive black hole (SMBH) sources:} The supermassive black holes in the center of galaxies or AGNs have a characteristic timescale, i.e. $r_s/c \sim 10^3 {\rm s} (M/10^8 M_\odot)$ (where $r_s=2 G M/c^2$ is the Schwarzschild radius of a black hole with mass $M$), much longer than milliseconds. So it is not straightforward to invoke a SMBH to power FRBs unless emission is confined in a region much smaller than event horizon. Nonetheless, some suggestions have indirectly made use of SMBHs to power FRBs. \citet{romero16,vieyro17} proposed that FRBs may be produced through interactions between a relativistic electron beam from an AGN jet and a turbulent plasma. The emitters (called cavitons) have a much smaller scale than the SMBH so that they can make millisecond-duration bursts. In this model, the coherent radio emission is produced through Langmuir-wave-driven intense electrostatic soliton emission, which may be broadly defined as one kind of ``bunching'' mechanisms discussed in Section \ref{sec:radiation}. Within the framework of the cosmic comb model \citep{zhang17}, \citet{zhang18b} invoked the episodic wind from an SMBH interacting with a neutron star to interpret the large RM and persistent radio emission associated with rFRB 20121102A. \citet{dasgupta18} applied the similar scenario to make episodic AGN winds to interact with a Kerr stellar-mass black hole to launch episodic jets that power FRBs. \citet{wada21} studied rFRB 20121102A within the framework of the binary comb model and constrained the allowed parameters of the companion of the FRB pulsar. They found that a SMBH could be a plausible companion of this FRB source. The host galaxy data of localized FRBs already rule out AGNs or galactic centers as the sources of the majority of FRBs. As a result, models attempting to interpret the bulk FRB population invoking AGNs or galactic centers are ruled out. Nonetheless, invoking AGNs or galactic centers for individual FRB sources within the scope of broader models (e.g. binary combs) remains possible.
    \item {\em Stellar flares:} For a short period of time, flares from Galactic stars were considered as the sources of at least some FRBs \citep{loeb14}. The suggestion faced the issue of the free-free absorption constraint \citep{luan14} and the duration limit ($R/c \gg 1$ ms for stars). The localization of rFRB 20121102A in a distance galaxy \citep{chatterjee17,marcote17,tendulkar17} quickly put away this model and any model invoking origins inside the Galaxy.
\end{itemize}

\subsection{Exotic repeater models}\label{sec:exotic}

Many FRB models have invoked hypothetical  objects or phenomena to interpret FRBs, which we summarize in this subsection. The confirmation of the existence of any of these objects/phenomena would have profound implications for astrophysics and physics in general. However, since some aforementioned models (e.g. magnetars, isolated or interacting neutron stars) have provided reasonable interpretations to most of the FRB phenomenology, we regard these models exotic. Instead of critically commenting on the validity of each model, we simply list them below. The only comment on all these models is the famous quote by Carl Sagan: ``extraordinary claims require extraordinary evidence'' . 

\begin{itemize}
    \item {\em Strange quark stars:} Strange quark stars are hypothetical compact stars made up of three flavor (u, d, s) quarks \citep{alcock86}. \citet{ouyed20,ouyed21} suggested that conversion from neutron stars to quark stars would make quark novae that can account for an array of astrophysical transients such as GRBs and FRBs. In particular, FRBs are produced when the quark nova ejecta chunks collide with the ambient medium. Strange stars may have a thin normal-matter crust \citep{alcock86}. Episodic accretion induced collapses of the crust have been also suggested to power repeating FRBs \citep{zhangy18b,geng21}. 
    \item {\em Primordial black holes:} Primordial black holes (PBHs) are hypothetical black holes formed shortly after the Big Bang, which can carry a wide mass distribution not subject to stellar evolution including masses much smaller than $M_\odot$. \citet{abramowicz18} proposed a PBH-NS interaction model for repeating FRBs. After a PBH enters the center of a NS, the NS will be accreted and eventually swallowed by the PBH. During the process, the NS magnetosphere is continuously reconfigured, making repeating FRBs.
    \item {\em Superconducting cosmic strings:} Cosmic strings are hypothetical string-like topological structures in the universe which are the macroscopic manifestation of string solutions in field theories. Like elastic, current-carrying wires, cosmic strings are envisaged to carry energy, to be dynamically evolving and super-conducting.  \citet{vachaspati08} suggested that oscillations at the ``cusps'' (points on an idealized string that reach speed of light for a brief instant) would radiate FRB-like emission. Other suggestions include collisions of string structures (cusps and kinks, \citet{cai12a,cai12b,ye17}),  interaction of a current carrying loop in the local magnetic field \citep{yu14}, and  decay of string cusps \citep{brandenberger17}. 
    \item {\em Axion stars, axion clumps, and axion quark nuggets:} The axion is a hypothetical elementary particle and a promising candidate for cold dark matter in the universe. If axions exist, it is hypothesized that they can form gravitationally bound axion clumps or axion stars (typically with a mass of $\sim 10^{-12} M_\odot$). It has been suggested that FRBs could be generated when axion stars collide with neutron stars or black hole accretion disks \citep{iwazaki15,iwazaki21}. Other ideas include induced collapse of axion clumps (``miniclusters'') by the strong magnetic field of a compact star \citep{tkachev15}, a black hole laser powered by axion superradient instabilities or ``BLAST'' \citep{rosa18}), and even magnetic reconnection in a neutron star magnetosphere triggered by the falling of ``axion quark nuggets \citep{vanwaerbeke19}. 
    \item {\em Macroscopic dipole collisions:} \citet{thompson17a,thompson17b} conjectured macroscopic, superconducting magnetic dipoles might have formed around the time of cosmic electroweak symmetry-breaking. The collisions of these ``large superconducting dipoles'' (LSDs) may make tiny explosions to power FRBs. The collisions more preferably happen near massive black holes where LSDs have higher densities. Both repeaters and non-repeaters may be produced with this mechanism. 
    \item {\em Dicke's superradiance:} In quantum optics, Dicke's superradiance (DSR, \citet{dicke54}) is a phenomenon that occurs when a group of excited (population inverted) atoms or molecules interact with a triggering event (e.g. a light) to radiate coherently. The phenomenon was well tested in the laborotory \citep{skribanowitz73}. \citet{houde18} hypothesized that DSR can occur in the Galactic scale involving $\sim 10^{30}-10^{32}$ entangled molecules over distances spanning 100-1000 AU, which can power FRBs. \citet{houde19} further suggested that a pulsar located from $\sim 100$ pc away from the entangled molecules could serve as the trigger for DSR. 
    \item {\em Alien technology:} \citet{lingam17} speculated that FRBs may be artificial beam-powered light sails of extragalactic aliens.  \citet{zhang20c} suggested that FRBs we observe are of astronomical origins, but communicative extraterrestrial intelligences (CETIs) in the Milky Way galaxy may choose to emit FRB-like signals if they want to broadcast their existence. The non-detection of any artificial FRB-like signals from the galaxy in a decade with all-sky radio monitors may place a meaningful upper limit on the average emission rate of such signals by CETIs in the Galaxy. To produce a 1 ms-Jy signal on Earth, the required emission power for aliens at a typical distance of 10 kpc is $\sim (10^{22} \ {\rm W}) f_{b,-3} (d / 10 \ {\rm kpc})^2$, where $f_b \sim 10^{-3}$ is the beaming angle. 
\end{itemize}

\subsection{Cataclysmic progenitor models}\label{sec:cataclysmic}

Even though the majority of detected FRBs are not observed to repeat, the cataclysmic progenitor models for FRBs are not taken as seriously as repeater models. The main arguments against these ideas to become the main stream FRB models include: (1) Since the energy budget of FRBs is much smaller than the energy available in cataclysmic events, and since repeaters have been detected, it is essentially impossible to prove that the apparently one-off FRBs will never repeat; (2) The FRB event rate density is much greater than the rate densities of all known cataclysmic events \citep{ravi19b}. The most common cataclysmic events in the universe is core-collapse supernovae with $R_{\rm CC} \sim 10^5 \ {\rm Gpc^{-3} \ yr^{-1}}$, whereas the FRB event rate density above $10^{37} \ {\rm erg \ s^{-1}}$ is a few $R_{\rm FRB} (L>10^{37} \ {\rm erg \ s^{-1}}) \sim (10^7-10^8) \ {\rm Gpc^{-3} \ yr^{-1}}$ \citet{lu20}. If any cataclysmic channels are relevant, they must only account for a small fraction of FRBs, maybe above a particular luminosity (where the rate density becomes smaller, e.g. $R_{\rm FRB} (L>10^{42} \ {\rm erg \ s^{-1}}) \sim (3.5\times 10^4) \ {\rm Gpc^{-3} \ yr^{-1}}$, \cite{luo20b}), or spread out in a wider luminosity range but with negligible contribution to the observed event rate density.

Nonetheless, some cataclysmic models are quite attractive, since these events are destined to produce brief electromagnetic radiation signals, whether or not they are FRBs. Two leading models include the ``blitzar'' scenario invoking implosion of supramassive neutron stars and various compact binary coalescence (CBC) models invoking mergers of neutron stars (NSs) and black holes (BHs). We highlight these models below.

\begin{itemize}
    \item {\em Blitzars:} Supramassive neutron stars (SMNSs) are spin-supported massive neutron stars whose non-spinning mass already exceeds the maximum NS mass allowed by the NS equation of state. The existence of an SMNS is therefore temporary. The NS will inevitably collapse to a BH as it is spun down via magnetic dipolar radiation or even gravitational wave radiation. As the bulk of the NS enters the horizon during its collapse to a BH, the closed magnetic field lines would detach from the star and get ejected (the open field lines penetrating the hole may stay longer). \citet{falcke14} suggested that such a magnetosphere ejection process would power an FRB and termed the phenomena ``blitzars''. The process was numerically simulated \citep{most18}, and a millisecond-duration episode of significant Poynting flux injection was indeed observed, suggesting the robustness of the mechanism. \citet{falcke14} envisaged that a significant amount of SMNSs may be produced from a few percent of core collapse supernovae. Assuming that these SMNSs do not carry a strong magnetic field, they suggested that collapse happens thousands to million years after the birth of the SMNSs, so that no bright counterpart is expected in association with FRBs. \citet{zhang14} pointed out that the so-called ``internal X-ray plateaus'' observed in both long and short GRBs are best interpreted as collapse of SMNSs born during the GRB events. He therefore suggested that FRBs should be produced hundreds to thousands of seconds after some GRBs if the blitzar mechanism is valid. One concern is whether the produced FRB can escape the messy environment near a GRB. \citet{zhang14} suggested that this is not a concern since the relativistic GRB jet has cleared a funnel to facilitate the propagation of the FRB. Since internal plateaus are more commonly observed following short GRBs \citep[e.g.][]{rowlinson10,rowlinson13,lv15}, \citet{zhang14} suggested that there could be intriguing tripple associations among FRBs, short GRBs and gravitational waves. Follow-up radio observations to search FRB-like events have been carried out for some FRBs \citep[e.g.][]{bannister12,palaniswamy14,rowlinson19,bouwhuis20}, even though no confirmed association has been reported (but see \citet{bannister12} for two untriggered events whose occurring epochs are consistent with the suggested epoch of \citet{zhang14}). An intriguing association between a non-repeating FRB 20190425A and a BNS merger gravitational wave event GW190425 was claimed by \citet{moroianu22} with a chance coincidence of $\sim 1.9\times 10^{-4}$. The FRB is delayed by 2.5 h from the GW event. Such an association, if indeed physical, is consistent with the suggested scenario by \cite{zhang14}. The potential host galaxy of FRB 20190425A is also found consistent with that of a BNS merger \cite{panther23}.
    \item {\em NS-NS mergers:} There are many suggested associations between one-off FRBs and NS-NS mergers. Most of these suggested processes occur right before the merger. \citet{hansen01} considered  possible electromagnetic precursor emission before the merger caused by magnetospheric interactions between the two NSs and estimated the X-ray and radio luminosities. \citet{piro12} and \citet{lai12} studied the pre-merger magnetospheric interaction processes using the unipolar inductor model and estimated a brief EM signal with luminosity up to $10^{46} \ {\rm erg \ s^{-1}}$ within 1 s before the merger. Prompted by the discovery of four more FRBs \citep{thornton13} that established the possible astronomical origin of FRBs,    \citet{totani13} suggested that an FRB may be made right before the merger as the magnetospheres of the two merged NSs synchronize to orbital motion. The unipolar inductor model was specifically applied to interpret FRBs by \citet{wang16}, who showed that many of the observed FRB properties could be reproduced within this model. \citet{sridhar21b} proposed a pre-merger FRB model invoking specifically the synchrotron maser model. \citet{cooper22} discussed a merger-induced pulsar magnetospheric emission mechanism to produce a one-off FRB before an NS-NS merger. The general charged compact binary coalescence (cCBC) model \citep{zhang16a,zhang19} (see below) also applies to NS-NS mergers since NSs are generally charged, even though its signal may be outshone by other signals discussed above. One common issue of all the pre-merger NS-NS models for FRBs is that the neutron-rich ejecta launched due to tidal effect would make the environment ``dirty'' so that the FRB emission may escape only in a small solid angle. This further reduces the detection rate of these events, making NS-NS mergers incapable to interpret the majority of FRBs.
    
    It is worth restating that the post-merger blitzar scenario \citep{zhang14} gives another possibility of NS-NS merger association with one-off FRBs. Short GRB observations and theoretical modeling suggest that the collapse of the post-merger SMSN happen 100 to $10^4$ s after the merger \citep{lasky14,ravi14,lv15,gao16}. This provides a time window of interest for the search of FRBs associated with NS-NS mergers. The GW190425/FRB 20190425A association with a 2.5 hr time difference is consistent with this scenario \cite{moroianu22}.
    
    Finally, if a NS-NS merger leaves behind a stable massive magnetar. The standard magnetar mechanism may operate and powers repeating FRBs \citep{yamasaki18,margalit19,wang20}.
    \item {\em WD-WD mergers}: Even though the size of a white dwarf is too large to accommodate the millisecond-duration of FRBs, \citet{kashiyama13} proposed that mergers of two white dwarfs would lead to synchronization of the magnetic fields and produce millisecond radio bursts from the polar region of a post-merger magnetized white dwarf. This model predicts Type Ia supernova - FRB associations, which has never been observed.      
    \item {\em NS-BH mergers:} For CBCs, if at least one of the members is charged, one naturally gets a Poynting flux with luminosity rising sharply towards merger \cite{zhang19}. Such cCBCs would naturally give rise to an FRB-like signal in association with the merger (with the FRB observationally delayed due to the plasma dispersion). Since NSs (and all spinning magnetized objects) are globally charged \citep{michel82}, the cCBC signal must exist for neutron star mergers. Since the NS-NS merger systems are messier (see discussion above), NS-BH mergers, especially those ``plunging events'' without tidal disruption of the NS, are ideal systems to observe these cCBCs. \citet{zhang19} estimated that the total  cCBC electromagnetic luminosity of these systems can reach $5\times 10^{42} \ {\rm erg \ s^{-1}}$ for a dimensionless charge (charge normalized to the critical charge defined by the mass of the merging member) $\hat q \sim 10^{-7}$. Another channel to power an EM counterpart in the plunging NS-BH merger systems is to invoke a charged BH due to its interaction with the magnetic field of the companion NS, making the system a black hole battery \cite{mingarelli15,levin18}. \citet{dai19} showed that such a mechanism can produce a detectable EM transient (probably in the X-ray band), especially if the BH carries a rapid spin. The post-merger system of such NS-BH mergers may also release the BH spin energy to power a brief EM transient \citep{pan19,zhong19}. 
    
    For non-plunging events, the standard magnetospheric interaction effect may not be important (unless the BH is charged, see below). However, if jet-like materials can be released before the merger, the synchrotron maser mechanism may still operate to produce FRB-like events \citep{sridhar21b}.
    \item {\em BH-BH mergers:} BH-BH mergers are not supposed to produce any EM counterparts unless they are either surrounded by matter or electromagnetic fields. For the former case, the EM signals should typically have long durations unless the matter density is close to the nuclear density\footnote{The free fall timescale, which is the shortest timescale in an accretion system is proportional to $\rho^{-1/2}$ where $\rho$ is the mass density. For a typical stellar density, this timescale is or the order of $10^2-10^3$ s \citep{zhang18}, much longer than milliseconds relevant to FRBs.}. For the latter case, brief, FRB-like events may be emitted if at least one of the BHs is charged through the cCBC process \citep{zhang16a}. Such a process has been robustly supported from numerical relativity simulations \citep{liebling16}. Other related ideas include merger-induced discharge of Kerr-Newmann BHs \citep{liu16} and direct electric dipole radiation from merging charged primordial BHs \citep{deng18}. One commonly asked question is how BHs attain and retain significant charges. One interesting fact is that collapse of a spinning neutron star would leave behind a spinning BH with charge, i.e. a Kerr-Newmann BH, and that the charge does not appear to rapidly deplete \citep{nathanail17}. The charge may be retained, if a force-free magnetosphere is formed around the KN BH. Another possibility is that two BHs are merging in a magnetized environment (e.g. an AGN disk). The BHs will gain charges via the Wald mechanism \citep{wald74} and launch a  Poynting flux whose luminosity rapidly increases towards the merger \citep{kelly17}.
\end{itemize}

There are other one-off FRB models. They are either exotic or have been significantly constrained by the observational data. We list some examples in the following.

\begin{itemize}
    \item {\em Schwinger pairs at the birth of magnetars:} \citet{lieu17} suggested that at the birth of a rapidly spinning magnetar, abundant pairs would be produced from the polar cap region with the Schwinger mechanism, i.e. pairs are drawn from vacuum by strong electric fields. The pairs produce FRBs by bunched coherent curvature radiation. The star would be quickly spun down in milliseconds, and the source is not expected to repeat. The main difficulty of such a model is that a new magnetar born from massive star core collapse is buried inside the exploding star and the source is highly opaque when the suggested process happens. Even though it is not discussed in the original paper, one way to produce a naked magnetar might be through a NS-NS merger. Such a model would then falls into the broad category of NS-NS merger models discussed above.
    \item {\em Primordial black hole evaporation:} Primordial BHs with mass $M_c \sim 5\times 10^{14}$ g \citep{rice17} are supposed to evaporate now. Besides making $\gamma$-rays \citep{hawking74}, these events were suggested to emit radio waves as well \citep{rees77}. \citet{keane12} suggested that this mechanism could be one possibility to interpret FRBs. Since the total energetics of such an event is $\sim 10^{21} M_c \sim 5\times 10^{35}$ erg \citep{rees77}, this model is relevant only if FRBs are nearby (e.g. within the Galaxy). The cosmological distance of FRBs and their much greater isotropic energies rule out this mechanism to interpret FRBs.
    \item {\em White hole explosions:} White holes (WHs) are hypothetical objects in general relativity that have opposite properties as black holes. Some quantum gravity theories predict black-to-white transition as a vast amount of energy falls into a black hole reaching the Planck density. The quantum gravity pressure would push the matter backwards making a white hole. Primordial BHs with mass $\sim 10^{26}$ g are expected to explode today as WHs, which may generate non-repeating FRBs \citep{barrau14,barrau18}.
\end{itemize}

\subsection{Summary}
Even if there have been more than 50 FRB source models discussed in the literature, current observational constraints and ``Occam's razor'' principle have actually narrowed down the model options quite significantly. One may summarize the state-of-the-art of the source models as follows:
\begin{itemize}
    \item Repeaters are very likely powered by neutron stars that can provide a large enough energy budget and frequent enough triggers, either from isolated systems or interacting systems. 
    \item Among isolated neutron star sources, the leading candidate is magnetars. However, it is unclear whether younger (rapid rotators) or older (slow rotators) objects are more favorable to produce FRBs. Arguments in favor of both cases have been discussed in the literature. More data are needed to draw a conclusion.
    \item Certain interaction processes may play a role in defining the observed properties of FRBs and probably even in triggering the bursts.
    \item Non-neutron-star repeating sources are not needed, but not excluded. If these sources exist, likely they involve stellar-mass BHs.
    \item The existence of a small population of cataclysmic FRBs is not robustly established. If they exist, blitzars and CBCs are the best guesses.
\end{itemize}

\section{Environmental models}\label{sec:environment}

Since FRBs are extragalactic phenomena, their local environments are not well observed. Nonetheless, the association with a persistent radio source for rFRB 20121102A \citep{chatterjee17} and rFRB 20190520B \citep{niu22} and the large $\rm DM_{host}$ in these two and several other sources suggest that there could be compact nebulae near some FRB sources. This led to the speculation of the association of a supernova remnant or a pulsar/magnetar wind nebula with at least some FRB sources. Also the apparent periodicity observed in rFRB 20180916B \citep{chime-periodic} and probably rFRB 20121102A \citep{rajwade20} raised the speculation of a binary environment at least for some FRB sources. Rapid RM variations in some active repeaters rFRB 20201124A \cite{xuh22} and rFRB 20180520B \cite{anna-thomas22,dais22} suggested a dynamically evolving magnetized environment of these FRBs. In this section, we discuss several environmental models for FRBs.

\subsection{Persistent radio sources}

Persistent radio sources (PRSs) are associated with at least two active repeaters, rFRB 20121102A \citep{chatterjee17} and rFRB 20190520B \citep{niu22}. These two sources also possess the highest absolute values of RM among FRBs: $\sim 10^5 \ {\rm rad \ m^2}$ for the former \cite{michilli18} and $\sim 10^4 \ {\rm rad \ m^2}$ for the latter \cite{anna-thomas22,dais22}. Leading scenarios to interpret the PRSs include supernova remnants (SNRs), FRB-heated sources, or pulsar (magnetar) wind nebulae (PWNe or MWNe), all even mini-AGNs, which will be discussed in next subsections. Regardless of the detailed models, it is possible to present a generic discussion of the emission properties of PRSs, which gives a relation between the specific luminosity of the PRS $L_\nu^{\rm PRS}$ and the $\rm RM$ associated with the FRBs \citep{yang20b,yang22}. 

The radiation mechanism of PRSs is very likely synchrotron radiation of relativistic particles from a nebula in the vicinity of the FRB source. Since the PRS emission is not rapidly varying, it is likely that the PRS does not possess a relativistic bulk motion (unlike GRB afterglows). One interesting property of synchrotron emission is that the specific emission power of each particle only depends on the magnetic field strength $B$ (this is because the total emission power $P_e \sim (4/3) \gamma_e^2 \sigma c \beta (B^2/8\pi)$ and the characteristic frequency $\nu_{\rm SR} \sim (3/4\pi) \gamma_e^2 (e B/m_e c)$ (Eq.(\ref{eq:omega_SR})) are both proportional to $\gamma_e^2$ so the dependence on electron Lorentz factor $\gamma_e$ is canceled out), e.g. \citep{rybicki79}
\begin{equation}
    P_\nu \simeq \sqrt{3}\phi \frac{e^3}{m_e c^2} B,
\end{equation}
where $\phi$ is a factor of the order unity. The peak specific luminosity of the PRS can be then estimated as 
\begin{equation}
    L_{\rm \nu,max} \sim N_e^{R} P_\nu \sim \left(\frac{4\pi}{3} R_{\rm PRS}^3 n_e \zeta_e^{\rm R} \right) \left(\sqrt{3}\phi \frac{e^3}{m_e c^2} B \right),
\end{equation}
where $n_e$ is the total ionized electron number density, $\zeta_e^{\rm R}$ is the fraction that are accelerated to relativistic speeds, and the nebula is assumed as a filled sphere with radius $R_{\rm PRS}$ and a uniform magnetic field $B$. From Equation (\ref{eq:RM2}) and assuming that the observed RM of the source is mostly contributed from the nebula (which is reasonable since other RM components are typically much smaller than the RM of the PRS), one can approximately write RM as
\begin{equation}
    | {\rm RM} | \sim \frac{e^3}{2\pi m_e^2 c^4} (n_e \zeta_e^{\rm NR}) (b_{\parallel} B) R_{\rm PRS},
\end{equation}
where $\zeta_e^{\rm NR}$ is the fraction of ionized electrons that mainly contribute to RM and $b_\parallel = B_\parallel / B \lesssim 1$ is a fractional number to denote the parallel component of the magnetic field. One immediately sees that both $L_{\rm \nu,max}$ and $|{\rm RM}|$ linearly depend on $n_e$ and $B$, so that their ratio is independent of two key parameters of the PRS, i.e.
\begin{equation}
    \frac{L_{\rm \nu.max}}{|{\rm RM}|} \simeq \frac{8\pi^2 \phi}{\sqrt{3}}  \left(\frac{\zeta_e^{\rm R}}{\zeta_e^{\rm NR} b_\parallel}\right) (m_e c^2)  R_{\rm PRS}^2.
\end{equation}
Assuming that $R_{\rm PRS}$ does not differ significantly among sources, \citet{yang20b,yang22} suggested that the reason of a detectable PRS for rFRB 20121102A was because of its relatively large $|\rm RM|$. The non-detection of PRSs for the majority of repeating FRBs is simply due to their relatively small $|\rm RM|$ values. This suggestion is supported by the recent detection of a PRS from rFRB 20190520B, with a relatively large $|\rm RM|$ \citep{niu22}. 

\subsection{Supernova remnants}

The association of FRBs with SNRs was  suggested in the early FRB literature \citep[e.g.][]{connor16,piro16,murase16,kashiyama17}. Prompted by the discovery of the PRS of the first repeater rFRB 20121102A, \citet{metzger17} suggested that repeating FRBs are powered by new-born magnetars from extreme explosions such as long GRBs or superluminous supernovae (SLSNe). Within such a picture, an FRB source should be surrounded by an SNR, which itself makes radio emission and whose evolution dictates the secular DM and RM evolution of the FRB source \citep{metzger17,yangzhang17,piro18,metzger19,margalit18}.  Since such expected coordinated DM/RM evolution is not observed, it is now clear that most FRB sources are not associated with dwarf star forming galaxies or active star-forming regions within the host galaxies, which are typical for long GRBs and SLSNe \citep{lizhang20,bhandari20,heintz20}. The global FRB redshift distribution also seems not follow the star formation history of the universe (\citet{zhangzhang22}, see also \citet{hashimoto22,qiang22}, but see \citet{shin22}). So, probably most FRBs are not associated with SNRs. In any case, a small fraction of FRBs, especially the active repeaters \citep{chatterjee17,niu22}, may be  associated with SNRs. 

A dense SNR initially blocks FRBs due to various absorption/attenuation processes. The detailed optically-thinning conditions depend on the explosion parameters (ejecta energy, ejecta mass, ejecta speed) and the ambient medium density profile (a constant density  \citealt{yangzhang17,piro18} or a pre-explosion wind profile with $n \propto r^{-2}$ \citealt{metzger17,piro18}). In any case, the general condition is that the SNR's age needs to be of the order of year or decade in order to allow FRBs escape freely without suffering from various attenuation processes, as discussed in \S\ref{sec:attenuation}.

The interaction between an SNR blastwave and an ambient medium could be one source of synchrotron emission that powers the observed PRS emission as observed from rFRB 20121102A and rFRB 20190520B. \citet{metzger17} applied a parameterized self-absorbed synchrotron spectrum in the form of $F_\nu = F_0 (\nu/\nu_a)^{5/2} (1- \exp [-(\nu/\nu_a)^{-(p+4)/2}])$ to fit the observed spectrum of the PRS of rFRB 20121102A and showed that it can roughly interpret the data.

An SNR around an FRB source provides a testable prediction about the secular evolution of DM and RM, as well as their temporal evolution rates \citep{metzger17,yangzhang17,piro18}. The detailed scaling relations, on the other hand, depend on several factors, including the density profile of the ambient medium, whether the ejecta is fully ionized, the density profile of the ejecta itself, as well as the ionization status of the pre-shocked medium. In general, the evolution of an SNR includes four stages: 1. the free expansion stage when the ejecta velocity remains constant, i.e. $v \propto t^0$; 2. the Sedov-Taylor stage when the ejecta accumulates enough mass from the medium and adiabatically decelerates with the total energy in the blastwave conserved; 3. the snow-plow phase when the ejecta decelerates with significant radiative cooling, which is characterized by momentum conservation; and 4. the disappearance stage when the SNR is mixed with ISM. The transition radius $R_{\rm dec}$ between the free-expansion phase and the Sedov-Taylor phase occurs when the swept mass from the medium becomes comparable to the original mass in the ejecta, with the transition time defined by $t_{\rm dec} = R_{\rm dec} / v$ where $v=(2E/M)^{1/2}$ is the  velocity of the blastwave with kinetic energy $E$ and mass $M$. For a medium number  density $n$ and the mean molecular weight $\mu_m \sim 1.2$, one has
    \begin{eqnarray}
     R_{\rm dec} & =& \left(\frac{3M}{4\pi n \mu_m m_p}\right)^{1/3} \nonumber \\
     & \simeq & (0.43 \ {\rm pc}) \left(\frac{M}{M_\odot}\right)^{1/3} n_2^{-1/3}, \\
     t_{\rm dec} & = & \frac{R_{\rm dec}}{v}
     \nonumber \\
     & \simeq  & (42 \ {\rm yr}) E_{51}^{-1/2} \left(\frac{M}{M_\odot}\right)^{5/6} n_2^{-1/3}
    \end{eqnarray}
    for a constant density medium \cite{yangzhang17} and
    \begin{eqnarray}
     R_{\rm dec} & =& \left(\frac{M}{4\pi A}\right)^{1/3} \nonumber \\
      & \simeq & (100 \ {\rm pc}) \left(\frac{M}{M_\odot}\right) A_*^{-1}, \\
     t_{\rm dec} & = & \frac{R_{\rm dec}}{v}
     \nonumber \\
     & \simeq  & (1.0\times 10^4 \ {\rm yr}) E_{51}^{-1/2} \left(\frac{M}{M_\odot}\right)^{3/2} A_*^{-1}
    \end{eqnarray}
    for a wind medium (see also the expressions in \citet{metzger17} in terms of $v$ rather than $E$), where $A = \dot M_w/(4\pi v_w)$ is the wind parameter, $A_* \equiv A/(5\times 10^{11}) \ {\rm g \ cm^{-1}}$ is the typical value of $A$ \citep{chevalier99}. Note that in reality the wind profile would not extend to infinite distances. It is very likely that the medium density profile already returns to the constant case way before reaching $R_{\rm dec}$ of the wind model. The transition from the Sedov-Taylor phase to the snow-plow phase occurs in thousands years after the explosion \citep{draine11,yangzhang17}. If FRBs can be only made when the neutron star engine is young, only the transition from the free expansion phase to the Sedov Taylor phase is relevant.

In general, an SNR may be separated in four regions. From outer to inner, they are: 1. unshocked medium (ISM or wind); 2. shocked medium; 3. shocked ejecta; 4. unshocked ejecta or the inner boundary of the ejecta if the reverse shock already crosses the shell. Denote regions with their respective numbers and the separation radii using the two adjacent numbers (i.e. $R_{12}$ as the forward shock radius, $R_{23}$ as the contact discontinuity radius, $R_{34}$ as the reverse shock radius or the inner boundary of the ejecta). The total DM from an SNR system can be in general calculated as
\begin{equation}
    {\rm DM_{SNR}} = \int_{R_{34}}^{R_{23}} n_{3} dr + \int_{R_{23}}^{R_{12}} n_2 dr + f \int_{R_{12}}^{R_{i}} n_1 dr,
\end{equation}
where $R_i$ is the ionization front in the unshocked medium, $n_i$ is the total electron number density in region $i$, and $f$ is the ionization fraction in region 1. After delineating how $R_{12}$, $R_{23}$, $R_{34}$, $n_2$ and $n_3$ evolve with time, one can derive the $t$-dependence of $\rm DM_{SNR}$.

The strengths of the magnetic field in regions 2 and 3 can be also estimated by assuming that a fraction $\epsilon_B$ of the internal energy in the respective region is stored in (ordered) magnetic fields. Making one additional assumption that $\left< B_\parallel \right>$ is of the same order as $B$ in the respective region, one can then calculate the total absolute value of RM in the SNR system via
\begin{eqnarray}
    {\rm |RM_{SNR}|} & = & \int_{R_{34}}^{R_{23}} \left<B_{\parallel,3}\right> n_{3} dr + \int_{R_{23}}^{R_{12}} \left<B_{\parallel,2}\right> n_2 dr \nonumber \\
    & + & f \int_{R_{12}}^{R_{i}} \left<B_{\parallel,1}\right> n_1 dr
\end{eqnarray}
and delineate its temporal evolution. 

The predicted scaling laws by various authors and their assumptions can be summarized as follows:
\begin{itemize}
    \item For a constant density medium, \citet{yangzhang17} assumed that the entire region 3 is ionized and obtained
    \begin{equation}
        {\rm DM_{SNR}^{FE}} \propto t^{-2}, ~~~ d{\rm DM_{SNR}^{FE}}/dt \propto t^{-3}
        \label{eq:DMFE}
    \end{equation}
    for the free-expansion phase, and
    \begin{equation}
        {\rm DM_{SNR}^{ST}} \propto t^{2/5}, ~~~ d{\rm DM_{SNR}^{ST}}/dt \propto t^{-3/5}
    \end{equation}
    for the Sedov-Taylor phase. Note that the DM evolution scaling does not depend on the medium profile during the free expansion phase, so that the same scaling Eq.(\ref{eq:DMFE}) also applies to the case of a wind medium profile with $n \propto r^{-2}$ \citep{metzger17}. The assumption of fully ionized region 3 may be reasonable in view of the existence of a repeating FRB source at the center so that any remaining neutral materials between the engine and $R_4$ should have been ionized by X-ray emission associated with the repeated bursts (see \S\ref{sec:PWN} for more discussion). One interesting finding is that $\rm DM_{SNR}^{ST}$ increases with time. This is because the DM increase rate in shocked medium (Region 2) is larger than the DM decrease rate in the unshocked medium (Region 1) during the self-similar deceleration phase. 
    \item \citet{piro18} argued that not the whole ejecta is fully ionization. Rather, only the region between the reverse shock and the forward shock is ionized\footnote{This assumption needs scrutiny because a new-born SNR is likely very hot that regions outside the shocked region are also likely ionized. FRB-associated X-rays will also ionize any neutral atoms in the region.}. Properly following the evolution of the reverse shock and assuming an ordered magnetic field in the ejecta, they considered the DM and RM evolution relations for both a constant density medium and a wind medium. For the constant density (ISM) case, they obtained     
    \begin{eqnarray}
        {\rm DM_{SNR}^{FE,ISM}} & \propto t^{-1/2}, ~~~ d{\rm DM_{SNR}^{FE,ISM}}/dt \propto t^{-3/2}, \\
        |{\rm RM_{SNR}^{FE,ISM}}| & \propto t^{-1/2}, ~~~ d|{\rm RM_{SNR}^{FE,ISM}}|/dt \propto t^{-3/2}
     \end{eqnarray}
    in the free expansion phase, and
    \begin{eqnarray}
        {\rm DM_{SNR}^{ST,ISM}} & \propto t^{2/5}, ~~~ d{\rm DM_{SNR}^{ST,ISM}}/dt \propto t^{-3/5}, \\
        |{\rm RM_{SNR}^{ST,ISM}}| & \propto t^{-1/5}, ~~~ d|{\rm RM_{SNR}^{ST,ISM}}|/dt \propto t^{-6/5}
     \end{eqnarray}
     in the Sedov-Taylor phase. Note that the scaling in the ST phase is the same as \citet{yangzhang17} who assumed full ionization, since in the ST phase, the shocked medium (Region 2) is the dominant region to contribute to the observed DM. 
    \item For a wind medium, \citet{piro18} obtained 
     \begin{eqnarray}
        {\rm DM_{SNR}^{FE,wind}} & \propto t^{-1}, ~~~ d{\rm DM_{SNR}^{FE,wind}}/dt \propto t^{-2}, \\
        |{\rm RM_{SNR}^{FE,wind}}| & \propto t^{-2}, ~~~ d|{\rm RM_{SNR}^{FE,wind}}|/dt \propto t^{-3}
     \end{eqnarray}
    in the free expansion phase, and
    \begin{eqnarray}
        {\rm DM_{SNR}^{ST,wind}} & \propto t^{-2/3}, ~~~ d{\rm DM_{SNR}^{ST,wind}}/dt \propto t^{-5/3}, \\
        |{\rm RM_{SNR}^{ST,wind}}| & \propto t^{-4/3}, ~~~ d|{\rm RM_{SNR}^{ST,wind}}|/dt \propto t^{-7/3}
     \end{eqnarray}
     in the Sedov-Taylor phase.
\end{itemize}

\subsection{Pulsar wind nebulae \& FRB-heated nebulae}\label{sec:PWN}

The FRB source, likely a young magnetar, would eject a wind through spindown and may eject even stronger winds during flaring activities. The wind would interact with the surrounding supernova remnant to form a pulsar wind nebula (PWN). Such a PWN may play an important role in powering the FRB emission itself through synchrotron maser emission \citep{lyubarsky14,metzger19}, may contribute to the observed DM or RM \citep{metzger19,margalit19}, and may contribute to the emission of PRS as well.

\citet{dai17} argued that a repeating FRB source does not necessarily need to have a surrounding supernova remnant to generate a PRS. The wind from the FRB pulsar may interact with the surrounding medium to form a pulsar wind nebula and power persistent radio emission. However, in order to power a detectable PRS as observed from FRB 20121102A, the central pulsar needs to be rapidly spinning (e.g. $P \lesssim 10$ ms, to allow a large energy budget) and does not possess a strong magnetic field (to allow a long spindown timescale). 

\citet{yang16} noticed that the interaction between the FRB ejecta and a surrounding synchrotron nebula could play an important role in both  nebular emission and FRB emission. In particular, for certain parameters, the FRB frequency could be below the synchrotron self-absorption frequency of the nebula. These FRBs would be absorbed and could not reach the observer. Rather, they would heat up the synchrotron nebula and make a bump in the synchrotron spectrum near the absorption frequency. This prediction was found suitable to interpret the spectrum of PRS of FRB 20121102A after the latter was discovered \citep{liqc20}.

Within the framework of an accreting black hole central engine, \citet{sridhar22} proposed a ``hypernebula'' model for FRBs. The intense mass loss from a super-Eddington accretion disk produces an energetic expanding bubble, which acts similarly as a magnetar wind to produce a synchrotron-emitting nebula. The model was found suitable to interpret the PRSs of some repeating FRBs. 

\subsection{Binary systems}

A widely discussed FRB source environment is binary systems, in which the companion (a massive star,  another neutron star, or even a massive black hole) of the FRB source (likely a pulsar or magnetar) plays a noticeable role in shaping the properties of the detected bursts. Binary interaction was invoked as one of the mechanisms to trigger FRBs within the cosmic comb model \citep{zhang17}. The discussion of binary systems becomes popular after the discovery of the $\sim 16$d period of rFRB 20180916B \citep{chime-periodic} as the observed period may be interpreted as the orbital period of the binary system. It was quickly realized that the massive companion of the FRB pulsar could provide a strong, opaque wind to block FRBs in certain directions, so that repeated bursts could be only observed in certain orbital phases \citep{ioka20,lyutikov20}. More generally, \citet{wada21} discussed three possible modes for companion - FRB source interactions: (1) The {\em funnel mode} is the mode in which companion wind is stronger than the FRB pulsar wind, so that the latter can only open a funnel as the pressures of the two winds balance. The funnel is visible by the observer at certain orbital phases \citep{ioka20,lyutikov20}; (2) The {\em $\tau$-crossing mode} is the mode in which the active window is defined by the orbital phases where the optical depth of FRB against Thomson scattering, free-free absorption and induced Compton scattering becomes less than a few (the photosphere radius due to induced Compton scattering is usually defined by $\tau \sim 10$ rather than $\tau \sim 1$). The FRB source pulsar crosses the photosphere twice during the orbibal motion and only when the orbit is above the photopshere could the FRB emission be observed; (3) The {\em inverse funnel mode} is the opposite case of the funnel mode, in which the FRB pulsar wind is stronger than the companion wind and the active phase is greater than half of the period.  \citet{zhanggao20} studied various binary systems including one NS companion using population synthesis models and found that a 16-d period is common and the companion is likely a B-type star. The frequency-dependent periofic window  of rFRB 20180916B has been raised as evidence against the simple binary comb scenario \cite{paster-marazuela21}. However, several scenarios have been proposed to account for the observations within various binary scenarios \cite{wada21,liqc21,lidz21}. 

The complicated RM evolution as well as apparent Faraday conversion observed in rFRB 20201124A \citep{xuh22} does not directly point toward a binary system (due to the lack of periodicity). However, a detailed modeling of the polarization properties of the system seems to require multiple layers of plasma to contribute to RM and radio wave absorption and a binary system is a likely possibility to account for the data \citep{lidz22,wangfy22,yang22b}. 

An extreme version of binary systems is to have the FRB pulsar orbiting a massive or even a supermassive black hole. \citet{zhang18b} suggested that rFRB 20121102 may reside near a supermassive black hole whose AGN-like-activities may be powering the persistent radio emission of the source. It is interesting that the parameter space allowed for the binary comb model to interpret its $\sim 157$-d period also prefers a supermassive black hole as the companion \citep{wada21}. The large absolute RM value and sign change  observed in rFRB 20190520B may be also interpreted by invoking a massive black hole in the vicinity of the source \citep{dais22}.

\section{Propagation effects}\label{sec:propagation}

Besides the standard dispersion and Faraday rotation, FRB radio waves undergo additional interesting propagation effects before being detected on Earth. The propagation effects may leave imprints on the observed signals and observed information may in turn be used to diagnose the physical properties of the medium where FRB waves propagate through. 

\subsection{Multi-path effects: scattering, scintillation, and RM scatter}\label{sec:scattering}

One important feature of radio wave propagation is that the observed radio waves at a particular time is likely the superposition of rays from multiple paths. This is because the frequency-dependent propagation speed of radio waves depend on plasma density the waves propagate through and because the densities along the multiple lines of sight likely have fluctuations, mostly likely because of turbulence that is ubiquitous in astrophysical environments. These fluctuations would spread the rays, blur the image, broaden the radio pulse, and smear the bandwidth. All these effects are characterized as scattering (describing pulse broadening) and scintillation (describing intensity fluctuation and bandwidth smearing) \citep{rickett77,rickett90}. 

Scattering is often manifested as a temporal scattering tail in FRB pulses. Let the FRB and a thin plasma screen (lens) be located at the angular diameter distances $D_s$ and $D_l$ from Earth, respectively. Let the angular diameter distance between the source and the screen be $D_{sl}$, which is close to but not equal to $D_s - D_l$ for cosmological sources. The scattering half angle $\theta_s$ and the scattering timescale $\tau_s$ can be calculated as \citep[e.g.][]{rickett77,rickett90,macquart13,cordes16b,xu16,yang22}, i.e.
\begin{eqnarray}
    \theta_s & \simeq & \frac{D_{ls}(\lambda/2\pi)}{D_s r_{\rm diff}},  \\
    \tau_s & \simeq & \frac{\lambda}{2\pi c} \left(\frac{r_{\rm F}}{r_{\rm diff}}\right)^2=\frac{D_l D_{s} \theta_s^2}{c D_{ls} (1+z_l)} \nonumber \\
    & = & \frac{D_l D_{ls} (\lambda/2\pi)^2}{c D_s r_{\rm diff}^2 (1+z_l)} \myeq \frac{D_{ls} (\lambda/2\pi)^2}{c r_{\rm diff}^2 (1+z_l)}, \label{eq:taus}
\end{eqnarray}
where $\lambda$ is the observed wavelength (longer by a factor $(1+z_l)$ than that at the scattering screen), and $z_l$ is the redshift of the screen (lens). Because in the FRB case the screen is usually in the host galaxy, when relevant we also write the simpler expression in the last equation for the case of $D_s = D_l$. Here there are two important length scales. One is the Fresnel scale
\begin{equation}
    r_{\rm F} = \left[ \frac{D_l D_{ls} (\lambda/2\pi)} {D_s (1+z_l)} \right]^{1/2} \myeq \left[ \frac{D_{ls} (\lambda/2\pi)} {1+z_l} \right]^{1/2},
\end{equation}
which is the geometric mean of the effective distance $D_{\rm eff} = D_l D_{ls} / D_s$ and the rest-frame reduced wavelength $\lambdabar_s = \lambda/[2\pi (1+z_l)]$. For a spherical wave, this is the transverse scale of the wave front where the light path difference is $\lambdabar_s$ at a distance of $D_{\rm eff}$. 

A more important distance scale is the so-called diffractive lengthscale, $r_{\rm diff}$, which is the transverse scale of the wave front where the {\em root-mean-square} difference between the two rays is $\lambdabar_s$. Let us assume that the scattering effect is introduced by electron density fluctuations that arise from a turbulent cascade and the relevant spectrum takes the power-law form in wave number $k$ \citep[e.g.][]{rickett77,cordes85,cordes02,macquart13,xu16}
\begin{equation}
    P_{\delta n_e}(k) = C_n^2 k^{-\beta}, ~~ 2\pi/L \leq k \leq 2\pi/l_0,
\end{equation}
where $l_0$ and $L$ are the inner (dissipation) and outer (injection) scales of the turbulent energy, $C_n^2$ is the spectral coefficient (the amplitude of turbulence) that describes the significance of the density fluctuations, and $\beta$ is the spectral index, which equals 11/3 for the ``Kolmogorov'' turbulent spectrum but can take a more general value. The turbulence is short-wave-dominated when $\beta > 3$ and long-wave-dominated when $\beta < 3$. From the density variance $\left< (\delta n_e)^2 \right> = \int P_{\delta n_e}(k) d^3 {\bf k}$ and $L \gg l_0$, one can write \citep{xu16}
\begin{eqnarray}
 C_n^2  \sim  \frac{\beta-3}{2(2\pi)^{4-\beta}}(\delta n_e)^2 L^{3-\beta}, & ~~\beta > 3, \\
 C_n^2  \sim  \frac{3-\beta}{2(2\pi)^{4-\beta}}(\delta n_e)^2 l_0^{3-\beta}, & ~~\beta < 3.
\end{eqnarray}
It is convenient to define a {\em scattering measure} as the line integration of $C_n^2$ along the line of sight \citep{cordes85,cordes02}
\begin{equation}
    {\rm SM} = \int_0^D C_n^2 dl \simeq C_n^2 \Delta,
    \label{eq:SM}
\end{equation}
where in the second equation we have assumed that scattering only happens in a thin screen with thickness $\Delta$. One can finally write the expression of $r_{\rm diff}$ in the two regimes \citep{xu16}
\begin{eqnarray}
 r_{\rm diff} \sim (\pi r_e^2 \lambda^2 {\rm SM} l_0^{\beta-4})^{-\frac{1}{2}}, & ~~r_{\rm diff} < l_0, \\
 r_{\rm diff} \sim (\pi r_e^2 \lambda^2 {\rm SM})^{\frac{1}{2-\beta}}, & ~~r_{\rm diff} > l_0,
\end{eqnarray}
where $r_e = e^2/m_e c^2$ is the classical radius of the electron. With all these preparations, one can finally derive the observed scattering timescale that has dependence as \citep{xu16,yang22}
\begin{eqnarray}
 \tau_{\rm sc}^{\rm obs} & = & (1+z_l) \tau_{\rm sc} \nonumber \\
 & \propto & \left\{
 \begin{array}{ll}
    \delta n_e^2 \Delta^2 \lambda^4 (1+z_l)^{-3},  &  r_{\rm diff} < l_0, \\ 
    \delta n_e^{\frac{4}{\beta-2}} \Delta^{\frac{\beta}{\beta-2}} \lambda^{\frac{2\beta}{\beta-2}} (1+z_l)^{-\frac{\beta+2}{\beta-2}},
      &  r_{\rm diff} > l_0
 \end{array}
 \right.
\end{eqnarray}
regardless of the regime of $\beta$. For Kolmogorov turbulence with $\beta=11/3$, the numerical value of the index is $2\beta/(\beta-2) = 22/5 = 4.4$. The value of $\tau_{\rm sc}^{\rm obs}$ depends on the SM and for FRB parameters, the contribution of $\tau_{\rm sc}^{\rm obs}$ from the host galaxy or the immediate environment of the FRB source is much greater than those from the IGM and from the Milky Way \citep{cordes16b,xu16}. 

With the scattering timescale, one can immediately define a scintillation bandwidth
\begin{equation}
    \Delta \nu_{s} \sim 1/\tau_{\rm sc}^{\rm obs} \sim (1 \ {\rm kHz}) \tau_{-3}^{-1},
\end{equation}
which is too small to be identified in the observing band of the telescopes. On the other hand, 
scintillation band smearing is detected in the radio band, which should have a very different origin. For FRBs, the detected scintillation bandwidth fringes are likely dominated by the multi-path propagation effect within the Milky Way galaxy.

The multi-path effect can also affect the observed polarization properties. For linearly polarized FRB emission, the multi-path effect can introduce an RM scatter \citep{feng22}, i.e. different lines of path undergo different Faraday rotations so that the final observed emission is depolarized \citep{beniamini22,yang22}. The RM scatter may be estimated as \citep{yang22}
\begin{eqnarray}
 \sigma_{\rm RM}&\simeq&\frac{e^3}{2\pi m_e^2c^4} (l_{\rm s}\Delta)^{1/2} \delta(n_eB_\parallel)_{l_{\rm s}}\nonumber\\
&=&0.81~{\rm rad~m^{-2}}\left(\frac{\sqrt{l_{\rm s}\Delta}}{{1~\rm pc}}\right)\left(\frac{\delta(n_eB_\parallel)_{l_{\rm s}}}{1~{\rm cm^{-3}\mu G}}\right),
\end{eqnarray}
where $\delta(n_eB_\parallel)_{l_{\rm s}}$ is $\delta(n_eB_\parallel)$ on the scale of $l_s$, and 
\begin{equation}
    l_s(\lambda) \simeq \frac{\lambda D_{ls}}{2\pi r_{\rm diff}}
\end{equation}
is the maximum transverse scale of the multi-paths. The effect of $\sigma_{\rm RM}$ is to introduce a frequency-dependent polarization degree, with the fractional reduction of the linear polarization amplitude defined by $f_{\rm RM,depol} \equiv 1 - \exp (-2\lambda^4 \sigma_{\rm RM}^2)$ \citep{osullivan12,feng22}. This effect presents an interpretion to the frequency-dependent linear polarization degree of a sample of repeating FRBs \citep{feng22}. 

One prediction of the RM scatter theory is that it is positively correlated to the observed scattering timescale, i.e. $\sigma_{\rm RM} \propto \tau_{sc}^\alpha$, with $\alpha \sim (0.5 - 0.8)$ \citep{yang22}. This is qualitatively consistent with the observational data \citep{feng22}.

Another mechanism to scatter FRB emission is through filamentation of the FRB waves in a magnetar wind. This may induce additional modulation in the emission with a $\tau_{sc} \propto \nu^{-2}$ scattering dependence, which is not widely observed \cite{sobacchi22}. On the other hand, such an effect may induce large scintillation bandwidths of $\sim 100$ MHz as observed, which corresponds to an undetectable ns-duration scattering timescale \cite{sobacchi22}.

\subsection{Plasma lensing and gravitational lensing}

An extreme version of the plasma multi-path effect is plasma lensing \citep{cordes17}. In general, a denser lens with a positive electron column density would serve as a diverging lens, but rays passing through different parts of the lens, especially from voids, may converge to generate caustics that amplify burst signals.  Since plasma lenses may be dynamically evolving, the lensed bursts can allow different spectral behaviors, in contrast to gravitational lensing that retains the spectral shape. 

A simplest model is a 1D Gaussian plasma lens model \citep{clegg98} that can be described as ${\rm DM}(x) = {\rm DM}_l \exp (-x^2/x_0^2)$, where $x_0$ is the characteristic transverse scale of the lens and $x$ is the transverse coordinate. Let the transverse coordinates in the source, lens, and observer's planes are $x_s$, $x$, and $x_{\rm obs}$, respectively, and define dimensionless coordinates $u_s = x_s/x_0$, $u=x/x_0$, and $u_{\rm obs}=x_{\rm obs}/x_0$, the lens equation in geometric optics could be expressed as 
\begin{equation}
    u (1+\alpha e^{-u^2}) = u'
\end{equation}
through the Kirchhoff diffraction integral of the Gaussian lens \citep{cordes17}. Here 
\begin{equation}
    u' = (D_{l} / D_s) u_s + (D_{ls} / D_s) u_{\rm obs},
\end{equation}
and 
\begin{equation}
    \alpha = \frac{\lambda^2 r_e {\rm DM}_l}{\pi x_0^2} \left( \frac{D_{ls} D_l}{D_s} \right)
\end{equation}
is a dimensionless parameter. The amplification factor can be written as 
\begin{equation}
    G= |1+\alpha(1-2u^2) e^{-u^2} |^{-1},
\end{equation}
which has a maximum 
\begin{equation}
    G_{\rm max} \sim x_0 / r_{\rm F}
\end{equation}
at the caustics where $\alpha = \alpha_{\rm min}$. \citet{cordes17} constrained the lens parameters required to have caustics, which reads ${\rm DM}_l D_{ls} / x_0^2 \gtrsim 0.65 \ {\rm pc^2 \ AU^{-2} \ cm^{-3}}$. They argued that the apparently more active repetition behavior of rFRB 20121102A compared with other sources may be a consequence of significant plasma lensing. The discoveries of several more active repeaters cast the doubt to interpret all of them with the plasma lensing effect and tend to suggest that different FRBs may have different active levels and some of them (maybe young magnetars) are intrinsically more active than others. Nonetheless, plasma lensing may leave certain imprints in FRB observations. For example, \citet{er20} argued that frequency-dependent delay due to the geometric effect could be comparable to the dispersion delay, so that the measured DM could be overestimated if sigmals propagate through a high-density gradient clump of plasma. 

Similar to other astronomical objects, FRBs can undergo gravitational lensing. The high event rate of FRBs makes it plausible that lensed FRBs can be detected as the detected sample increases quickly with time \citep{lili14}. Since  gravitational lenses are not dynamically evolving, multi-images of the lensed bursts would be more analogous with each other with the a strict delay timescale for all the bursts from the same repeater source. The combination of observing multiple images with VLBI and the time delay of the images would allow a direct probe of the proper motion of a repeating FRB, which will directly constrain the physical conditions at the source \citep{dailu17}. 

\subsection{Large-amplitude wave effects}\label{sec:large-a} 

One unique property of FRB waves, thanks to their very high luminosites in radio frequencies, is that at a small enough radius from the engine, the amplitude of the electromagnetic waves is so large that electrons interacting with the waves would move with a relativistic speed. For an FRB with luminosity $L$, the Poynting flux at a distance $r$ from the source is $F=L/(4\pi r^2)$, which can be also written in terms of the EM wave amplitude $F=c E_w^2/(8\pi) \simeq c B_w^2/(8\pi)$. As a result, the wave amplitude can be written as
\begin{equation}
    E_w \simeq B_w =\sqrt{ \frac{2 L}{c r^2}} = (820 \ {\rm esu ~ or~ G}) L_{42}^{1/2} r_{13}^{-1}. 
\end{equation}
One can define a dimensionless parameter
\begin{equation}
    a \equiv \frac{eE_w}{m_e c \omega} = \frac{\omega_{B_w}}{\omega}
\end{equation}
of a wave for its amplitude (where $\omega_{B_w} = e B_w / m_e c = e E_w / m_e c$), which is essentially the dimensionless oscillation velocity $v_{\rm osc}/c$ of an electron in response to the wave when $a<1$. Plugging in the typical FRB parameters, one has
\begin{equation}
    a = 2.3 L_{42}^{1/2} r_{13}^{-1} \nu_9^{-1}.
\end{equation}
One can see that for an $L=10^{42} \ {\rm erg ~ s^{-1}}$ FRB, the amplitude factor is $a \gg 1$ when $r \ll 10^{13}$ cm. In such a large-amplitude wave regime, a series of propagation effects not shared by low-amplitude radio waves are introduced. Similar effects apply to laboratory lasers which can have very large intensities to reach the relativistic regime. The importance of the large-amplitude effects within the context of FRBs was first pointed out by \cite{luan14} and later discussed by various authors within various contents \citep[e.g.][]{gruzinov19b,beloborodov20,kumar20b,lu20c,yang20a}. In analogy to the large-amplitude wave effects for laboratory lasers, \citet{yang20a} systematically studied the large-amplitude effects for FRBs, which can be summarized as follows:
\begin{itemize}
    \item Enhancement of emission cross section. In the $a \gg 1$ regime, an electron moves in a ``figure-of-eight'' trajectory because besides the traditional harmonic motion due to the oscillating $E_w$, it is also affected by the Lorentz force from the oscillating $B_w$ \cite{sarachik70}. In the oscillation-center rest frame, the electron moves with a Lorentz factor $\gamma' = a/\sqrt{2}$. Similar to synchrotron radiation, the emission power of the electron is $P \sim a^2 P_T$, where $P_T = e^4 E_w^2/3 m_e^2 c^3$ is the received power given by the Thomson formula. Considering the Poynting energy flux in the waves is $S=c E_w^2/8\pi$ and that the cross section is defined as $\sigma = P / S$, one gets \citep[e.g.][]{yang20a}
    \begin{equation}
        \sigma = \frac{P}{S} \sim a^2 \sigma_T.
        \label{eq:sigma-a^2}
    \end{equation}
    
    With the existence of a background magnetic field $B$, as is the case of FRBs emitted from a magnetar magnetosphere, the problem becomes more complicated. In the inner magnetosphere where $B \gg B_w$ is satisfied, the large amplitude effect is suppressed, since the electron is confined by the much stronger background $B$. In a dipolar field, one has $B \propto r^{-3}$, which decays faster than $B_w \propto r^{-1}$. As a result,  the large-amplitude effect would become important when $B$ becomes smaller than $B_w$ \citep{beloborodov21,beloborodov21b}. Detailed numerical results suggest that $\sigma / \sigma_T$ is typically greater than $a^2$, with a dependence on the angle between the wave vector $k$ and the $B$ vector and the relationship between $\omega_B / \omega$ and $a$ \citep{qu22c}. When the plasma is streaming outwards relativistically, bright FRBs can propagate through it and escape the plasma successfully \citep{qu22c}.
    \item Transparency of strong waves. In the $a \gg 1$ and weak magnetic field (far away from magnetosphere) regime, the dispersion relation for a circularly polarized wave is modified as \citep{yang20a}
    \begin{equation}
        \omega^2 = k^2c^2 + \frac{\omega_p^2}{\gamma},
    \label{eq:large-a-dispersion}
    \end{equation}
    where $\gamma = (1+a^2/2)^{1/2}$. This effectively reduces the near-source plasma frequency by a factor of $\sqrt{\gamma}$, or reduce the plasma density by a factor of $\gamma$. This would reduce the DM contribution from the vicinity of the FRB source (e.g. within 1 AU for a $L=10^{42} \ {\rm erg \ s^{-1}}$ burst) by a factor of $\sim \gamma$, making the FRB more transparent \citep{lu20c,yang20a}. The FRB-induced medium filamentation  \citep{sobacchi22b} would further modify the dispersion relation (Eq.(\ref{eq:large-a-dispersion})) and further reduce the near-source DM. 
    \item Relativistic self-focusing. In the $a\gg 1$ regime, the non-linear refractive index is $n_r = c/v_p = \sqrt{1-\omega_p^2/\gamma(a)\omega^2}$, which is intensity-dependent. Consider a beamed FRB with a decreasing intensity from the center. The propagation effect naturally ``squeezes'' the light, making the FRB more beamed \citep{yang20a}. Such an effect is especially important for a high-density emitter, e.g. in the synchrotron maser scenario. The squeezing effect becomes negligible in a magnetosphere environment \citep{lyutikov20d}.
    \item Ponderomotive force electron acceleration in wakefield waves. An electromagnetic pulse with a non-uniform energy density (which is the case of an FRB) would exert a ponderomotive force (${\vec F}_p = -m_e c^2 \nabla (1+ \left< {\vec a}^2 \right>)^{1/2}$ in the relativistic regime, where ${\vec a}=e {\vec A}/m_e c^2$ ($\vec A$ is the vector potential, i.e. $\vec B = \nabla \times \vec A$) is a dimensionless vector whose amplitude is comparable to $a$) to the ambient plasma. Electrons would be more easily expelled away from equilibrium due to the radiation pressure, forming an oscillating electrostatic field in the plasma. This is the so-called wakefield wave. Such a field would accelerate electrons. However, such an effect is too small to be observational interesting \citep{yang20a}.
\end{itemize}

\section{FRBs as astrophysical and cosmological probes}\label{sec:probes}

Regardless of their physical origin(s), FRBs are tremendous cosmic probes that can be used to study various problems in astrophysics, cosmology, and even fundamental physics. In this section we summarize some proposed applications of FRBs as cosmological probes. Reviews on these subjects can be also found in \citet{bhandari21} and \citet{xiao21}. 

\subsection{Missing baryons: $\Omega_b$ and $f_{\rm IGM}$}

Most of the following probes make use of the salient feature of the $\left< {\rm DM_{IGM}} \right> - z$ relation (Eq.(\ref{eq:DM-z})), which makes a connection between two observables, DM and $z$. The complication is that there are multiple components that contribute to DM (Eq.(\ref{eq:DMterms})). However, in most cases, ${\rm DM_{IGM}}$ is the dominant term. If one can properly deduct other components, one can directly measure $\Omega_b f_{\rm IGM}$ from the data (Eq.(\ref{eq:DM-z})). This has been done with a small sample of FRBs \citep{macquart20}. The results are consistent with indirectly inferred $\Omega_b$ from cosmic microwave background and Big Bang nucleosynthesis measurements \citep{Planck18,BBN}. This solves the long-standing ``missing baryon problem'' and suggests that the majority of the missing baryons are in the intergalactic medium. If one adopts the best-fit $\Omega_b$ from the CMB measurements, one can directly constrain $f_{\rm IGM}$. The results inferred from FRBs (\citet{lizx20} and Figure \ref{fig:DM-z} right panel of this review) are generally in agreement with the previous results using other methods \citep{fukugita98}. 

\subsection{IGM inhomogeneity}

Equation (\ref{eq:DM-z}) is an average relationship. For individual lines of sight, the measured DM at the same $z$ could be very different because the IGM is inhomogeneous. Numerical simulations \citep{mcquinn14} showed that the standard deviation $\sigma [{\rm DM}]$ of the DM distribution ranges from 180 to 400 $\rm cm^{-3} \ pc$ at $z=1$ pending on whether the ``missing'' baryons lie around the virial radius of $10^{11}-10^{13} M_\odot$ halos or further out. \citet{jaroszynski19} showed $\sim 13$\% scatter of DM at $z=1$ and $\sim 7$\% scatter at $z=3$ using Illustris simulation, see also \citet{takahashi21}. \citet{macquart20} presented a sample of 8 FRBs with $z$ measurements, which indeed showed a large scatter and the authors expected that the range of scatter should increase with redshift. The current data with 21 $z$-known FRBs (Fig.\ref{fig:DM-z} right panel of this review) do not show such a trend. \citet{li19} reconstructed the DM-$z$ relation for nearby FRBs using the observed optical galaxy data and the halo baryon distribution models and found that the inferred $\rm DM_{IGM}$ values for individual FRBs indeed deviate significantly from the predicted values based on the average relation Eq. (\ref{eq:DM-z}). A more detailed study of FRB 20190608 making use of both optical and radio data led to a reconstruction of the cosmic web along the line of sight \citep{simha20}.
With a much larger sample of localized FRBs with $z$ measurements, the scatter of the ${\rm DM_{IGM}} -z$ relation will be mapped directly from the data. This scatter is also very important to decide how good FRBs are to serve as other types of probes as discussed below.

\subsection{Circum-galactic medium}

Individual galaxies are surrounded by a circum-galactic medium (CGM), which is the gas surrounding the galaxies outside their disks or ISM but inside the virial radii. The properties of the CGM are poorly studied. The amount of mass in the CGM would affect the scatter of the ${\rm DM_{IGM}} -z$ relation. FRBs can probe the CGM directly, either for the  halo of our own Milky Way Galaxy or the halo of foreground galaxies along the line of sight of some FRBs. Low DM FRBs from nearby galaxies can be used to directly constrain $\rm DM_{halo}$ of the Milky Way \citep{prochaska19b}. Analyses of the radio data of FRB 20181112 posed strong constraints on the properties of the halo of a foreground galaxy, which has low net magnetization and turbulence \citep{prochaska19}. The studies in this direction will flourish as more data are accumulated. 

\subsection{FRB host galaxy and the surrounding medium}

Another uncertainty that hinders the application of the ${\rm DM_{IGM}} -z$ relation to probe the universe is the DM contribution from the FRB host galaxy as well as the immediate medium around the FRB source. Both are poorly known and difficult to measure because they are degenerate with $\rm DM_{IGM}$, which itself has a large uncertainty. Nonetheless, $\rm DM_{host}$ and $\rm DM_{src}$ have been studied from different aspects. Theoretically, \citet{xu15} simulated the DM distributions for three types of FRB hosts and different viewing angles. The DM contribution from a dense medium (e.g. supernova remnant) around FRBs has been extensively modeled \citep{metzger17,yangzhang17,piro18}.  Observationally, some FRBs with an apparent excess DM (e.g. rFRB 20121102A, \citet{tendulkar17}; and rFRB 20190520B, \citet{niu22}) have shown evidence of a large $\rm DM_{host}+DM_{src}$. Information of the host galaxy type and relative position of the FRB in its host galaxy \citep[e.g.][]{tendulkar17,bannister19,bhandari20,xuh22} can also help to estimate the DM contribution to the host galaxy. If one assumes that the $\rm DM_{host}+DM_{src}$ of a large sample of FRBs follow a normal distribution (which may be the case if the outliers such as rFRB 20190520B are removed), the average DM contribution from the host/source may be inferred statistically using the observed DM-fluence relation \citep{yang17} or DM-$z$ relation \citep{lizx20}. With 5 FRBs with $z$ measurements, \citet{lizx20} estimated the local value of $\rm DM_{host} + DM_{src}$ as $\sim 107^{+24}_{-45} \ {\rm pc \ cm^{-3}}$ (the measured value is smaller by a factor of $(1+z)$). The larger sample of the current 21 FRBs leads to the similar constraint (see Fig.\ref{fig:DM-z} right panel and related discussion). With a large enough sample, $\rm DM_{host} + DM_{src}$ can be also directly inferred through differential increase of the observed $\rm DM_E$ with $z$ \citep{yangzhang16}. From cosmological simulations, it was found that $\rm DM_{host}$ is redshift-dependent and the median value ranges from $\sim 35 \ {\rm pc \ cm^{-2}}$ at $z=0.1$ to $\sim 106 \ {\rm pc \ cm^{-2}}$ at $z=1.5$ \citep{zhanggq20}.

\subsection{Dark energy}

Suppose a large sample of FRBs are localized and $z$ measured, the IGM inhomogeneity and host/source DM contribution can be better quantified. This would open an  opportunity to compare the data with different $\left< {\rm DM_{IGM}} \right> - z$ models and constrain relevant model parameters. The first exciting prospect is to use FRBs to constrain the evolution of the universe, in particular, the nature of dark energy as delineated by the $E(z)$ function in Equation (\ref{eq:E(z)}). Simulations \citep{zhou14,gao14,walters18} suggest that depending on the degree of IGM inhomogeneity, meaningful constraints on dark energy may be achieved with a large enough sample, especially in combination with other cosmological probes such as Type Ia supernovae, cosmic microwave background (CMB), and baryon acoustic oscillations (BAO). The challenges for robustly extracting distance and the quantitative estimates of the systematics control needed for FRBs to be competitive distance probes have been discussed by \citet{kumarlinder19}.

\subsection{Reionization history}

Another prospect of using the $\left< {\rm DM_{IGM}} \right> - z$ relation as cosmological probes is to probe the reionization history of the universe. This is because the observed DM is only contributed by free electrons. The relation (Eqs.(\ref{eq:DM-z}) and (\ref{eq:chiz})) carries the ionization fraction for both H and He \citep{deng14,zheng14}. Theoretical modeling and observational constraints suggest that He might be fully ionized at $z \sim 3$ \citep{zheng14}, whereas H is ionized at $z > 6$ \citep{fan06}. The detailed ionization history, especially that of H ionization in the so-called ``dark ages'', is not well constrained, and FRBs can potentially probe it directly. Detailed simulations \citep{caleb19b,bhattacharya21} showed that He ionization from $z=3$ to $z=6$ can be differentiated with $1.6 \times 10^3 - 10^4$ FRBs. For H reionization, the epoch of reionization may be constrained via an observed $\rm DM_{max}$ or 40 FRBs detected at redshifts $z \in (6, 10)$ \citep{beniamini21}. 

\subsection{Large-scale structure and turbulence}

With the DM and spatial distribution of a large sample, one can perform a study of the angular correlation of DMs for FRBs, extracting their structure function and correlation function to probe the large-scale structure \citep{shirasaki22} or even turbulence at very large scales. The pre-CHIME sample showed a preliminary evidence of possible large-scale turbulence \citep{xuzhang20}, which is not confirmed with the larger CHIME sample \citep{xu21}. Nonetheless, the results are broadly consistent with the statistical modeling of the cosmological DM from numerical simulations \citep{takahashi21}.  \citet{rafiei-ravandi21} found a statistically significant cross-correlation between CHIME FRBs and galaxies in the redshift range $z \in (0.3, 0.5)$. 

\subsection{Source and intergalactic magnetic fields}

Besides using DM to perform various constraints, a combination of DM and RM may place a constraint on magnetic fields under ideal situations. Similar to Eq.(\ref{eq:DMterms}), one may decompose the observed RM to several terms
\begin{equation}
   \rm RM = RM_{ion} + RM_{MW} +RM_{IGM} + \frac{RM_{host} + RM_{src}}{(1+z)^2},
\end{equation}
where $\rm RM_{ion}$ is the contribution from the Earth ionosphere that gives a measurable small contribution, and the $(1+z)^2$ correction factor in the last two terms comes from the $\theta = \lambda^2 {\rm RM}$ relation, where $\theta$ is the polarization angle. In general, the observed RM is likely dominated by the near-source medium, which is likely a dynamically evolving magnetized environment \citep{michilli18,luo20,xuh22,feng22}. The observed DM, on the other hand, is dominated by the IGM term. As a result, $\rm RM/DM$ is not a good probe of the average $B_\parallel$ along the line of sight (unlike pulsars). One may remove the Milky Way and IGM contributions to DM (with the caveat of a large uncertainty in $\rm DM_{IGM}$), and estimate the average line-of-sight magnetic field in the host and source (most likely in the source region) as 
\begin{equation}
    B_\parallel^{\rm src} \sim (1.23 \mu{\rm G}) (1+z)
    \left| \frac{\rm RM_{host,obs}+RM_{src,obs}}{\rm DM_{host,obs}+DM_{src,obs}} \right|,
\end{equation}
with the observed RM as a proxy of the numerator. The derived $\left< B_\parallel \right>$ values for FRBs are on average consistent with those of pulsars and magnetars observed in the Milky Way \citep{wangwy20}, with the exception of rFRB 20121102A, which has a much higher value \citep{hilmarsson21}.

Another way of estimating $B_\parallel$ near the FRB source is to make use of the observed variations of DM and RM, i.e. 
\begin{equation}
    B_\parallel^{\rm src} \sim (1.23 \mu{\rm G}) (1+z)
    \left| \frac{\rm \Delta RM}{\rm \Delta DM} \right|.
\end{equation}
This already assumed that the variation of $B_\parallel$ is not the dominant factor for RM variations. For rFRB 20201124A, the detection of significant $\rm \Delta RM$ and the non-detection of $\Delta {\rm DM}$ led to a constraint of $B_\parallel^{\rm src} > 0.2 {\rm mG}$ \citep{xuh22}. Some repeating FRBs (e.g. rFRB 20190520B) show significant RM reversals, suggesting the reversal of the magnetic field directions. In such cases, one has \cite{yang22b}
\begin{equation}
{\rm \frac{\Delta RM}{RM} \simeq \frac{\Delta DM}{DM}} + \frac{\Delta B_\parallel}{B_\parallel}.
\end{equation}
Since both $\rm \Delta RM / RM$ and $\Delta B_\parallel / B_\parallel$ are of the order of unity, the value of $B_\parallel$ cannot be constrained.

If the RM contribution from the host and source is small, or its behavior can be well quantified for a large FRB sample, one may combine the observed DM and RM information to make a constraint on the poorly known IGM magnetic field. \citet{hackstein19} showed that less than 100 FRBs from magnetars in a stellar-wind environment hosted by starburst dwarf galaxies at $z \gtrsim 0.5$ would be able to differentiate different IGM magnetic field models. Recent observations of more complicated FRB surrounding medium in terms of RM variations \citep{michilli18,luo20,xuh22} and RM scatter \citep{feng22} make it difficult to correct for the dominant RM contribution from the near-source region, rendering constraining the IGM magnetic fields much more challenging. 

\subsection{Additional probes with gravitationally lensed FRBs: $H_0$, $\Omega_k$, and dark matter}

The high event rate of FRBs makes it likely to detect gravitationally lensed FRBs in the future. These lensed sources, especially the lensed repeating sources, offer new opportunities to probe cosmology using FRBs. Thanks to their very short durations, the time delays between the images can be measured with an unprecedented precision. Since the gravitational lensing geometry involves the measurements of the angular diameter distances of the source and lens, which depend on the Hubble constant $H_0$ through $z$ and the curvature of the universe $\Omega_k$, lensed FRBs can be used to directly measure $H_0$ and $\Omega_k$ \citep{li18}. Simulations showed that with about 10 lensed repeating FRB systems, $H_0$ can be measured to a sub-percent precision level and $\Omega_k$ can be measured to a precision of $\sim 0.076$ in a model-independent manner \citep{li18}. 

FRBs can be micro-lensed by massive compact halo objects (MACHOs) which have been proposed as one type of contributor to the dark matter. For $M_{\rm MACHO} \gtrsim 20 M_\odot$, the delay time would be longer than one millisecond. If such lensed events are observed, one FRB with a single pulse would be observed as a double-pulse (lensed by one MACHO object) or triple-pulse (lensed by a MACHO binary) bursts. The non-detection of these events would place an upper limit on the abundance of these MACHOs \citep{munoz16,wangfy18,laha20}. As a type of MACHO, the abundance of primordial black holes is already loosely constrained using the CHIME catalog database \citep{zhouh22}. The constraints will be further improved as the FRB sample continues to grow. 

A search for lensed FRBs has been carried out with the 1st CHIME FRB catalog with no detection \cite{kader22}.  This posed a novel constraint on the abundance of primordial black holes \cite{leung22}. \citet{connor22} forecast the detection rates of gravitational lensing of FRBs with delay timescales from microseconds to years, corresponding to a wide range of the lens mass spanning fifteen orders of magnitude. 

\subsection{Neutron star equation of state}

The equation of state (EoS) close to or at the nuclear density is still poorly constrained. This leaves a large uncertainty in the neutron star (or even quark star) EoS \cite{lattimer07,li-ang20}. Some FRB observations may offer clue to the unknown EoS. For example, if the suggested GW190425/FRB 20190425A association is real, the production of FRB 20190425A would demand a relatively large neutron star maximum mass ($M_{\rm TOV}$), which would eliminate some EoSs \citep{moroianu22}. For another example, if FRB bursts carry information of NS crustal oscillations, with a large enough sample, one may offer a constraint on the neutron star EoS based on FRB burst morphology \citep[e.g.][]{wadiasingh20b}. 

\subsection{Fundamental physics: Weak equivalence principle, photon mass, and Lorentz invariance violation}

Thanks to their very short durations, FRBs have been also suggested as probes for fundamental physics because of the lack of spreading in time in contrast to the predictions of some theories. 

The first test is Einstein's weak equivalence principle (WEP), which states that all point-like structureless particles fall along the same path within a gravitational field. This is the foundation of the general theory of relativity, a geometric description of gravitation. According to this principle, photons with different energies from the same source should travel with the same trajectory with the same speed to reach the observer. In the parameterized post-Newtonian (PPN) description, the deviation from the WEP is the PPN parameter $\gamma$ deviating from 1. FRBs cannot be used to directly constrain $\gamma$, but can be used to test the difference of $\gamma$ values between two frequencies $\nu_1$ and $\nu_2$, which are usually the boundaries of the detection frequency band \citep{wei15}. Thanks to their large distances and short durations, one can constrain $\Delta \gamma$ with FRBs to be as small as  $10^{-15} - 10^{-20}$ \citep{wei15,tingay16,xing19,hashimoto21}. 

Another interesting constraint FRBs can offer is the photon mass \citep{wu16b,bonetti16}. If photons indeed have a non-zero rest mass, the lower-frequency photons (with a lower ``Lorentz factor'') should travel slightly slower than higher-frequency photons. The duration of an FRB therefore presents an absolute maximum delay due to such an effect. With a more sophisticated method by combining non-zero photon mass delay and the plasma dispersion delay (it turns out that the two dispersion relations have the similar forms with different normalization factors and slightly different $z$-dependences), a more stringent constraint can be reached with FRBs of known redshifts, especially with a sample of $z$-known FRBs using a Bayesian approach \citep{shao17}. The most stringent upper limit of the photon mass posed by FRBs already reached $m_\gamma \lesssim 5\times 10^{-48}$ g \citep{wu16b,bonetti16,shao17,xing19}.

Another widely discussed fundamental physics constraint is Lorentz invariance violation (LIV) due to the delay of high-energy photons as they travel through the foam-like space in very small scales. The effect is most significant at high-energies, so that short-duration GRBs are much more suitable to pose meaningful constraints than FRBs, which have very low photon energies.

\section{Problems and prospects}\label{sec:prospects}

The rapid progress in the FRB field is accompanied by many open questions, which continue to drive the field forward. We discuss three most pressing questions at the time when this review is written. 

\subsection{Do all FRBs repeat?}

This question is interesting from both observational and theoretical aspects. Observationally, it is much more difficult to prove that an FRB does NOT repeat than it does. If you have not detected a repeated burst from the source yet, it could well be that (1) it repeated but the telescope has missed it; (2) it repeated, but the burst is below the telescope sensitivity; or (3) it simply has not repeated and the waiting time is longer than the observing time. If one adopts the two sub-bursts of FRB 20200428 from the Galactic magnetar SGR 1935+2154 \citep{CHIME-SGR,STARE2-SGR} as one burst, then the source may not be regarded as a repeating FRB source yet (many repeated radio bursts from the source are not bright enough to be detected as FRBs at cosmological distances) even though we are certain that it should be an FRB repeater, because the magnetar source itself did not show significant difference before and after the FRB and there is no reason why the physical conditions to make FRB 20200428 would not be satisfied again to make another one. Theoretically, this question is very interesting because it is related to whether any of the cataclysmic FRB models are relevant. 

There have been great efforts in addressing this question. 1. From the observational side, although repeater bursts are found to display some interesting characteristics (e.g. longer duration, down-drifting subpulses, narrower spectra \cite{chime-repeaters}), there is still no definite clue to suggest that apparent non-repeaters are indeed different. 2. Machine-learning methods have been proposed to differentiate repeaters and non-repeaters \citep{chen22,luojw23,zhuge23}, and the results seem to suggest that most apparently non-repeating FRBs are indeed different from the repeating bursts. 3. A statistical study of the observational properties of repeaters and apparent non-repeaters suggested that there might be two populations \citep{zhong22}.  4. With limited data in the pre-CHIME era, arguments have been made that rFRB 20121102A is much more active than any other non-repeaters \citep{palaniswamy18,caleb19a} and so there might be two distinct classes (or at least two classes of repeaters with distinct activity levels). These arguments need to be revisited with the uniform, much larger database from CHIME. 5. One interesting test is to study the {\em observed} fraction of repeating sources from all FRBs, $F_{\rm r,obs}$. \citet{ai21} showed that if there indeed exist non-repeaters and if repeaters repeat forever, $F_{\rm r,obs}$ should approach a maximum after a certain observing time (when most repeaters are discovered) and then decline with time afterwards. However, uncertainties in the repetition rate and its distribution in repeaters make this criterion not clean. In some cases, the required time to reach the maximum is longer than astronomers' timescale (e.g. longer than 1000 yr). When the lifetime of the repeaters is considered, there is essentially no achievable maximum within astronomers' timescale. In any case, continuously monitoring $F_{\rm r,obs}$ may provide important clues to address this open question. The long-term CHIME observations seem to suggest a constant $F_{\rm r,obs}$ over time (Z. Pleunis, 2022, Cornell FRB workshop), which may suggest the existence of non-repeating FRBs.

In long terms, besides refining the above analyses with a much larger data set, a detection of an FRB robustly associated with a cataclysmic event (e.g. a gravitational wave event) would offer a strong support to the existence of these special types of FRBs. The plausible GW190425/FRB 20190425A association \cite{moroianu22} might be the first such case. Based on the event rate density arguments, these FRBs must only be a small fraction of all FRBs and may have some special properties. Another caveat is for individual cases, the robustness of the association must be addressed through various (e.g. temporal, spatial and distance) chance coincidence probabilities as well as theoretical arguments \cite{moroianu22}. The new population may be established only after a sample of such association events are detected. 

\subsection{Are there more than one class of repeating FRBs?}

This question actually has two aspects: First, observationally, do we see different clustering properties among the observed repeaters? Second, physically, are there more than one type of engine sources that power different repeaters? From the observational side, I'd argue that there have been already three types: 1. regular active repeaters in the cosmological distances (e.g. rFRB 20121102A, rFRB 20180916B, rFRB 20190520B, rFRB 20180301A, rFRB 20201124A), which have not been found in the Milky Way galaxy yet; 2. less energetic and less active magnetar repeaters such as SGR 1935+2154 that produced two sub-pulses in FRB 20200428; and 3. the globular cluster FRB 20200120E in M81, which has a high activity level but produces bursts with much lower luminosities than other cosmological active repeaters \citep{nimmo22}. Since the central source of the second type is already identified as a magnetar, the general trend in the community is to attribute all three observationally identified types to magnetars, with different evolutionary stages and probably different formation channels as well. For example, the first type (active cosmological repeaters) might be younger magnetars formed from recent supernova explosions and the third type may be magnetars produced from older formation channels such as WD-WD and NS-NS mergers or AIC of WDs. Even though this ``magnetars make them all'' hypothesis is theoretically attractive and passes some observational constraints, it nonetheless suffers from some drawbacks. For example, the detection of FRB 20200120E from the M81 globular cluster suggests that these systems are quite common \citep{lu22,kremer21}. This seems to be inconsistent with the fact that none of the 30 discovered magnetars from the Milky Way or LMC/SMC are associated with globular clusters. The fact that the CHIME DM distribution demands a dominant delayed population of FRBs with respect to star formation \citet{zhangzhang22}, see also \citet{qiang22,hashimoto22}, but see \citet{shin22}) also suggests that if magnetars do it all, the old-population magnetar channel should be the dominant one, in the contrary to the known magnetar population data. So, the current data may have already suggested the existence of other non-magnetar FRB engines. 

\subsection{FRB radiation mechanisms: where and how?}

Within the magnetar model of FRBs, there exist uncertainties regarding the location of the emission region (e.g. magnetospheres vs. relativistic shocks) and the radiation mechanism (bunched emission, plasma instabilities vs. vacuum maser mechanisms). As discussed in Sects.\ref{sec:magnetars}, \ref{sec:magnetosphere} and \ref{sec:shocks}, active studies and intense debates exist in the field, and growing evidence suggests that the magnetospheric origin is relevant for at least some FRBs. It remains unclear whether more than one emission site and more than one coherent mechanism are operating in FRBs. It is foreseen that the investigations in this direction will continue for years to come and the debates may not be settled in the near future, as the history of the study of the radiation mechanism of radio pulsars suggested. 

\subsection{Prospects}

In a young and rapidly growing field, it is fun to make predictions. The three authors of \citet{petroff19}, Emily Petroff, Jason Hessels, and Duncan Lorimer, made their respective predictions about the field in five years in the original review paper and also in their later updated review paper \citep{petroff22}. It is amusing to see that even though some of the predictions were realized, some unpredicted surprising discoveries were made within less than a three-year period since the first predictions were made. These include a periodically modulated FRB (rFRB 20180916B) with a 16-d period, a MJy low-luminosity FRB (FRB 20200428) from a Galactic magnetar, and a repeating low-luminosity FRB (rFRB 20200120E) from a globular cluster in M81. The FRB field seems to discourage conservative predictions. Just for fun, I close this review with ten predictions for the next 5-10 years.
\begin{enumerate}
    \item The detected FRB number will continue to grow rapidly, reaching $\sim 10^4$ different sources (including both non-repeating and repeating FRB populations) from the survey programs such as the CHIME FRB Project and reaching $\sim 10^4$ bursts from a few active individual sources from dedicated observational campaigns such as the FAST FRB Key Project.
    \item The FRB community will continue to grow and the numbers of papers and citations per year will keep rising for another 5-10 years.
    \item Surprises will continue to arrive, which will shake the FRB theoretical framework a few times before a standard paradigm is established. 
    \item X-ray counterparts of FRBs from nearby galaxies will be discovered, which are consistent with an SGR origin of FRBs.
    \item Despite active searches, prompt optical flashes in coincidence with FRBs will NOT be discovered, because of the intrinsic faintness of the prompt optical emission.
    \item Claims about the associations between a progenitor of a magnetar (e.g. a long GRB, a superluminous supernova, a short GRB, or a regular Type II supernova) and a repeating FRB source will be made, but a firm association cannot be established because of the uncertainties in chance coincidences.
    \item More claims about the associations between non-repeating FRBs and gravitational wave sources will be made, but the sample is not large and consistent enough to draw a definite conclusion. 
    \item More Galactic FRBs will be detected, most likely from SGR 1935+2154 or other magnetars, but also possible from sources other than magnetars, such as the Galactic center, young or old neutron stars, or even black hole binary systems.
    \item Multiple channels of repeating FRBs will be widely accepted. The ansatz that ``all FRBs repeat'' still cannot be completely ruled out.
    \item Debates on the physical mechanisms of FRBs will continue among theorists, not only because ``a competent theorist can make any model to match any observational data'', but also because there might be indeed several physically plausible mechanisms that operate together.
\end{enumerate}

\begin{acknowledgments}
I thank Shunke Ai, Matthew Bailes, Edo Berger, Andrei Beloborodov, Paz Beniamini, Shivani Bhandari, Roger Blandford, Gabriele Bruni, Manisha Caleb, Shami Chatterjee, Connery Chen, Xuelei Chen, Liam Connor, Jim Cordes, Zi-Gao Dai, Wei Deng, Yi Feng, Bryan Gaensler, He Gao, Jin-Lin Han, Jason Hessels, Kunihito Ioka, Clancy James, Jin-Chen Jiang, Vicky Kaspi, Jonathan Katz, Kyle Kremer, Shri Kulkarni, Pawan Kumar, Sibasish Laha, Dong Lai, Casey Law, Kejia Lee, Di Li, Dongzi Li, Ye Li, Zhengxiang Li, Lin Lin, Duncan Lorimer, Wenbin Lu, Jia-Wei Luo, Rui Luo, Yuri Lyubarsky, Yun-Peng Men, Brian Metzger, Alex Moroianu, Khota Murase, Chen-Hui Niu, Jia-Rui Niu, Divya Palaniswamy, Fiona Panther, Ue-Li Pen, Emily Petroff, Sterl Phinney, Luigi Piro, Yuanhong Qu, Vikram Ravi, Lorenzo Sironi, Shriharsh Tendulkar, Chris Thompson, Tomoki Wada, Zorawar Wadiasingh, Fa-Yin Wang, Pei Wang, Wei-Yang Wang, Xiang-Gao Wang, Linqing Wen, Xue-Feng Wu, Zi-Wei Wu, Shaolin Xiong, Heng Xu, Ren-Xin Xu, Siyao Xu, Yuan-Pei Yang, Wenfei Yu, Bin-Bin Zhang, Chun-Feng Zhang, Rachel C. Zhang, Yong-Kun Zhang, Shuang-Nan Zhang, De-Jiang Zhou, Wei-Wei Zhu, Jia-Ming Zhu-Ge, and many others for stimulative collaborations or discussion on various subjects included in this review. Special thanks are due to Jonathan Katz and another anonymous referee for very careful reviews, Veronique Van Elewyck and Debbie Brodbar for editorial comments, as well as Sterl Phinney and the Caltech FRB theory group (Liam Connor, Dongzi Li, Kyle Kremer, Nadine Soliman and Nicholas Rui), Yuan-Pei Yang, Kunihito Ioka, Brian Metzger, Andrei Beloborodov, Yuanhong Qu, Edo Berger, Zorawar Wadiasingh, Navin Sridhar, Emanuele Sobacchi, Eduardo Vitral, Ranjan Laha, and Stefano Covino for providing numerous comments to improve this review. 
\end{acknowledgments}

\appendix

\section{Acronyms}

\begin{equation}
\begin{array}{ll}
{\rm AGN} & {\rm active~galactic~nuclei} \\
{\rm ALAE} & {\rm amplified~linear~ acceleration~ emission} \\
{\rm ASKAP} & {\rm Australian~square~kilometre~array} \\
& {\rm pathfinder} \\
{\rm AXP} & {\rm anomalous~ X-ray ~pulsar} \\
{\rm BAO} & {\rm baryon~acoustic~ oscillations} \\
{\rm BH} & {\rm black~hole} \\
{\rm CBC} & {\rm compact~binary~coalescence} \\
{\rm cCBC} & {\rm charged~compact~binary~coalescence} \\
{\rm CETI} & {\rm communicative~ extraterrestrial~intelligence} \\
{\rm CGM} & {\rm circum-galactic~medium} \\
{\rm CHIME} & {\rm Canadian~hydrogen~intensity~mapping} \\
& {\rm experiment} \\
{\rm CM} & {\rm conversion~measure} \\
{\rm CMB} & {\rm cosmic~microwave~background} \\
{\rm DM} & {\rm dispersion~measure} \\
{\rm DSA} & {\rm deep~synoptic~array} \\
{\rm DSR} & {\rm Dicke's ~superradiance} \\
{\rm EM} & {\rm emission~measure} \\
& {\rm or:~ electromagnetic} \\
{\rm FAST} & {\rm five-hundred-meter ~ aperture} \\
& {\rm spherical~telescope} \\
{\rm FEL} & {\rm free~electron~laser} \\
{\rm FRB} & {\rm fast~radio~burst} \\
{\rm GJ} & {\rm Goldreich-Julian} \\
{\rm GRB} & {\rm gamma-ray~burst} \\
{\rm GP} & {\rm giant ~pulse} \\
{\rm GW} & {\rm gravitational~wave} \\
{\rm ICS} & {\rm inverse~ Compton ~scattering} \\
{\rm IGM} & {\rm intergalactic~medium} \\
{\rm ISM} & {\rm interstellar~medium} \\
{\rm LGRB} & {\rm long~gamma-ray~burst} \\
{\rm LIV} & {\rm Lorentz ~invariance~violation } \\
{\rm LMC} & {\rm large~Magellanic~cloud} \\
{\rm L-mode} & {\rm left~mode} \\
{\rm LSD} & {\rm large~superconducting~dipole} \\
{\rm MACHO} & {\rm massive~compact~halo~object} \\
{\rm MW} & {\rm Milky~Way} \\
{\rm MWN(e)} & {\rm magnetar~wind~nebula(e)} \\
{\rm O-mode} & {\rm ordinary~mode} \\
{\rm PA} & {\rm polarization~angle} \\
{\rm PIC} & {\rm particle-in-cell} \\
{\rm PPN} & {\rm parameterized~ post-Newtonian} \\
{\rm PRS} & {\rm persistent~radio~source} \\
{\rm PWN(e)} & {\rm pulsar~wind~nebula(e)} \\
{\rm NS} & {\rm neutron~star} \\
{\rm rFRB} & {\rm repeating~fast~radio~burst} \\
{\rm RM} & {\rm rotation~measure} \\
{\rm R-mode} & {\rm right~mode} \\
{\rm RRAT} & {\rm rotating~radio~transients} \\
{\rm SGR} & {\rm soft~ gamma-ray~ repeater} \\
{\rm SGRB} & {\rm short~gamma-ray~burst} \\
{\rm SLSN(e)} & {\rm superluminous~supernova(e)} \\
{\rm SM} & {\rm scattering~measure} 
\end{array}
\nonumber
\end{equation}
\begin{equation}
\begin{array}{ll}
{\rm SMBH} & {\rm supermassive~black~hole} \\
{\rm SMC} & {\rm small~Magellanic~cloud} \\
{\rm SMNS} & {\rm supramassive~neutron~star} \\
{\rm SN(e)} & {\rm supernova(e)} \\
{\rm SNR} & {\rm supernova~remnant} \\
{\rm STARE2} & {\rm survey~for~transient~astronomical~radio} \\
& {\rm emission~2} \\
{\rm ULX} & {\rm ultraluminous~ X-ray} \\
{\rm VLA} & {\rm very~large~array} \\
{\rm WD} & {\rm white~dwarf} \\
{\rm WEP} & {\rm weak~equivalence~principle} \\
{\rm X-mode} & {\rm extraordinary~mode} \\
{\rm XRB} & {\rm X-ray~burst} \\
\end{array}
\nonumber
\end{equation}


\input{ms.bbl}

\end{document}

%% file: ms.bbl
%